\documentclass[11pt]{article}
\usepackage[margin=1in]{geometry}
\usepackage{xcolor}
\definecolor{lightblue}{HTML}{2970CC}
\definecolor{lightpurple}{HTML}{673147}
\definecolor{ForestGreen}{HTML}{FF5733}
\usepackage[colorlinks=true,linkcolor=lightblue,urlcolor=lightblue,citecolor=lightblue,breaklinks]{hyperref}
\usepackage[font=small,labelfont=bf]{caption}
\usepackage[numbers, sort&compress]{natbib}
\usepackage{color}
\usepackage{enumerate}
\usepackage{enumitem}
\usepackage{graphicx}
\usepackage{varwidth}
\usepackage{mathrsfs}
\usepackage{mathtools}
\usepackage{subcaption}
\usepackage{overpic}
\usepackage{multirow}
\usepackage{multicol}
\usepackage{amsmath,amsthm,amssymb,colonequals,etoolbox}
\usepackage{amsfonts}       
\usepackage{thmtools}
\usepackage{url}
\usepackage[capitalise,noabbrev]{cleveref}
\usepackage[T1]{fontenc}    
\usepackage{nicefrac}       
\usepackage{microtype}      
\usepackage{xcolor}         
\usepackage{duckuments}
\usepackage{booktabs}
\usepackage{makecell}
\usepackage{algorithm}
\usepackage{algorithmic}
\usepackage{wrapfig}
\usepackage{parskip}
\usepackage{authblk}

\DeclareMathOperator*{\argmin}{argmin}

\newcommand{\R}{\mathbb{R}}
\newcommand{\E}{\mathbb{E}}
\newcommand{\N}{\mathsf{N}}

\newcommand{\T}{\mathsf{T}}

\newcommand{\rev}{\mathsf{R}}

\newcommand{\calP}{\mathcal{P}}
\newcommand{\calL}{\mathcal{L}}
\newcommand{\Lsm}{\calL_{\text{sm}}}
\newcommand{\Ld}{\calL_{\text{d}}}
\newcommand{\sdottot}{\dot{s}_{\text{tot}}}
\newcommand{\sdotm}{\dot{s}_{\text{m}}}
\newcommand{\sdotsys}{\dot{s}_{\text{sys}}}
\renewcommand{\div}{\nabla\cdot}
\newcommand{\divx}{\nabla_x\cdot}
\newcommand{\divg}{\nabla_g\cdot}
\newcommand{\divgi}{\nabla_{g^i}\cdot}
\newcommand{\divxi}{\nabla_{x^i}\cdot}
\newcommand{\divi}{\nabla_{i}\cdot}
\newcommand{\vgi}{v_{g}^i}
\newcommand{\vx}{v_{x}}
\newcommand{\vg}{v_{g}}
\newcommand{\vxi}{v_{x}^i}
\newcommand{\vi}{v^i}
\newcommand{\calO}{\mathcal{O}}

\renewcommand{\paragraph}[1]{\textbf{#1}}
\newcommand{\rt}{r_t}
\newcommand{\Rt}{R_t}
\newcommand{\xt}{x_t}
\newcommand{\gt}{g_t}
\newcommand{\drt}{\dot r_t}
\newcommand{\dRt}{\dot  R_t}
\newcommand{\dxt}{\dot x_t}
\newcommand{\dgt}{\dot g_t}
\newcommand{\dt}{\Delta t}
\newcommand{\Id}{Id}
\newcommand{\domR}{\Omega}
\newcommand{\newchange}[1]{#1}

\DeclareMathOperator{\attn}{Attention}

\DeclareMathOperator{\multihead}{MultiHead}
\DeclareMathOperator{\head}{head}
\DeclareMathOperator{\softmax}{softmax}
\DeclareMathOperator{\concat}{Concat}
\DeclareMathOperator{\layernorm}{LayerNorm}
\DeclareMathOperator{\softplus}{softplus}

\newtheorem{theorem}{Theorem}[section]

\newtheorem{proposition}[theorem]{Proposition}

\title{Deep learning probability flows and\\ entropy production rates in active matter}
\author{Nicholas M.~Boffi}
\author{Eric Vanden-Eijnden}
\affil{Courant Institute of Mathematical Sciences}

\begin{document}
\maketitle

\begin{abstract}
Active matter systems, from self-propelled colloids to motile bacteria, are characterized by the conversion of free energy into useful work at the microscopic scale. 
They involve physics beyond the reach of equilibrium statistical mechanics, and a persistent challenge has been to understand the nature of their nonequilibrium states. The entropy production rate and the probability current provide quantitative ways to do so by measuring the breakdown of time-reversal symmetry. 
Yet, their efficient computation has remained elusive, as they depend on the system's unknown and high-dimensional probability density. 
Here, building upon recent advances in generative modeling, we develop a deep learning framework to estimate the score of this density. 
We show that the score, together with the microscopic equations of motion, gives access to the entropy production rate, the probability current, and their decomposition into local contributions from individual particles. 
To represent the score, we introduce a novel, spatially-local transformer network architecture that learns high-order interactions between particles while respecting their underlying permutation symmetry. 
We demonstrate the broad utility and scalability of the method by applying it to several high-dimensional systems of active particles undergoing motility-induced phase separation (MIPS). 
We show that a single network trained on a system of 4096 particles at one packing fraction can generalize to other regions of the phase diagram, including systems with as many as 32768 particles. We use this observation to quantify the spatial structure of the departure from equilibrium in MIPS as a function of the number of particles and the packing fraction.

\end{abstract}

\maketitle
Active matter systems are driven out of equilibrium by a continuous injection of energy at the microscopic scale of the constituent particles~\cite{marchetti_hydrodynamics_2013,fodor_irreversibility_2022,obyrne_time_2022}.
The nonequilibrium nature of their dynamics manifests itself in the breakdown of time-reversal symmetry (TRS), which can be quantified by the global rate of entropy production (EPR)~\cite{lebowitz_gallavotticohen-type_1999, parrondo_entropy_2009,jarzynski_equalities_2011, seifert_stochastic_2012}, and by the presence of  probability currents at statistical steady state~\cite{bodineau_cumulants_2007,battle_broken_2016,bertini_macroscopic_2015}.
Despite their wide recognition as quantities of fundamental importance, computing either the global EPR or the magnitude of the probability current has remained a long-standing challenge. 
At a fundamental level, both are defined via the microscopic density for the system~\cite{seifert_entropy_2005,gomez-marin_footprints_2008}, which is generically unknown outside of a few simplistic cases due to its high-dimensionality and its complexity~\cite{pietzonka_entropy_2018}.

The global EPR can in principle be computed directly from the microscopic equations of motion~\cite{fodor_how_2016,martin_statistical_2021,speck_active_2018, shankar_hidden_2018, dadhichi_origins_2018} by making use of the Crooks fluctuation theorem~\cite{crooks_entropy_1999}. 
However, this leads to a single number, which fails to quantify where TRS breaks down spatially in the system, and fails to reveal which particles are responsible.
This issue can be addressed for active matter field theories, where a similar approach leads to a \textit{local}, spatially-dependent definition of the EPR~\cite{nardini_entropy_2017, borthne_time-reversal_2020}, but only after a coarse-graining of the microscopic dynamics.
In general, methods based on the Crooks fluctuation theorem require a suitable definition of a time-reversed dyamics, which has been debated in the literature~\cite{chetrite_fluctuation_2008,mandal_entropy_2017,caprini_comment_2018}.
The use of a time reversal can be avoided via the stochastic thermodynamics definition of the \newchange{``entropy of the system''}~\cite{seifert_entropy_2005, seifert_stochastic_2012}, but doing so requires the logarithm of the system's microscopic density, which is unavailable outside of the simplest cases.
An orthogonal approach makes use of data compression algorithms to compute the global~\cite{martiniani_quantifying_2019} or local EPR~\cite{ro_model-free_2022}, but these methods are only valid asymptotically in the limit of infinite system size, and it is difficult to understand what they compute away from this limit.
Several methods have also been developed to infer a global measure of the probability current~\cite{li_quantifying_2019,otsubo_estimating_2020,otsubo_estimating_2022}, but thus far have been restricted to low-dimensional systems.
For a detailed coverage on the use of the EPR and steady-state currents to quantify nonequilibrium effects in active matter, we refer the reader to~\cite{fodor_irreversibility_2022}.

Here, building upon recent advances in generative modeling~\cite{hyvarinen_estimation_2005, song_score-based_2021,albergo_building_2023,albergo_stochastic_2023,boffi_probability_2023}, we tackle the challenging problem of estimating spatially-local probability currents and entropy production rates directly from their microscopic definitions.
To this end, we develop a machine learning method that estimates the gradient of the logarithm of the system's probability density, which can be characterized as a solution to the many-body stationary Fokker-Planck equation (FPE).
This quantity, known as the score function~\cite{hyvarinen_estimation_2005, song_score-based_2021}, enters the definition of both the probability current and the EPR.
We show how the method naturally decomposes the global EPR or probability current into microscopic contributions from the individual particles and their degrees of freedom, which enables us to identify spatial structure in the breakdown of TRS. 
To validate the accuracy of the learned solution, we develop diagnostics based on invariants of the stationary FPE that can be verified \textit{a-posteriori}.

\begin{figure*}[!t]
    \centering
    \includegraphics[width=\textwidth]{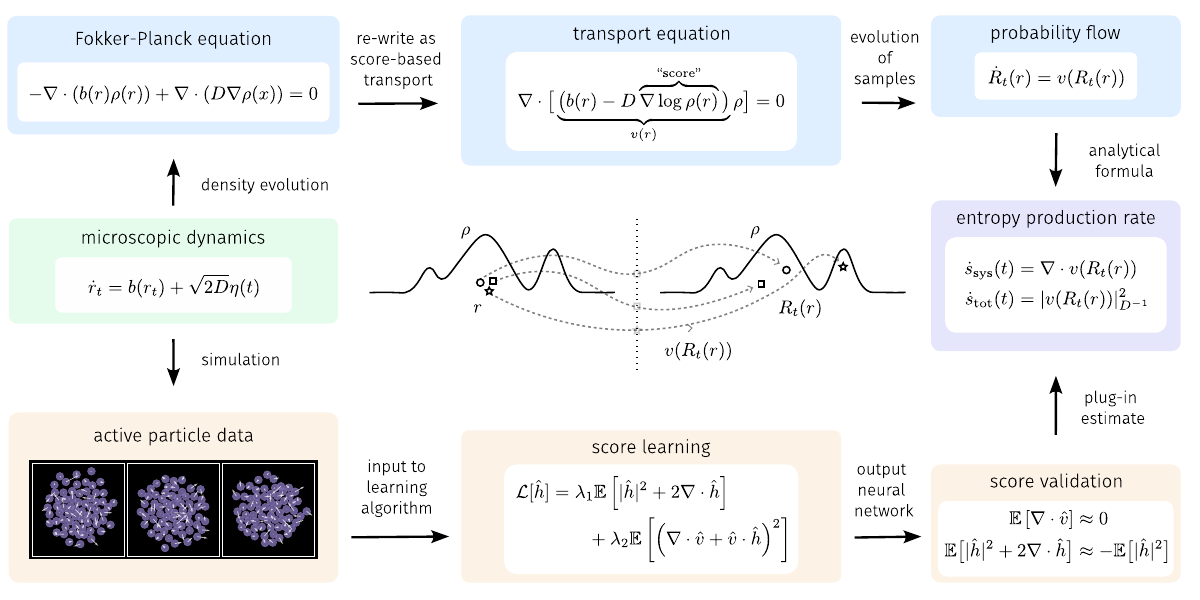}
    \caption{\textbf{Method overview.} (Green) The starting point for our approach is a microscopic dynamics describing the evolution of a set of interacting active particles.
    (Purple) The target is estimation of several definitions of the entropy production rate of the system, which we will accomplish by means of the probability flow.
    (Blue) Mathematically, our approach is built on viewing the system from the perspective of dynamical transport of measure. The microscopic stochastic dynamics induces a Fokker-Planck equation for a high-dimensional density describing the configuration of the system. This Fokker-Planck equation is equivalent to a transport equation that depends on the unknown ``score'' $\nabla\log\rho$ of the solution. The characteristics of this equation obey a probability flow ordinary differential equation, which gives immediate access to the entropy production rate.
    (Center) Illustration of nonequilibrium transport of measure at stationarity.
    (Orange) Algorithmically, our method approximates the unknown score by machine learning over a dataset of microscopic particle data. The learned approximation can be validated \textit{a-posteriori} by checking invariants of the stationary Fokker-Planck equation, and can be plugged in directly to the definition of the entropy production rate to obtain an estimate.}
    \label{fig:method}
\end{figure*}

We apply the method to several model systems involving active swimmers: two swimmers on the torus, where we can visualize the EPR and the probability flow across the entire phase space, a system of $64$ swimmers in a harmonic trap, and a system of $4096$ swimmers undergoing motility-induced phase separation (MIPS)~\cite{cates_motility-induced_2015}.
For the MIPS system, we learn using a novel spatially-local architecture that does not depend on the total number of swimmers, and we show that it can be extended to systems of up to $32,768$ swimmers at values of the packing fraction that differ from those seen during training.
Despite the high-dimensionality of these latter examples, our approach provides us with a microscopic description of both the current and the local EPR.
Importantly, this enables us to visualize the contributions of the individual particles directly without any need for averaging.
We use this property to confirm theoretical predictions about the spatial features of entropy production in MIPS, such as concentration on the interface between the dilute and condensed phases~\cite{nardini_entropy_2017,ro_model-free_2022}.
Our \textbf{main contributions} can be summarized as:
\begin{enumerate}[leftmargin=0.2in]
    \item We revisit the framework of stochastic thermodynamics and show how signatures of nonequilibrium behavior and lack of time-reversibility, such as the probability current and the EPR, can be related to the score of the system's stationary probability density. 
    \item We show how to use machine learning tools from the field of generative modeling to estimate the score function from microscopic data (Figure~\ref{fig:method}). To approximate this high-dimensional function accurately, we develop a new transformer neural network architecture that incorporates spatial locality and permutation symmetry. This enables transferability to systems with differing numbers of particles or packing fractions than seen at training.
    \item We illustrate the usefulness of the approach on systems involving active particles undergoing MIPS, where we show that the method can quantify the EPR at the individual particle level as a function of the activity, number of particles, or packing fraction. We confirm that entropy is dominantly produced at the interface between the cluster and the gas.
\end{enumerate}

These contributions continue in a line of work that seeks to apply methods based on machine learning to high-dimensional problems in scientific computing~\cite{noe_boltzmann_2019,khoo_solving_2018,khoo_solving_2020,raissi_physics-informed_2019,bar-sinai_learning_2019}, applied mathematics~\cite{gabrie_adaptive_2022, boffi_probability_2023,brunton_discovering_2016,lusch_deep_2018,rudy_data-driven_2017,nusken_solving_2021, e_algorithms_2021, e_deep_2018,kovachki_neural_2023,li_fourier_2020}, and the physical sciences~\cite{carleo_machine_2019,brunton_machine_2020,supekar_learning_2023,karniadakis_physics-informed_2021,kochkov_machine_2021}. 
In particular, considerable research effort has been spent designing machine learning methods to compute solutions of the many-body Schr\"odinger equation~\cite{carleo_solving_2017, hermann_deep-neural-network_2020,qiao_orbnet_2020,pfau_ab-initio_2020}; our work can be seen as an extension of this research effort to classical statistical mechanics and stochastic thermodynamics.

\section*{Stochastic thermodynamics}
\begin{figure*}[t!]
\centering
\begin{tabular}{c}
\multicolumn{1}{c}{%
\begin{overpic}[width=0.85\textwidth]{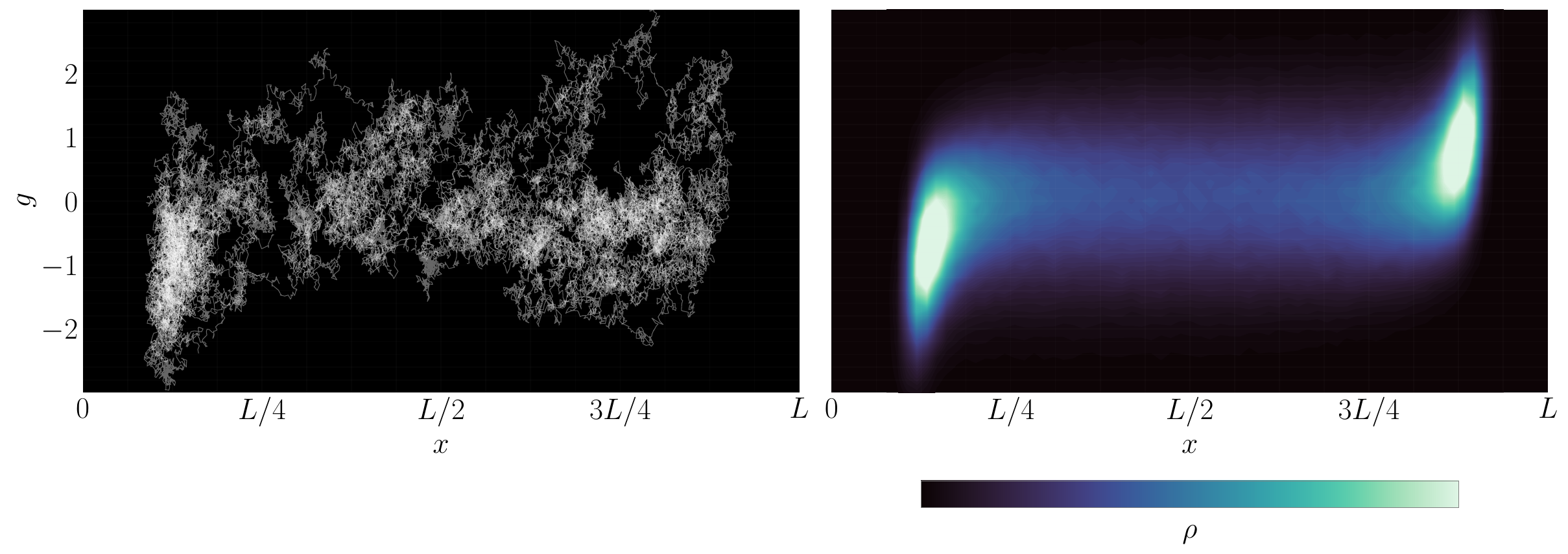}
\put(3, 36){\textbf{A}}
\put(53, 36){\textbf{B}}
\end{overpic}}\\\\
\begin{overpic}[width=0.85\textwidth]{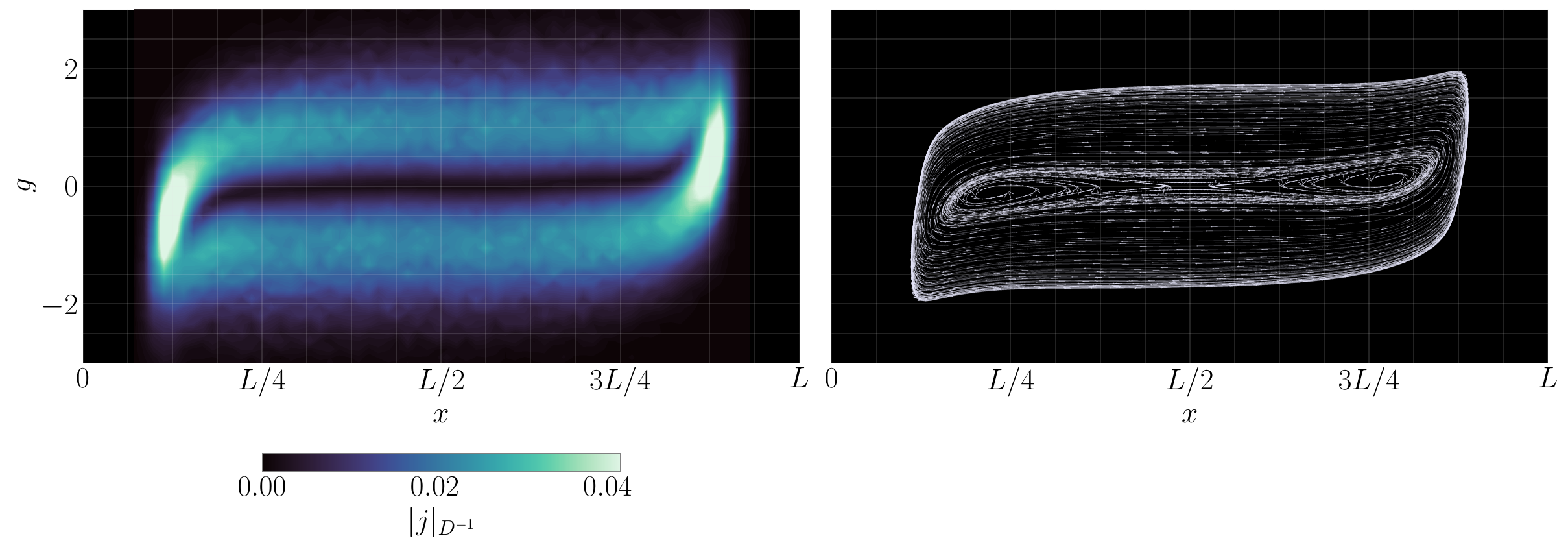}
\put(3, 35.5){\textbf{C}}
\put(53, 35.5){\textbf{D}}
\end{overpic}
\end{tabular}
\caption{\textbf{Stochastic dynamics and probability flows.} (A) Individual stochastic trajectories of~\eqref{eqn:interacting_particles} for $N=2$ and $d=1$ in the variables $x_t = x^2_t - x^1_t$ and $g_t = g^2_t - g^1_t$, with periodic boundary conditions on $[0,L]$. The trajectories $(x_t,g_t)$ tend to accumulate in two clusters corresponding to situations where particle 1 is just in front of particle 2 or vice-versa. This occurs because one particle catches up to the other in a typical trajectory (since either $|g_t^1|>|g_t^2|$ or $|g_t^1|<|g_t^2|$), but does not pass over it due to the short-range repulsive force between them. Random transitions between these modes occur when the magnitudes of $|g_t^1|$ and $|g_t^2|$ change order.
(B) Stationary probability density function $\rho$ of $(x_t, g_t)$ confirming the metastability observed in (A): $\rho$ is the solution of the stationary FPE~\eqref{eqn:fpe}.
(C) Visualization of the (diffusion-weighted) norm of the probability current $j$, defined in~\eqref{eqn:current}, over the phase space. The current is concentrated in the two modes, but is also nonzero along transition pathways between them.
(D) Phase portrait of the probability flow~\eqref{eqn:pflow}. Similar to the stochastic trajectories in (A), the flow lines preserve the density $\rho$ in (B), but are deterministic and interpretable, highlighting limit cycles within and between the two clusters. A movie of these limit cycles in a frame with one particle fixed is available \href{https://www.dropbox.com/scl/fi/q2ryj5u9nfl0yk3ovw3u9/pflow_movie.mp4?rlkey=98cqnwl72xa6vvq2dbqyhcxy0&dl=0}{at this link}.}
\label{fig:two_particle_intro}
\end{figure*}

\paragraph{Active swimmers.}
As an application of our approach, we will consider a suspension of $N$ self-propelled particles in $d=1$ or $d=2$ dimensions with translational degrees of freedom $\xt^i \in \R^{d}$ and orientational degrees of freedom $\gt^i \in \R^{d}$.
Their dynamics is given by the so-called active Ornstein-Uhlenbeck, or Gaussian colored-noise model~\cite{paoluzzi_critical_2016,szamel_self-propelled_2014,flenner_nonequilibrium_2016, caprini_entropy_2019,martin_statistical_2021,mandal_entropy_2017}:
\begin{equation}
  \label{eqn:interacting_particles}
  \begin{aligned}
    \dxt^i &= \mu\sum_{j\not=i} f(\xt^i-\xt^j) + v_{0} \gt^i + \sqrt{2\epsilon}\,\eta_x^i(t),\\
    \dgt^i &= -\gamma \gt^i + \sqrt{2\gamma}\,\eta_g^i(t).
  \end{aligned}
\end{equation}
In~\eqref{eqn:interacting_particles}, $\mu$ is the mobility, $f(x)$ is a short-range repulsive potential force whose specific form will be specified later, and $v_{0}\ge0$ is the self-propulsion speed of the particles.
$\eta^i_x(t)$ and $\eta^i_g(t)$ are independent white-noise sources, i.e., Gaussian processes with mean zero and covariances given by $\langle\eta_x^i(t)\eta_x^j(t') \rangle=\langle\eta_g^i(t)\eta_g^j(t') \rangle= \delta(t-t') \delta_{i,j} \Id$.
\newchange{The parameter} $\epsilon\ge0$ sets the scale of the thermal noise \newchange{(and need-not be small)}, while $\gamma>0$ tunes the persistence timescale of the self-propulsion. 
The orientational degrees of freedom $g_t^i$ introduce an active noise term with a finite correlation time $1/\gamma$ into the translational dynamics for $r_t^i$; the presence of this correlated noise drives the system out of equilibrium for any $v_0\not=0$ and $\gamma<\infty$.

Trajectories of~\eqref{eqn:interacting_particles} are shown in Fig.~\ref{fig:two_particle_intro}A, where we consider $N=2$ particles in dimension $d=1$ on the interval $[0,L]$ with periodic boundary conditions.
By translation invariance, we can define $x = x^2 - x^1$ and $g = g^2 - g^1$ to reduce dimensionality, which allows us to visualize the entire phase space.
We will use this low-dimensional system as a running illustrative example, while our main results consider~\eqref{eqn:interacting_particles} in higher-dimensional situations with up to $N=32,768$ particles in $d=2$ dimensions.

\paragraph{General microscopic description.}
Since the tools that we introduce to study~\eqref{eqn:interacting_particles} are transportable to other nonequilibrium systems, it is convenient to view these equations as an instance of the generic stochastic differential equation (SDE) for $\rt\in \domR$
\begin{equation}
  \label{eqn:microscopic_stochastic_general}
  \drt  = b(\rt) + \sqrt{2D}\, \eta(t),
\end{equation}
where $b(r)$ denotes the deterministic drift, $D$ denotes the diffusion tensor (assumed to be symmetric and positive semi-definite but not necessarily invertible), and $\eta(t)$ is a white noise process. 
Eq.~\eqref{eqn:interacting_particles} can be cast into the form of~\eqref{eqn:microscopic_stochastic_general} by setting $\rt =(\rt^1,\ldots, \rt^N)$ with $\rt^i = (\xt^i,\gt^i)\in \R^{2d}$ for $i=1,\ldots,N$ (so that $\domR = \R^{2Nd}$), along with proper identification of $b(r)$ and $D$.
For simplicity, we focus on drifts $b(r)$ that are independent of time, along with diffusion tensors $D$ that are constant in both space and time.
Importantly, we study systems that may not respect detailed-balance, so that $b(r) \neq - D\nabla U(r)$ for some potential $U(r)$. 
 
\paragraph{Many-body Fokker-Planck equation.}
The probability density function $\rho_t$ of the solution $\rt$ to \eqref{eqn:microscopic_stochastic_general} satisfies a many-body Fokker-Planck Equation (FPE) that can be written as a continuity equation
\begin{equation}
    \label{eqn:fpe}
     \partial_t \rho_t(r) + \nabla \cdot j_t(r) = 0,
\end{equation}
where we have defined the probability current $j_t(r)$
\begin{equation}
    \label{eqn:current}
     j_t(r) = b(r)\rho_t(r)-D\nabla\rho_t(r).
\end{equation}
We study systems that have reached statistical steady state, so that $\rho_t(r) = \rho(r)$ and $j_t(r) = j(r)$.
Then~\eqref{eqn:fpe} reduces to $\div j(r) = 0$ with $j(r) = b(r)\rho(r)-D\nabla\rho(r)$.
Since we do not assume that the system is in detailed-balance, its stationary density $\rho$ and current $j$ are in general unknown.
In particular, the system can sustain a nonequilibrium stationary current $j \neq 0$.
We visualize the stationary density $\rho$ and the steady-state current $j$ for our low-dimensional illustrative system in Figures~\ref{fig:two_particle_intro}B and~\ref{fig:two_particle_intro}C, respectively.

\paragraph{Current velocity and probability flow.}
At stationarity, assuming that $\rho(r)>0$ everywhere in $\Omega$, we may re-write~\eqref{eqn:fpe} as a time-independent transport equation 
\begin{equation}
  \label{eqn:transport}
  0 = \nabla\cdot\left(v(r)\rho(r)\right) ,
\end{equation}
where $v$ is the current (aka mean local) velocity field~\cite{nelson1967dynamical,seifert_entropy_2005} defined as
\begin{equation}
  \label{eqn:pflow_vel}
  v(r) = j(r)/\rho(r) = b(r) - D\nabla\log\rho(r).
\end{equation}
The current velocity $v$ is a fundamental object, and we will show that various definitions of the EPR can be computed from it~\citep{aurell_refined_2012, chennakesavalu_unified_2023}.
It contains strictly more information than $\rho$ alone, because it is always possible to construct an equilibrium system with the same $\rho$.
Calculation of the EPR requires access to the steady state currents captured by $v$, which arise through an interplay between both the system's stationary density and structural information about its dynamics.

To gain access to $v$ without explicit knowledge of $\rho$, we will develop a learning algorithm that estimates  the high-dimensional $\nabla\log\rho$ from data from the SDE in~\eqref{eqn:microscopic_stochastic_general}: $\nabla\log\rho$ is known as the Hyv\"arinen ``score'' function in the machine learning literature~\cite{hyvarinen_estimation_2005}.
In addition to enabling the computation of various definitions of the EPR, $v$ allows us to directly interrogate the flow of probability in the system.
To do so, we may study the characteristics of~\eqref{eqn:transport} via solution of the ordinary differential equation (ODE)
\begin{equation}
  \label{eqn:pflow}
    \dRt(r) = v(\Rt(r)), \qquad R_{t=0}(r) = r.
\end{equation}
We refer to~\eqref{eqn:pflow} as the probability flow equation, as it describes the transport of samples in phase space according to the probability current $j$. In particular, at stationarity, the flow map $R_t(r)$ leaves the density $\rho$ invariant, so that the density of $R_t(r)$ is $\rho$ \newchange{when $r$ is drawn randomly from $\rho$.}
This means that for any observable $A(r)$, we have
\begin{equation}
  \label{eqn:pflow:cons}
   \forall t \in \R \ : \quad  \int_{\domR} A(\Rt(r)) \rho(r) dr = \int_{\domR} A(r) \rho(r) dr.
\end{equation}
We stress that for $j\neq 0$, transport can occur even at stationarity; the condition in~\eqref{eqn:pflow:cons} ensures that this transport preserves $\rho$.
We visualize the phase portrait of~\eqref{eqn:pflow} for our low-dimensional illustrative example in Figure~\ref{fig:two_particle_intro}D.
The resulting ordered limit cycles may be contrasted with trajectories of the equivalent stochastic dynamics~\eqref{eqn:microscopic_stochastic_general} in Figure~\ref{fig:two_particle_intro}A; despite their striking qualitative differences, both leave $\rho$ invariant.

\section*{Entropy production rates}
In this work, we are primarily interested in nonequilibrium systems, and we will study how their nonequilibrium dynamics arises spatially from nonzero $v$ and $\div v$.
To this end, we now relate $\div v$ and $|v|^2$ to several definitions of the EPR.

\paragraph{Gibbs entropy and system EPR.}
\begin{figure}[t!]
\centering
    \includegraphics[width=0.6\linewidth]{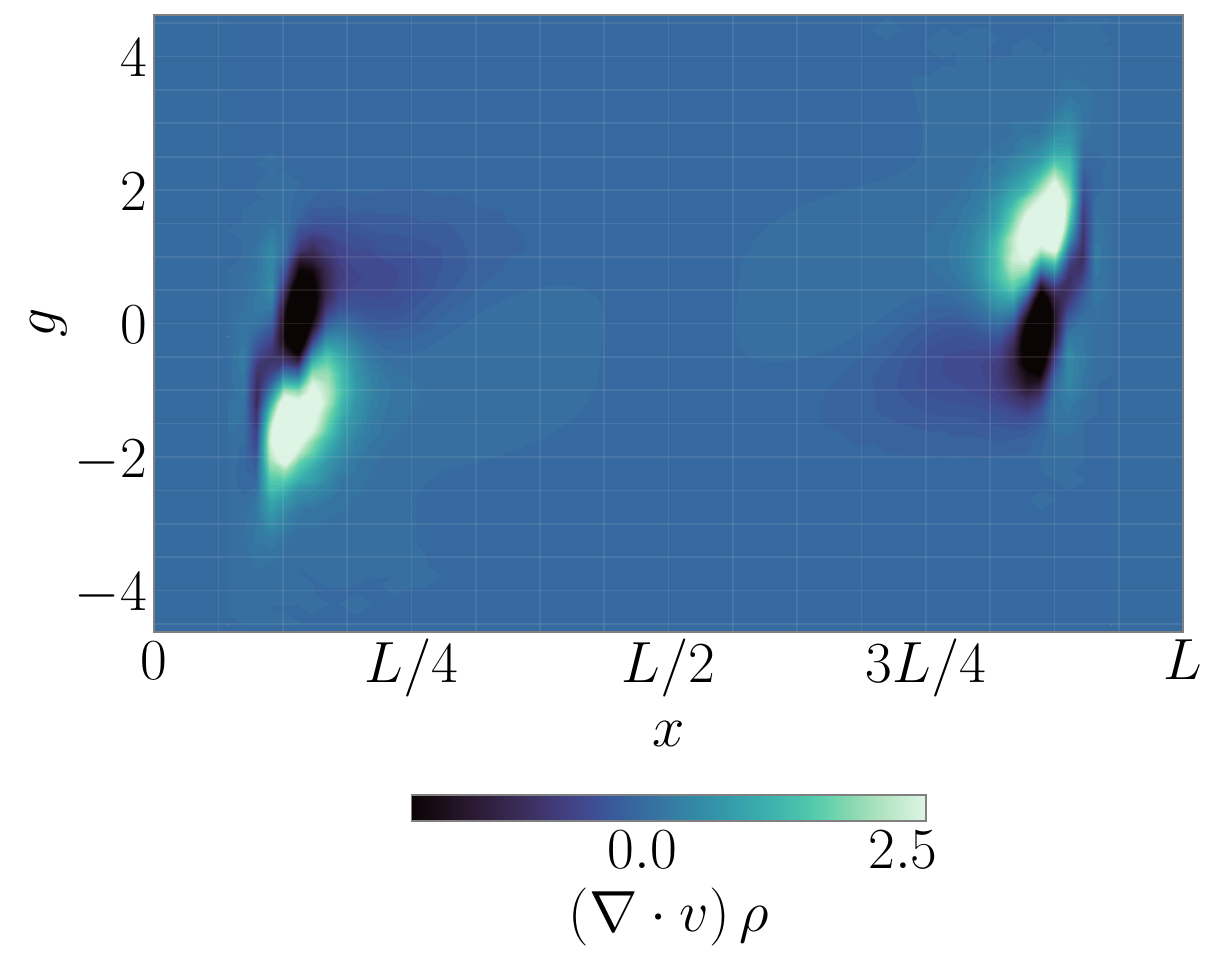}
    \caption{\textbf{System EPR.} Visualization of $\div v(r)$ across the phase space for~\eqref{eqn:interacting_particles} with $N=2$ and $d=1$ in the variables $x_t = x_t^2 - x_t^1$ and $g_t = g_t^2 - g_t^1$.
    The system EPR along a trajectory $R_t(r)$ of the probability flow~\eqref{eqn:pflow} can be written as $\sdotsys(t) = \div v(R_t(r))$, so that $\div v(r)$ gives insight into how entropy is generated locally by the system.
    Even though $\E_\rho[\div v]=0 $ at stationarity, $\div v \neq 0$ pointwise when the system is out of equilibrium.
    Here, system entropy is locally produced when the two particles collide, and released when they separate.}
    \label{fig:twop_div_v}
\end{figure}
\newchange{Given the stationary solution $\rho$ to~\eqref{eqn:fpe}, the Gibbs entropy of the system is defined as
\begin{equation}
\label{eqn:Gibbs}
 S_{\text{sys}} = -\int_{\domR} \log \rho(r) \rho(r) dr.
\end{equation}
At stationarity, $S_{\text{sys}}$ is time-independent, and hence must be preserved by the dynamics.
To see how this occurs at the level of the individual degrees of freedom, following Seifert~\cite{seifert_entropy_2005,seifert_stochastic_2012}, we can study the evolution of the \textit{stochastic entropy of the system} along trajectories of the SDE~\eqref{eqn:pflow}
\begin{equation}
  \label{eqn:stoch_entropy}
  \tilde s_{\text{sys}}(t) = -\log\rho(r_t).
\end{equation}
Taking the time derivative of~\eqref{eqn:stoch_entropy} gives
\begin{equation}
  \label{eqn:stoch_entropy:dot}
  \Dot {\tilde s}_{\text{sys}}(t) = -\nabla \log\rho(r_t) \circ \dot r_t,
\end{equation}
where $\circ$ denotes the Stratonovich product. The quantity defined in~\eqref{eqn:stoch_entropy:dot} is a stochastic function of time that can be evaluated along any trajectory. To obtain a deterministic function of~$r$ that conveys local information about the EPR, we can take the expectation of~\eqref{eqn:stoch_entropy:dot} conditioned on the event $r_t = r$~\cite{seifert_entropy_2005,seifert_stochastic_2012}. 
%
%
The current velocity defined in~\eqref{eqn:pflow_vel} can be expressed in terms of this conditional expectation as $v(r) = \langle \dot{r}(t) | r(t) = r\rangle$ (SI Appendix), so that $ \sdotsys(r) = \langle\Dot {\tilde s}_{\text{sys}}(t)|r_t=r\rangle $ is given by
\begin{equation}
  \label{eqn:stoch_epr_alt}
  \begin{aligned}
  \sdotsys(r) = -\nabla \log\rho(r)\cdot v(r) = \nabla\cdot v(r),
  \end{aligned}
\end{equation} 
where the last equality follows from~\eqref{eqn:transport} after division by $\rho>0$.}

The function defined in~\eqref{eqn:stoch_epr_alt} is referred to as the (local) system EPR: it is visualized over the phase space of our low-dimensional illustrative example in Figure~\ref{fig:twop_div_v}, which highlights alternating regions of system entropy production and consumption in the two modes.
Because $S_{\text{sys}}$ is a constant of motion at stationarity, we arrive at the condition
\begin{equation}
  \label{eqn:stoch_epr_alt_expect}
  \begin{aligned}
  \dot{S}_{\text{sys}} = \int_{\domR} \left(\nabla\cdot v(r)\right) \rho(r) dr = 0.
  \end{aligned}
\end{equation}
Later, we will make use of~\eqref{eqn:stoch_epr_alt_expect} as a quantitative test to measure convergence of our learning algorithm. To understand how $\sdotsys$ is distributed spatially in systems with a high-dimensional phase space, we may decompose $\div v$ into a \textit{local} sum of contributions from individual particles using $v(r)=(v^1(r),\ldots, v^N(r))$ to obtain
\begin{equation}
  \label{eqn:div_v_particle}
  \div v(r) = \sum_{i=1}^N \nabla_i\cdot v^i(r),
\end{equation}
where $\nabla_i$ denotes the gradient with respect to $r^i$.

\paragraph{Total EPR.}
\newchange{Assuming that $D$ is invertible, we can use \eqref{eqn:pflow_vel}  written as $\nabla \log \rho = D^{-1} (v-b)$ to decompose \eqref{eqn:stoch_epr_alt} as
\begin{equation}
  \label{eqn:stoch_epr}
  \begin{aligned}
  \sdotsys(r) &= \underbrace{|v(r)|_{D^{-1}}^{2}}_{\mathclap{\sdottot(r)}} - \underbrace{b(r)\cdot D^{-1} v(r)}_{\mathclap{\sdotm(r)}}.
  \end{aligned}
\end{equation}
The quantity $\sdottot$ is non-negative, and can be identified as the (local) total entropy production rate~\cite{seifert_entropy_2005,seifert_stochastic_2012, pigolotti_generic_2017}. 
The quantity $\sdotm$ is of indefinite sign, and can be identified as the (local) entropy production rate of the medium~\cite{marconi_heat_2017,puglisi_clausius_2017,chetrite_fluctuation_2008,crooks_entropy_1999,seifert_entropy_2005,lebowitz_gallavotticohen-type_1999}. }
Similar to~\eqref{eqn:div_v_particle}, assuming that $D$ is made of $N$ diagonal blocks $D_i$, we may decompose 
\begin{equation}
  \label{eqn:sdot_tot_particle}
  \sdottot(r) = |v(r)|_{D^{-1}}^2 = \sum_{i=1}^N |v^i(r)|_{D_i^{-1}}^2,
\end{equation}
into local contributions from individual particles.

\paragraph{Global EPR.}
\newchange{
The global EPR is defined as the Kullback-Leibler divergence between the forward and reverse path measures~\cite{crooks_entropy_1999,lebowitz_gallavotticohen-type_1999} 
\begin{equation}
  \label{eqn:global_epr}
  \dot{S}_{\text{tot}} = \frac1T\left\langle\log\left(\frac{\calP(\phi_{T})}{\calP^{\rev}(\phi_{T})}\right)\right\rangle,
\end{equation}
where $T>0$ is arbitrary and where $\phi_{T} = \{\rt\}_{{0 \leq t \leq T}}$ denotes a path of the SDE~\eqref{eqn:microscopic_stochastic_general} with initial condition drawn from $\rho$.
$\calP$ denotes the path measure of $\phi_T$, $\calP^\rev$ denotes the path measure of a reverse-time path, and the angular brackets denote an average over $\phi_T$ drawn from~$\calP$.
The global EPR can be challenging to compute because it requires a choice of reverse-time dynamics (to set $\calP^\rev$), and the correct choice has been a subject of debate~\cite{mandal_entropy_2017,caprini_comment_2018, caprini_entropy_2019, dabelow_irreversibility_2019, chetrite_fluctuation_2008, chernyak_path-integral_2006}.
Interestingly, there is  a  way to construct a reverse-time dynamics such that~$\dot{S}_{\text{tot}}$ can be written as an expectation of $\dot{s}_{\text{tot}}(r)$ over $\rho$, but it again requires knowledge of~$v(r)$. }
This reverse-time dynamics is the SDE whose solutions have the same statistical properties as the solutions to~\eqref{eqn:microscopic_stochastic_general} played in reverse~\cite{anderson1982reverse}. It reads (SI Appendix)
\begin{equation}
  \label{eqn:anti_sde}
  \dot{r}_t^\rev = b(r_t^\rev) - 2v(r_t^\rev) + \sqrt{2D}\, \eta(t),\end{equation}
where $\eta(t)$ is the same Gaussian white noise process as in the forward SDE~\eqref{eqn:microscopic_stochastic_general}. \newchange{It is easy to check that the stationary density of~\eqref{eqn:anti_sde} is also $\rho$.} By a standard path integral argument~\cite{onsager_fluctuations_1953} or an application of the Girsanov theorem~\cite{oksendal2003sde}, when $D$ is invertible we may compute (SI Appendix)
\begin{equation}
  \label{eqn:epr_girsanov}
  \begin{aligned}
  \dot{S}_{\text{tot}} &= \int_{\domR}|v(r)|_{D^{-1}}^2\rho(r)dr,
  \end{aligned}
\end{equation}
which, by Eq.~\eqref{eqn:pflow:cons}, may be understood as the the total EPR $\sdottot$ averaged over $r$ drawn from $\rho$.
\newchange{\eqref{eqn:epr_girsanov} highlights that a system is at equilibrium if and only if $v = 0$, so that $\dot{S}_{\text{tot}} = 0$. An analogous relation first appeared in~\cite{seifert_entropy_2005} for driven colloidal particles, where the definition of the reverse-time process is less ambiguous than for the active systems we study here.}

\section*{Learning algorithm}
\paragraph{Score.}
The expressions for $v$, the system EPR, and the total EPR depend on the score $\nabla\log\rho(r)$, which is a high-dimensional function we typically do not have access to.
In this section, we develop a machine learning algorithm to approximate it: a graphical summary of the method is given in Figure~\ref{fig:method}. 
In addition to providing access to $v$, and therefore to the total and system EPRs, $\nabla\log\rho$ has the important advantage that it is independent of the normalizing constant of $\rho$, which is typically unknown and intractable. 
This enables us to exploit expressive function classes that need-not represent normalized probability distributions.

\paragraph{Score matching.}
The score $\nabla\log\rho$ can be shown to be the unique minimizer of the loss
\begin{equation}
  \label{eqn:score_matching}
  \begin{aligned}
    \calL_{\mathsf{sm}}[\hat h] &= \mathbb{E}_{\rho}\left[\left|\hat h\right|^2 + 2\nabla\cdot \hat h\right],\\
    \nabla\log\rho &= \argmin_{\hat h} \mathcal{L}_{\text{sm}}[\hat h],
  \end{aligned}
\end{equation}
where $\E_\rho$ denotes expectation over $\rho$.
Eq.~\eqref{eqn:score_matching} is known as the ``score matching'' loss in the machine learning literature~\cite{hyvarinen_estimation_2005}.
We provide a derivation of this loss and demonstrate the uniqueness of its minimizers in SI Appendix.

\paragraph{Exploiting the stationary FPE.}
While a useful loss function,~\eqref{eqn:score_matching} is valid for \textit{any} data distribution, and does not make use of the fact that $\rho$ solves the stationary FPE~\eqref{eqn:fpe}; it is therefore agnostic to the underlying physics.
Intuitively, exploiting our prior knowledge that $\rho$ solves~\eqref{eqn:fpe} should impose additional structure that can be leveraged to improve the quality of the learned score.
As written,~\eqref{eqn:fpe} is an equation for $\rho$, while we are interested in estimating $\nabla\log\rho$. Dividing by $\rho$ yields a nonlinear equation for the score
\begin{equation}
  \label{eqn:stationary_fpe_score}
  \nabla\cdot v + v\cdot\nabla\log\rho = 0.
\end{equation}
Equation~\eqref{eqn:stationary_fpe_score} may be used to construct a physics-informed loss based on the squared residual~\cite{raissi_physics-informed_2019,karniadakis_physics-informed_2021}
\begin{equation}
  \label{eqn:stationary_fpe_score_loss}
  \calL_{\mathsf{FPE}}[\hat h] = \E_{\rho}\left[\left(\nabla\cdot \hat v + \hat v\cdot \hat h\right)^2\right],
\end{equation}
where $\hat v(r) = b(r) - D\hat h(r)$. We propose minimization of the composite loss
\begin{equation}
  \label{eqn:loss}
  \mathcal{L}[\hat h] = \lambda_1 \mathcal{L}_{\mathsf{sm}}[\hat h] + \lambda_2 \mathcal{L}_{\mathsf{FPE}}[\hat h],
\end{equation}
which consists of both the physics-agnostic score matching loss $\calL_{\mathsf{sm}}$ and the physics-informed loss $\calL_{\mathsf{FPE}}$.
In our experiments, we find best performance incorporating both terms, and we set $\lambda_1 = \lambda_2 = 1$ throughout unless otherwise indicated.

\paragraph{Empirical loss.}
In practice, we minimize an empirical approximation of~\eqref{eqn:loss}
\begin{equation}
  \label{eqn:sm_empirical}
  \begin{aligned}
  \hat{\calL}[\hat h] &= \frac{\lambda_1}{n}\sum_{\alpha=1}^n \left(|\hat h(r_\alpha)|^2 + 2\nabla\cdot \hat h(r_\alpha )\right) \\
  &\qquad + \frac{\lambda_2}{n}\sum_{i=1}^n \left(\nabla\cdot \hat v(r_\alpha ) + \hat v(r_\alpha )\cdot \hat h(r_\alpha )\right)^2
  \end{aligned}
\end{equation}
over a dataset of samples $\{r_\alpha\}_{\alpha=1}^n$ with each $r_\alpha \sim\rho$. We can generate such a dataset by simulating the SDE in~\eqref{eqn:microscopic_stochastic_general} with a numerical integration scheme like the Euler-Maruyama method. 
To make the optimization computationally tractable for high-dimensional systems of particles, we can perform the estimation over an expressive parametric class for $\hat h(r)$ such as neural networks, and can use a first-order optimization scheme such as Adam~\cite{kingma_adam_2017} to optimize the parameters.
To increase the diversity of the dataset, we can take steps of the SDE~\eqref{eqn:interacting_particles} between steps of the optimization algorithm, which is similar to online learning and helps prevent overfitting.

\paragraph{Quantitative validation.}
There are several metrics that we can use to verify the accuracy of the learned approximation $\hat{h}$ to $\nabla\log\rho$.
The loss in~\eqref{eqn:stationary_fpe_score_loss} is exactly the squared residual for the stationary score-based FPE~\eqref{eqn:stationary_fpe_score}, and hence provides a quantitative measure of how well the learned score satisfies its governing equation.
At optimality,~\eqref{eqn:score_matching} satisfies $\calL_{\mathsf{sm}}[\nabla\log\rho] = -\E_\rho\left[|\nabla\log\rho|^2\right]$ (SI Appendix); deviation from this relation also provides a measure of convergence. 
Last, we can verify the constraint~\eqref{eqn:stoch_epr_alt_expect} to ensure that the global EPR is a constant of the motion.

\section*{Neural network architecture}
\begin{figure*}[!t]
    \centering
    \includegraphics[width=\textwidth]{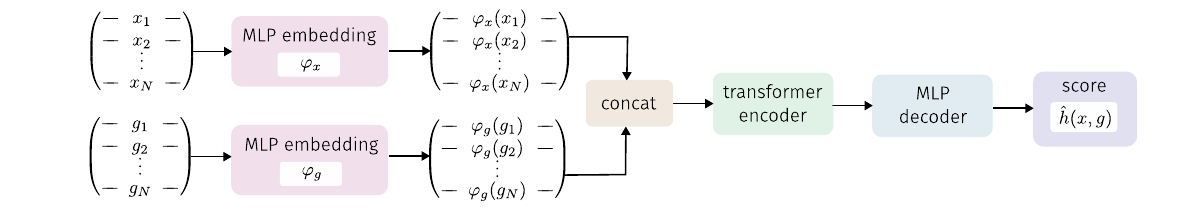}
    \caption{\textbf{Network architecture.} Depiction of the transformer architecture introduced in this work.
    The particle positions and orientations are fed into separate multi-layer perceptrons (MLPs) that embed the input into a latent space of higher dimensionality.
    The embeddings are concatenated particle-wise and fed into a transformer encoder block (SI appendix for further details), where multiple layers of multi-head attention modules learn relevant interactions between particles.
    The output of the encoder block is decoded by a shared MLP applied to each particle state to obtain the score.}
    \label{fig:network}
\end{figure*}

\paragraph{Permutation symmetry.}
An important ingredient in our learning algorithm is a proper choice of the neural network used to estimate $\nabla\log\rho$.
One guiding principle that can be used for physical problems is to build the symmetries of the system into the network~\cite{carleo_machine_2019}.
In addition to its conceptual motivation, this approach has been shown to be statistically advantageous~\cite{bietti_sample_2021}.
In~\eqref{eqn:interacting_particles}, the most relevant symmetry group is permutation invariance amongst the particles, which generates complex multi-modal structure in the stationary density $\rho$. 
Generically, all the configurations generated by permutations will not be present in a given dataset, and this makes it crucial to use a representation of $\nabla \log \rho(r)$ where the permutation invariance is build-in.

\paragraph{Invariance and equivariance.}
Permutation invariance at the level of $\rho$ gives rise to permutation \textit{equivariance} at the level of $\nabla\log\rho$. 
In a numerical implementation, we can choose to parameterize $\log\rho$ and take its gradient or to parameterize $\nabla\log\rho$ directly. 
While it seems physically natural to parameterize $\nabla\log\rho$ as a gradient field, state of the art results in diffusion-based generative modeling directly parameterize the score without this added constraint~\cite{song_score-based_2021,song_maximum_2021,karras_elucidating_2022}, and we follow this approach here.
Doing so reduces the number of gradients that must be computed via automatic differentiation during training, and tends to improve performance.

\begin{figure*}[!t]
    \includegraphics[width=\textwidth]{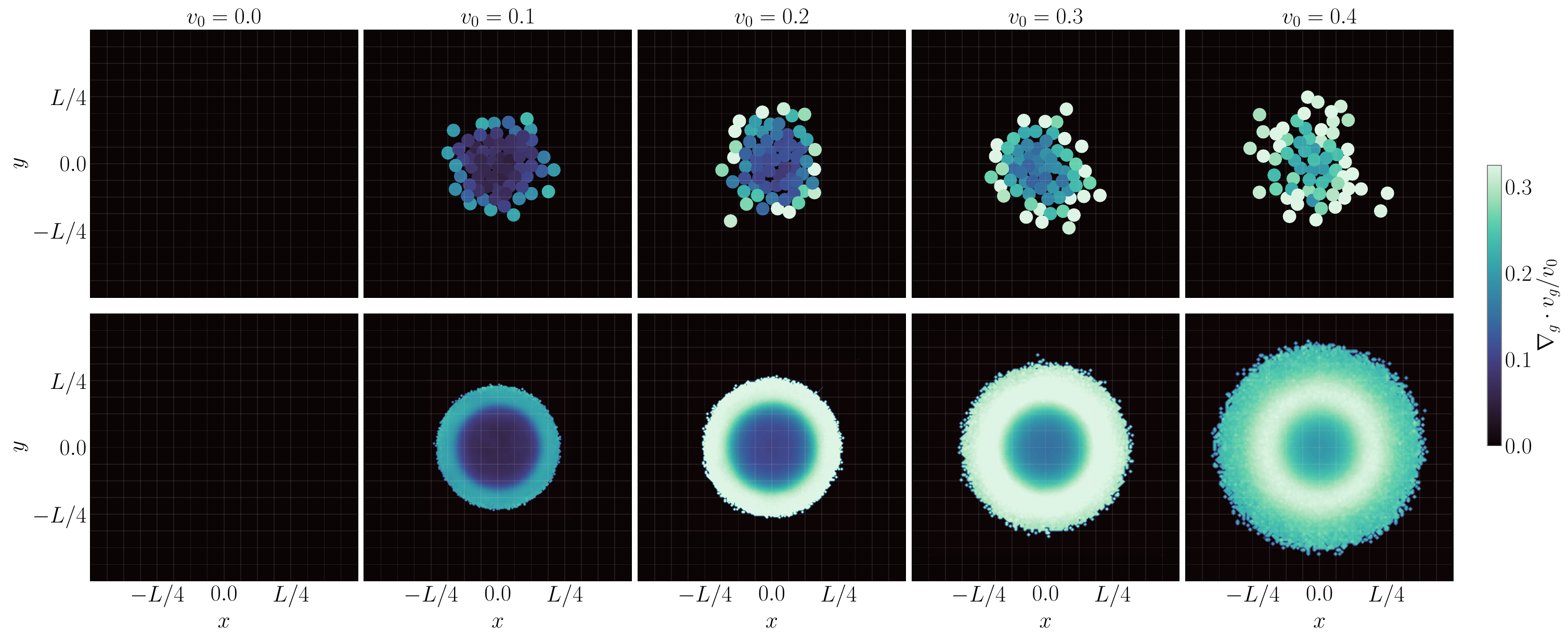}
    \caption{\textbf{64 swimmers in a harmonic trap: system EPR.}
    (Top) The contribution of the per-particle orientational degrees of freedom to the system EPR $\divgi\vgi$ as a function of the activity $v_0$, visualized directly on the particles.
    For $v_0 = 0$, the system is at equilibrium and the network learns that the system EPR vanishes.
    As $v_0$ increases, nonequilibrium effects emerge, and the particles on the boundary display the highest contribution to the EPR.
    (Bottom) A spatial map visualizing the typical contribution of a particle at position $(x, y)$ to the system EPR, obtained by averaging the data in the top row over many system snapshots.
    The map highlights the role of interfacial contributions, and displays a prominent ring at the boundary of the cluster.}
    \label{fig:64_div_vg}
\end{figure*}


\paragraph{Transformers.}
Perhaps the most natural way to proceed is to employ an architecture that can directly learn the relevant order of the interactions in the system.
The transformer architecture~\cite{vaswani_attention} has emerged as a powerful tool for learning complex interactions in language~\cite{zaheer_big_2021,wang_learning_2019}, and is built upon operations (self-attention and token-wise mappings) that are naturally permutation equivariant.
In addition to language modeling, transformers currently achieve state of the art in image classification~\cite{parmar_image_nodate,dosovitskiy_image_2021}, and have been applied to problems such as protein structure prediction~\cite{chandra_transformer-based_2023} and quantum chemistry~\cite{von_glehn_self-attention_2023}.
Yet, to our knowledge, they have not been used to study interacting particle systems in active matter and stochastic thermodynamics.
Transformers also have the advantage that their attention maps can be inspected \textit{a-posteriori} for insight into the learned interactions~\cite{abnar_quantifying_2020}.
 
We introduce a transformer architecture that learns interactions between embeddings of the particle positions $x^i$ and orientations $g^i$ (Figure~\ref{fig:network}).
The output of a series of transformer encoder blocks consisting of self-attention, LayerNorm~\cite{ba_layer_2016}, and particle-wise multi-layer perceptrons (MLPs) is decoded by an additional particle-wise MLP to obtain the score $\hat{h}$.
For large numbers of interacting particles (as we will study in the MIPS system), we introduce a modification of this architecture that exploits a spatially-local ansatz to define the score $\hat{h}^i$ at the level of the individual particles.
Remarkably, in addition to a large gain in memory efficiency, this architecture enables transfer learning to datasets with a larger numbers of particles, where we find physically meaningful predictions without any additional training.
We provide an overview of the relevant features of the transformer architecture, including more detail on the constituent elements of the encoder blocks in SI Appendix.

\section*{Active swimmers in a trap}
\begin{figure*}[!t]
    \includegraphics[width=\textwidth]{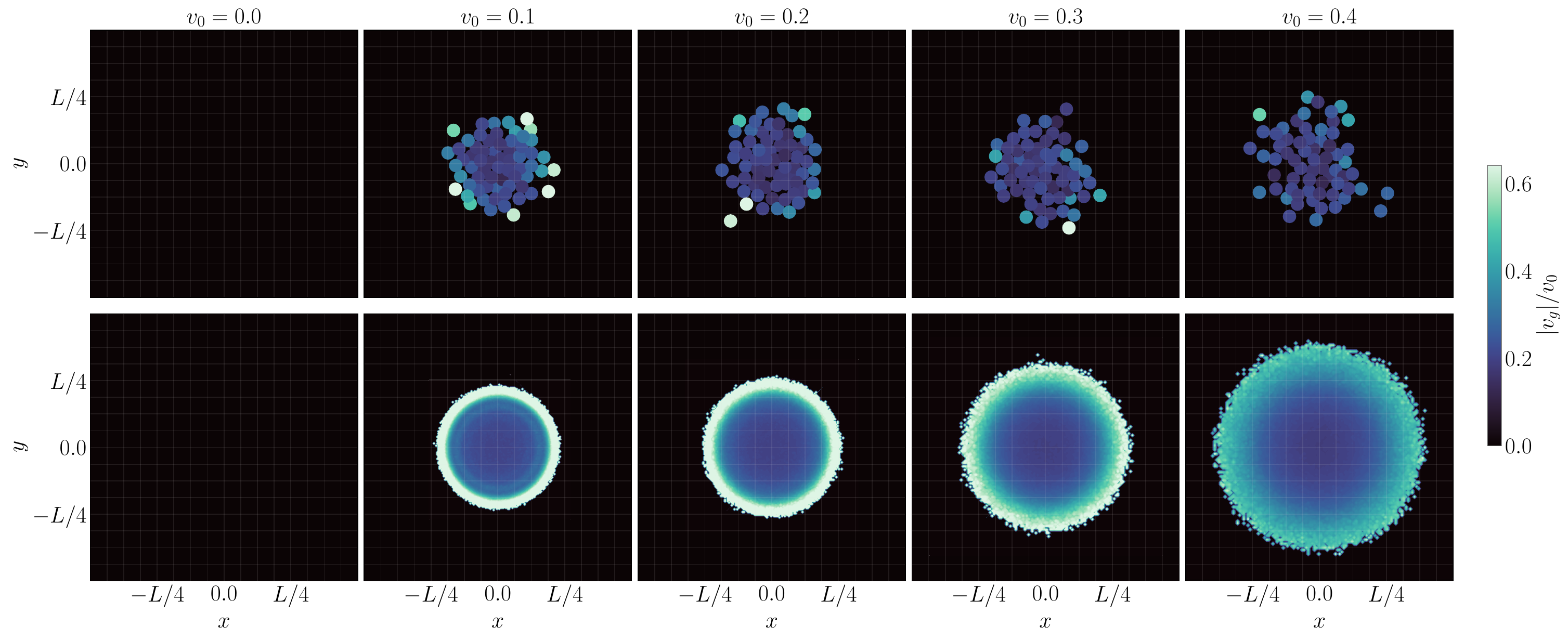}
    \caption{\textbf{64 swimmers in a harmonic trap: total EPR.}
    (Top) The contribution of the per-particle orientational degrees of freedom to the total EPR $|\vgi|^2$ as a function of the activity $v_0$, visualized directly on the particles.
    As in Figure~\ref{fig:64_div_vg},  the network learns that the system is at equilibrium for $v_0 = 0$, and the total EPR vanishes.
    As $v_0$ increases, the total EPR is dominated by outlier contributions from particles on the edge of the cluster.
    (Bottom) A spatial map visualizing the typical contribution of a particle at position $(x, y)$ to the total EPR, obtained by averaging the data in the top row over many system snapshots.
    The map distills the signal present in the outliers in the top row, and displays a concentrated ring of entropy production at the interface.}
    \label{fig:64_gdot}
\end{figure*}

\paragraph{Dynamics.}
As a first application of our method, we now consider a system of $N=64$ interacting active Ornstein-Uhlenbeck particles in a harmonic trap, similar to what was studied by~\cite{martin_statistical_2021}; this gives rise to a $256$-dimensional many-body FPE in~\eqref{eqn:fpe}.
In this case, the system under study is given by~\eqref{eqn:interacting_particles} with $\epsilon = 0.1$, $\gamma = 0.1$, $\mu = 1$, \newchange{and where $f$ is a conservative force governed by the many-body potential} 
\begin{equation}
    \label{eqn:trap_potential}
    \Phi(x) = \frac{A}{2}\sum_{i=1}^N |x_i|^2 + \frac{1}{2}\sum_{\substack{i,j = 1\\ i\not = j}}^NV(x_i - x_j).
\end{equation}
Above, $A=0.05$ and $V(x) = \tfrac{k}{2}\left(2a - |x|\right)^2\Theta(2a-|x|)$ with $a=1$ the particle radius, $k=2.5$, and where $\Theta$ denotes the Heaviside step function.
To make the system more amenable to learning, we smooth the force slightly to avoid the hard cutoff mediated by $\Theta$ (see SI Appendix for details).
Due to the presence of the trap, the system assembles into an active cluster with a dense core and a motile boundary that is similar to the phase separation observed in MIPS.
The trap makes the presence of these features more robust to variations in the parameters~$\gamma, \epsilon$, and~$v_0$, which enables us to study the structure of the total and system EPRs as a function of the activity~$v_0$ at fixed persistence~$\gamma$ and bath temperature~$\epsilon$. Localizing the cluster also allows us to perform spatial averaging of the EPR, which connects our results with the field-theoretic approach developed in~\cite{nardini_entropy_2017}. We stress, however, that our approach does not require this spatial averaging, and we will remove the trap when we study MIPS.

\paragraph{EPR decomposition.}
The contribution of each particle to the system EPR $\nabla_i \cdot v^i(r) = \nabla_{x^i} \cdot v_x^i(r) + \nabla_{g^i} \cdot v_g^i(r)$ can be decomposed additively into local contributions from the orientational degrees of freedom $g^i$ and the translational degrees of freedom $x^i$ to understand how they independently contribute to the system EPR.
Similarly, the contribution to the total EPR decomposes as $|v^i(r)|_{D_i^{-1}}^2 = \tfrac{1}{\gamma}|v_g^i(r)|^2 + \tfrac{1}{\epsilon}|v_x^i(r)|^2$.
In the following, we make use of these decompositions to isolate further how the total and system EPRs are built up from individual particle contributions.

\paragraph{Orientational contributions.}
We first focus on the contribution of the orientational degrees of freedom to the total EPR $|v_g|$ and the system EPR $\divg v_g$ (Figures~\ref{fig:64_div_vg}~\&~\ref{fig:64_gdot}, top).
For $v_0 = 0$, the system is at equilibrium and hence the EPR vanishes.
As $v_0$ is increased, the system becomes increasingly nonequilibrium, and spatial structure begins to emerge in the EPR.
Consistent with theoretical predictions~\cite{nardini_entropy_2017}, we find that both quantities concentrate on the boundary of the cluster.
The contribution to the EPR of the system $\divg v_g$ increases smoothly with radial distance from the center of the cluster.
The contribution to the total entropy production $|v_g|$ is similar, but is more dominated by a few outliers on the boundary.

In Figures~\ref{fig:app:64_omitted_1}--\ref{fig:app:64_omitted_2} (SI Appendix), we visualize the translational contributions $|v_x|$ and $\divx v_x$, as well as the EPR $|v|$ and the system EPR $\div v$.
$|v_x|$ displays similar features to $|v_g|$ with slightly lower contrast between the core and the boundary, so that $|v|$ also displays concentration on the interface.
We find that $\divx v_x \approx -\divg v_g$, which causes $\div v$ to roughly vanish pointwise per-particle. 
Small-scale fluctuations in the particles are present around zero, which together average so that $\E_\rho[\div v] \approx 0$ as required by stationarity.
We find high accuracy as measured by the residual of the score-based FPE in~\eqref{eqn:stationary_fpe_score} and the stationarity condition~\eqref{eqn:stoch_epr_alt_expect} (Figure~\ref{fig:64_stats} SI Appendix).

\paragraph{A spatial map of entropy production.}
The presence of the trap constrains the shape and location of the cluster, which facilitates averaging in space and in time.
To build up a spatial map of the entropy production, we discretize space into a $256\times 256$ grid.
We can then sum the particle-wise contributions $\nabla_{g^i}\cdot v_g^i$ and $|v_g^i|^2$  in each grid cell over a dataset of samples, normalizing by the number of particles that appear in each cell in the dataset.
The result is a spatial map that describes the typical value of the EPR a particle would attain at a given spatial position.
We visualize these spatial maps in Figures~\ref{fig:64_div_vg}~\&~\ref{fig:64_gdot} (bottom), where we find a distinct ring of entropy production at the boundary of the core.

\paragraph{Attention map.}
An advantage of the transformer architecture is that we can visualize the \textit{attention map}, which gives us insight into which \textit{other} particles are used to compute the score of a given particle.
We employed attention rollout~\cite{abnar_quantifying_2020} to propagate the flow of attention across all heads from layer to layer (see SI Appendix).
The resulting attention map reveals that the network learns a physically-intuitive spatially-local attention pattern, where each particle is primarily influenced by its nearest neighbors (Figure~\ref{fig:64_attention}).
Interestingly, the interactions are still significantly longer-range than those present in the interaction potential for the system.

\section*{Motility-induced phase separation}
\begin{figure*}[b!]
\centering
\begin{tabular}{c}
\begin{overpic}[width=\textwidth]{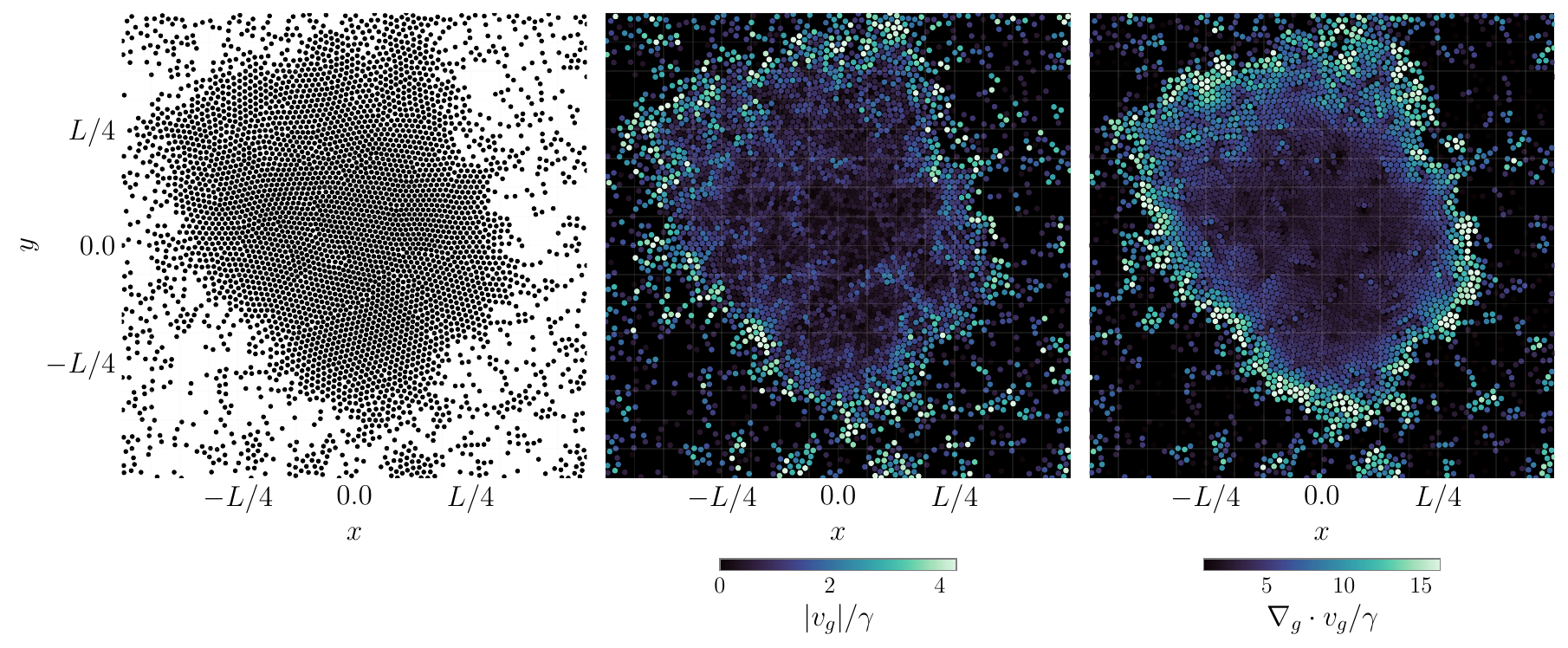}
\put(7,  41){\textbf{A}}
\put(38, 41){\textbf{B}}
\put(69, 41){\textbf{C}}
\end{overpic}
\end{tabular}
\caption{\textbf{Motility-induced phase separation.}
(A) Reference depiction of the MIPS cluster.
(B/C) Particle-wise contributions to the total EPR $|\vgi|^2$ and the system EPR $\divgi \vgi$.
Both quantities concentrate on the boundary of the cluster, with sporadic contributions throughout the dilute phase when particles collide.
$|\vgi|$ vanishes in the center of the cluster, indicating a nontrivial phase dependence in the velocity field.
A movie visualizing the evolution of these quantities along stochastic trajectories can be found \href{https://www.dropbox.com/scl/fi/qh3hlxp1omoz4xorzo4k1/sde_movie_N-4096.mp4?rlkey=7fjd18nos0n0k5ongsllimsvw&dl=0}{at this link}.}
\label{fig:mips}
\end{figure*}

\begin{figure*}[t!]
\centering
\begin{tabular}{c}
\begin{overpic}[width=\textwidth]{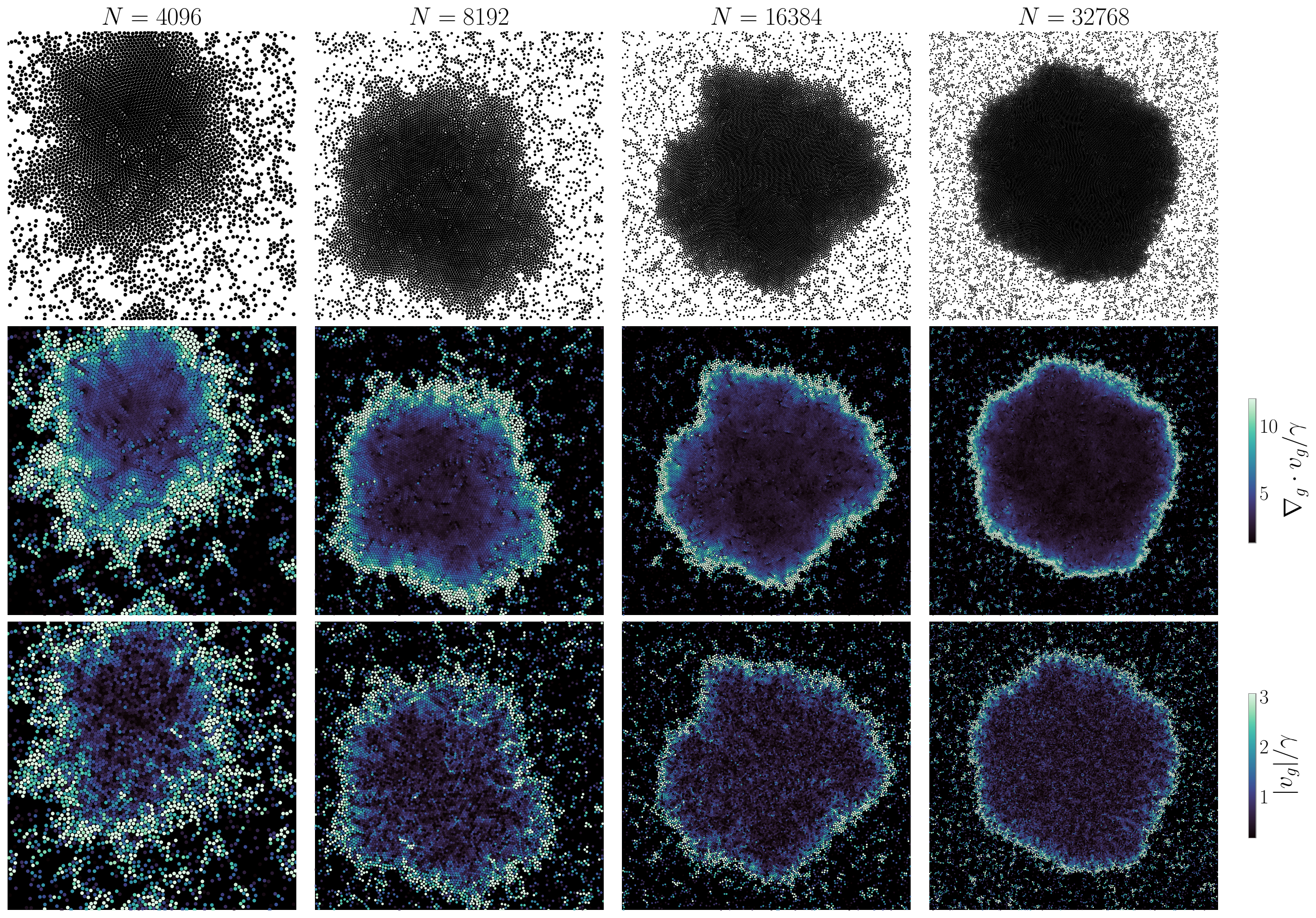}
\end{overpic}
\end{tabular}
\caption{\textbf{Motility-induced phase separation: transfer learning towards larger systems.}
Our network architecture is defined at the level of individual particles, and depends only on local neighborhoods.
This enables us to extend the learned solution from the $N=4096$ training set to larger values of $N$.
We consider values of $N$ up to $8\times$ larger, and find physically-consistent predictions in all cases.
Movies visualizing the evolution of the EPR along stochastic trajectories can be found at the following links:
\href{https://www.dropbox.com/scl/fi/2xn5xuwdnym0wihw4nr8o/sde_movie_N-8192.mp4?rlkey=ro4vskdqrsufnwnohlmbodqbs&dl=0}{$N=8192$},
\href{https://www.dropbox.com/scl/fi/q3y69aztjdsglc98hkuwm/sde_movie_N-16384.mp4?rlkey=5u342mkvneffcs94ecw2rqwkj&dl=0}{$N=16384$},
\href{https://www.dropbox.com/scl/fi/ym78mwr5qxbrrrss394kq/sde_movie_N-32768.mp4?rlkey=7yfpr59v3n19igebeblfjo8tx&dl=0}{$N=32768$}.}
\label{fig:mips_N_transfer}
\end{figure*}

\begin{figure*}[t!]
\centering
\begin{tabular}{c}
\begin{overpic}[width=0.7\textwidth]{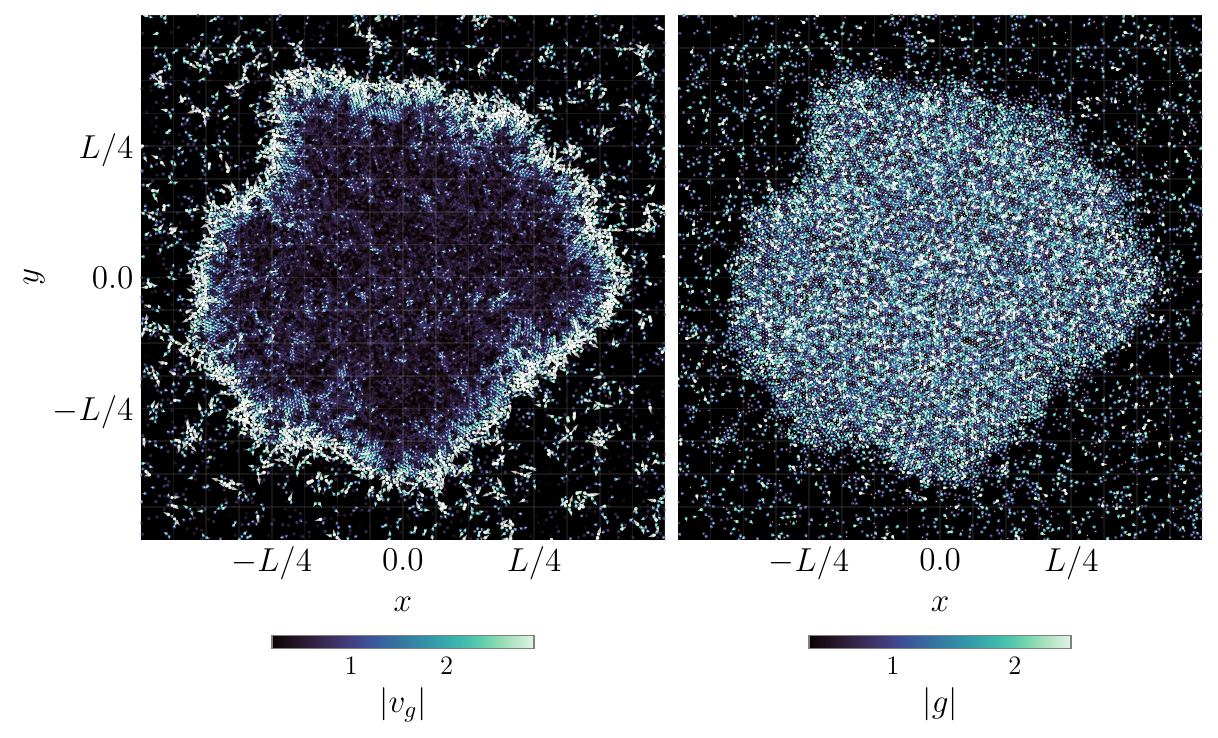}
\put(11,  60){\textbf{A}}
\put(55, 60){\textbf{B}}
\end{overpic}
\end{tabular}
\caption{\textbf{Motility-induced phase separation: probability flow.}
Particle values of $\vgi$ (A) and $g^i$ (B), with directionality visualized as arrows for $N=16384$.
$\vgi$ vanishes in the interior of the solid, but points outward near the solid side of the interface.
There is also a layer of particles pointing both inward and outward directly at the interface, corresponding to particles that are exiting and leaving the cluster.
The values of $g^i$ appear random, and do not have a clear phase dependence by eye, in contrast with $\vgi$.}
\label{fig:vg}
\end{figure*}

\begin{figure*}[t!]
\centering
\begin{tabular}{c}
\begin{overpic}[width=\textwidth]{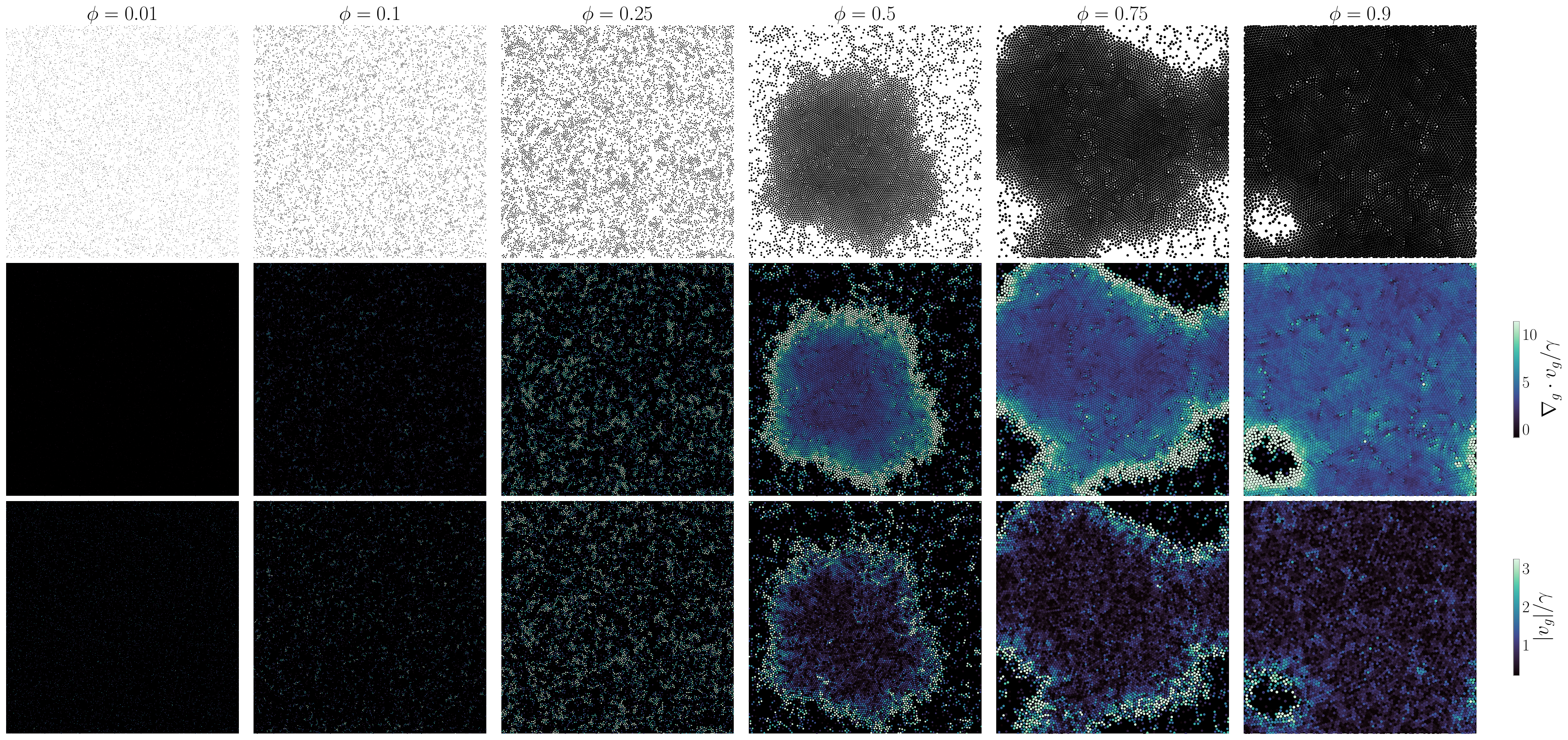}
\end{overpic}
\end{tabular}
\caption{\textbf{Motility-induced phase separation: transfer learning to other packing fractions.} %
Similar to Figure~\ref{fig:mips_N_transfer}, we find that the learned solution generalizes to other packing fractions $\phi$.
Here, we vary $\phi$ at resolution $N=8192$ by varying the size $L$ of the box; for presentation, we rescale the results to the same square.
The solution identifies contributions to the EPR from particle-particle collisions in the gaseous phase and at the interface of the gaseous and solid phases.
For low and high packing fraction, the EPR diminishes, as the system becomes dominated by a gas or a solid.
}
\label{fig:mips_phi_transfer}
\end{figure*}

We now consider a system of $N=4096$ interacting particles undergoing motility-induced phase separation, given by~\eqref{eqn:interacting_particles} with $v_0 = 0.025$, $\mu = 1$, $\gamma = 10^{-4}$, $\epsilon = 0$, and with periodic boundary conditions.
In this case, the corresponding many-body FPE is $16,384$-dimensional, and its solution poses a formidable task.
Because we consider the athermal, hypoelliptic setting with $\epsilon=0$, the velocity field defined in~\eqref{eqn:pflow_vel} only depends on the score in the $g$ variables $\nabla_g\log\rho$.
To target $\nabla_g \log\rho$ directly, we consider only the score matching loss~\eqref{eqn:score_matching} and set $\lambda_2 = 0$ in the combined loss~\eqref{eqn:loss}. 
The resulting loss decouples into equivalent losses for $\nabla_x\log\rho$ and $\nabla_g\log\rho$, while the physics-informed loss~\eqref{eqn:stationary_fpe_score_loss} couples the two scores, so they cannot be learned independently.
\newchange{Because $\epsilon = 0$, the diffusion tensor $D$ is no longer invertible, and the connection between $v$ and the total EPR $\sdottot(t) = |v(R_t(r))|_{D^{-1}}^2$ breaks down.}
\newchange{By contrast, the identity $\sdotsys(t) = \nabla \cdot v(R_t(r))$ still holds, and $v$ remains a fundamental object that describes the transport of the particles according to the probability current.}
%
%
For further details, including an overview of a variant of the denoising score matching loss function~\cite{vincent_connection_2011} we use to reduce computational expense, \newchange{along with a discussion of the technical issues that arise for $\epsilon = 0$}, see (SI Appendix).
We learn $\nabla_g\log\rho$ on a single dataset with $N=4096$ particles and with a packing fraction $\phi=0.5$, but make use of an architecture that enables us to transfer learn to higher values of $N$ and other values of $\phi$, as we now describe.

\paragraph{Network architecture.}
The large number of particles makes it computationally intractable to use the same transformer architecture we used for $N=64$: the self-attention mechanism has time and memory complexities that scale as $\mathcal{O}\left(N^2\right)$, which quickly both become prohibitive for large $N$.
Nevertheless, Figure~\ref{fig:64_attention} shows that the learned attention map is spatially-local in the $N=64$ case, and we expect the same behavior to hold true for the MIPS system.
To exploit spatial locality, we developed a transformer architecture defined at the level of an individual particle, which restricts the attention mechanism to a local neighborhood (SI Appendix).
This approach has the additional advantage of increasing the effective size of the dataset, because there are many distinct local neighborhoods in a given snapshot of the system.
As the size of the attention window is increased, the architecture used for $N=64$ is recovered.
Because we define the transformer at the single particle level, our network can be extended to systems with larger $N$ or with a different packing fraction $\phi$, which we \newchange{also} demonstrate in this section.

\paragraph{System EPR and magnitude of the probability flow.}
Figure~\ref{fig:mips}A shows the MIPS cluster for reference, while Figures~\ref{fig:mips}B~\&~\ref{fig:mips}C display the orientational contributions to \newchange{the magnitude of the probability flow} $|v_g|$ and the system EPR $\divg v_g$, respectively.
Both quantities are visualized as individual particle contributions without any averaging in space or in time.
Consistent with the results for $N=64$ from the previous section, we find that the dominant source of entropy production is at the interface between the gas and solid phases.
There are also pockets of entropy production spread sporadically throughout the gas in regions with particle-particle collisions.
A movie of a stochastic trajectory, colored as in Figure~\ref{fig:mips}, can be found \href{https://www.dropbox.com/scl/fi/8pqw964xr89is4iq1ukks/sde_movie_N-4096.mp4?rlkey=m1qccgywblevuplj2lwjmstr8&dl=0}{at this link} (we recommend downloading the movie for high-resolution viewing).

\paragraph{Transfer learning toward larger systems.}
Because our neural network architecture is defined at the particle level, it is agnostic to the number of particles $N$ in the system. 
This means that we can take a single network trained with $N=4096$ and investigate whether it can make reasonable predictions for higher values of $N$ without any re-training.
\newchange{Physically, because the possible local environments for a given particle should be roughly independent of $N$ for $N$ sufficiently large, we expect our learned network to generalize beyond the training data.}
In Figure~\ref{fig:mips_N_transfer}, we show predictions of $|v_g|$ and $\divg v_g$ as a function of $N$, ultimately scaling up to $32768$ particles.
As the number of particles increases, the cluster becomes more well-defined, and the \newchange{signals} in the \newchange{probability flow and the} EPR seen for $N=4096$ become increasingly high-resolution.
These results highlight the remarkable fact that the local environment learned with $N=4096$ -- where dataset generation is significantly cheaper -- can be used to make predictions about systems with a larger number of interacting particles.

\paragraph{Probability flow.}
We can use our ability to scale to larger $N$ to investigate the probability flow near the thermodynamic limit: in Figure~\ref{fig:vg}A, we visualize the directionality of $v_g^i$, and we compare it to $g^i$ in Figure~\ref{fig:vg}B.
The result reveals a surprising set of observations: $v_g^i$ vanishes in the solid phase, typically points outwards at the solid edge of the interface, and typically points inwards at the gaseous edge of the interface.
While it follows by force-balance that $|v_x^i|$ must vanish in the solid, an equivalent for $|v_g^i|$ is nontrivial.
This is highlighted when contrasting the particle-wise values of $|v_g^i|$ with those of $|g^i|$.
Unlike the $|\vgi|$, The $|g^i|$ appear random, and by eye, uncorrelated with their phase.
This is consistent with the fact that their dynamics is decoupled from the translational degrees of freedom in~\eqref{eqn:interacting_particles}.

Together, these observations reveal a simple picture for the probability flow.
Particles in the solid are frozen with $v_x^i = v_g^i = 0$. 
Free particles in the gas have $v_x^i = v_0 g^i$ and $v_g^i = 0$.
At particle-particle collisions, $v_g^i$ becomes nonzero, and entropy is produced.
These events are mostly concentrated at the interface between the gas and the solid, where there are particles both exiting and entering the cluster, but also occur sporadically throughout the gas.

\paragraph{Packing fraction transfer.}
In addition to transferring to larger values of $N$, we can investigate the ability of the learned network to transfer to other regimes of the phase diagram, so long as the system parameters defining the particles are fixed.
For example, because the score is defined at the level of the local neighborhood of each particle, \newchange{and because these local environments should be statistically similar in some regions of the phase diagram}, we expect the score to be able to transfer to other packing fractions $\phi$.
In Figure~\ref{fig:mips_phi_transfer}, we show that a single network trained on a dataset of $N=4096$ particles with $\phi=0.5$ can make physically consistent predictions for a range of values from $\phi=0.01$ to $\phi=0.9$ on a dataset with $N=8192$ particles.
For very low packing fraction, there are few particle-particle collisions and no cluster, and the EPR is essentially zero everywhere.
As the packing fraction increases, particle-particle collisions begin to occur, so that pockets of entropy production become spread throughout the gas.
As a cluster forms for intermediate $\phi$, the EPR becomes concentrated at the interface.
As $\phi$ increases further, the cluster becomes dominant, again leaving few regions of entropy production.
Analogous figures for $N=4096, N=16384,$ and $N=32768$ are shown in Figures~\ref{fig:mips_phi_transfer_N4096}-\ref{fig:mips_phi_transfer_N32768} (SI Appendix).

\section*{Discussion}
In this work, we demonstrated the capability of machine learning algorithms to learn the entropy production rate and the probability flow of complex interacting particle systems, even in high-dimensional scenarios typically plagued by the curse of dimensionality.
In addition to uncovering structure in the EPR, we highlighted that a network trained with a given number of particles $N$ and a fixed packing fraction $\phi$ can generalize to other values of $N$ and $\phi$.
As a result, our method paves the way to investigating questions about active systems in the thermodynamic limit.

Physically, we focused on active particles without alignment interactions.
A natural extension of this work is to consider more complex models such as the Vicsek model~\cite{vicsek_novel_1995}. 
Numerically, we considered transformer architectures based on standard self attention modules, but could likely scale to larger systems with less local interactions by incorporating recent advances such as FlashAttention~\cite{dao_flashattention_2022, dao_flashattention-2_2023}.

\section*{Acknowledgments}
We thank Grant Rotskoff, Michael Albergo, and Stephen Tu for many useful discussions.
NMB is funded by the ONR project entitled Mathematical Foundation and Scientific Applications of Machine Learning.
EVE is supported by the National Science Foundation under Awards DMR-1420073, DMS-2012510, and DMS-2134216, by the Simons Collaboration on Wave Turbulence, Grant No. 617006, and by a Vannevar Bush Faculty Fellowship.

\appendix
\setcounter{equation}{0}
\setcounter{figure}{0}
\renewcommand{\thefigure}{S\arabic{figure}}
\renewcommand{\theequation}{S\arabic{equation}}

\newpage
\section{Short review of the calculus of stochastic thermodynamics}

In this section, we review some key formulas used in stochastic thermodynamics~\cite{seifert_stochastic_2012} to derive the expressions for the EPR given in Eqs.~\eqref{eqn:stoch_epr_alt} and~\eqref{eqn:stoch_epr}. To do so, we reconsider~\eqref{eqn:microscopic_stochastic_general} as a proper stochastic differential equation (SDE) and write it as:
\begin{equation}
    \label{eqn:microscopic_sde}
    dr_t = b(r_t)dt + \sqrt{2D}dW_t
\end{equation}
where $W_t$ is a Wiener process. We also assume that $b$ is continuously differentiable, but we do not necessarily assume that $D$ is invertible.

\subsection{Two stochastic characterizations of $v(r)$} Since $r_t$ is not differentiable, the relation $v(r) = \langle\dot r_t|r_t=r\rangle$ requires some justification. In this section we provide two; some additional estimators for $v(r)$ are also discussed in SI Appendix~\ref{sec:estim:v} below. Our first result is:
\begin{proposition}
    \label{prop:1}
    Given any vector-valued test function $\phi(r)$ and any $\tau> 0$, the current velocity $v(r) = b(r) - D \nabla \log \rho(r)$ satisfies
\begin{equation}
    \label{eq:v:exp}
    \begin{aligned}
        \frac{1}{\tau}\left\langle\int_0^\tau\phi(r_t)\circ dr_t \right\rangle = \int_\Omega \phi(r)\cdot v(r) \rho(r)dr,
    \end{aligned}
\end{equation}
where  $\circ$ denotes the Stratonovich product, and where the angular brackets denote an expectation over both the noise $W_t$ in~\eqref{eqn:microscopic_sde} and the initial condition $r_0$ drawn from the stationary density $\rho$.
\end{proposition}
We stress that \eqref{eq:v:exp} holds whether or not $D$ is invertible. 
We will use this equation below with various choices of $\phi$ (some of which involve $D^{-1}$, in which case we assume $D$ is invertible) to derive the EPR.

Our second result characterizes $v(r)$ using the variational formulation of the conditional expectation:

\begin{proposition}
    \label{prop:cond:expect}
    The current velocity $v(r) = b(r) - D\nabla \log \rho(r)$ is the unique minimizer over all functions $\hat v(r)$ of the objective function
\begin{equation}
    \label{eq:obj}
    G[\hat v]=\frac1\tau \left\langle \int_0^\tau \left( |\hat v(r_t)|^2 dt - \hat v(r_t) \circ dr_t\right) \right\rangle 
\end{equation} 
where $\tau>0$ is arbitrary, $\circ$ denotes the Stratonovich product, and where the angular brackets denote an expectation over both the noise $W_t$ in~\eqref{eqn:microscopic_sde} and the initial condition $r_0$ drawn from the stationary density $\rho$.
\end{proposition}

Through optimization over an expressive parametric class of functions such as neural networks,~\eqref{eq:obj} can be used to learn an approximation of $v(r)$.
This alternative learning strategy gives a way to estimate $v$ directly, and will be exploited in future work.

\begin{proof}[Proof of Proposition~\ref{prop:1}] Observe that
\begin{equation}
    \begin{aligned}
       &  \frac{1}{\tau}\left\langle\int_0^\tau \phi(r_t)\circ dr_t\right\rangle \\
    &=  \frac{1}{\tau}\left\langle\int_0^\tau  \left(\phi(r_t)\cdot b(r_t)dt + \phi(r_t)\circ \sqrt{2D}dW_t\right)\right\rangle\qquad \text{(using the SDE~\eqref{eqn:microscopic_sde})}\\
        & = \frac{1}{\tau}\left\langle\int_0^\tau  \left(\phi(r_t)\cdot b(r_t)dt + \text{tr}[D\nabla\phi(r_t)]dt + \phi(r_t)\cdot \sqrt{2D}dW_t\right) \right\rangle\quad \text{(Stratonovitch to It\^o)}\\
        & =\frac{1}{\tau}\left\langle\int_0^\tau  \left(\phi(r_t)\cdot b(r_t) + \text{tr}[D\nabla\phi(r_t)]\right)dt\right\rangle \qquad \text{(It\^o integrals have zero expectation.)}\\
        &= \int_\Omega\left(\phi(r)\cdot b(r) + \text{tr}[D\nabla\phi(r_t)]\right)\rho(r)dr \qquad \text{(the density of $r_t$ is $\rho$ for all $t\ge0$)}\\
        &= \int_\Omega\phi(r)\cdot \left(b(r) -D\nabla\log\rho\right)\rho(r)dr \qquad \text{(integrating by parts and using $\nabla \rho = \nabla \log \rho \, \rho$)}\\
        &\equiv \int_\Omega\phi(r)\cdot v(r)\rho(r)dr \qquad \text{(by definition of $v=b - D\nabla\log\rho$)}
    \end{aligned}
\end{equation}
This completes the proof.
\end{proof}

\begin{proof}[Proof of Proposition~\ref{prop:cond:expect}]
Using \eqref{eq:v:exp} with $\phi=\hat v$  to treat the second term under the integral in~\eqref{eq:obj} we can express $G[\hat v]$ as
\begin{equation}
    \label{eq:obj:2}
    G[\hat v]=\int_\Omega \left( |\hat v(r)|^2 - \hat v(r) \cdot v(r)\right) \rho(r) dr
\end{equation} 
where $v(r) = b(r) - D\nabla \log \rho(r)$ is the current velocity. The unique minimizer of this quadratic objective is $\hat v(r) = v(r)$.
\end{proof}

\subsection{Entropy production rates}
\label{app:epr}

\subsubsection{System EPR}
Equation~\eqref{eqn:stoch_entropy} implies that
\begin{align}
    \label{eqn:stochastic_entr_sde_2}
     d \tilde s_{\text{sys}}(t)= -d\log\rho(r_t) &= -\nabla\log\rho(r_t)\circ dr_t.
\end{align}
Using \eqref{eq:v:exp} with $\phi = -\nabla \log \rho$, we deduce that 
\begin{align}
    \label{eqn:stochastic_entr_sde_3}
     \frac1{\tau} \left\langle\int_0^\tau  d \tilde s_{\text{sys}}(t) \right\rangle = -\frac1{\tau} \left\langle \int_0^\tau  -\nabla\log\rho(r_t)\circ dr_t \right\rangle= \int_{\Omega } \underbrace{\left[-\nabla \log \rho(r) \cdot v(r)\right] }_{\mathclap{\sdotsys(r)}}\rho(r) dr,
\end{align}
consistent with \eqref{eqn:stoch_epr_alt}. This also implies that $\sdotsys = -\nabla \cdot v$ since $\nabla \log\rho \cdot v + \nabla\cdot v=0$ by~\eqref{eqn:transport}. Note that \eqref{eqn:stochastic_entr_sde_3} holds whether $D$ is invertible or not. 

\subsubsection{Total EPR and EPR of the medium} By definition of $v=b-D \nabla \log \rho$, when $D$ is invertible, we have
\begin{equation}
\begin{aligned}
    -\nabla\log\rho_t(r_t)\cdot b(r_t) &= D^{-1}(v(r_t) - b(r_t))
\end{aligned}
\end{equation}
As a result, we can use \eqref{eq:v:exp} with $\phi = D^{-1} v$ and $\phi= D^{-1} b$ along with \eqref{eqn:stochastic_entr_sde_3}, to deduce that 
\begin{equation}
\begin{aligned}
    \label{eqn:stochastic_entr_sde_4}
  \frac1{\tau} \left\langle \int_0^\tau  d \tilde s_{\text{sys}}(t) \right\rangle &= \frac1{\tau}\left\langle\int_0^\tau  D^{-1}(v(r_t) - b(r_t)) \circ dr_t  \right\rangle,\\
  &= \int_{\Omega } \overbrace{\big[\underbrace{|v(r)|^2_{D^{-1}} }_{\mathclap{\sdottot(r)}} - \underbrace{b(r) \cdot D^{-1} v(r)}_{\mathclap{\sdotm(r)}}\big]}^{\mathclap{\sdotsys(r)}}\rho(r) dr,
\end{aligned}
\end{equation}
consistent with~\eqref{eqn:stoch_epr}.

\subsection{EPR from the probability flow ODE~\eqref{eqn:pflow}}
We now consider two analogous derivations along trajectories of the probability flow~$R_t(r)$. These derivations are simpler because $R_t(r)$ is differentiable and $\dot R_t(r) = v(r_t(r)) $ by definition.

\subsubsection{System EPR}
By the chain rule we have
\begin{align}
    \label{eqn:pflow_system_EPR_1}
    -\frac{d}{dt}\log\rho(\Rt(r)) = -\nabla\log\rho(\Rt(r))\cdot v(\Rt(r))
\end{align}
where we have used that $\dRt(r) = v(\Rt(r))$ by the probability flow ODE~\eqref{eqn:pflow}. Now, by the score-based stationary FPE~\eqref{eqn:stationary_fpe_score}, we have that $\nabla\log\rho(\Rt(r))\cdot v(\Rt(r)) = -\div v(\Rt(r))$, so that
\begin{align}
    \label{eqn:pflow_system_EPR_2}
    -\frac{d}{dt}\log\rho(\Rt(r)) = \nabla\cdot v(\Rt(r)).
\end{align}
This leads to the identification
\begin{equation}
    \label{eqn:app:sdotsys}
    \sdotsys(R_t(r)) = \nabla\cdot v(R_t(r)).
\end{equation}
consistent with \eqref{eqn:stoch_epr_alt}.

\subsubsection{Total EPR and EPR of the Medium}
Using the relation $v = b - D\nabla\log\rho$ to write $\nabla\log\rho = D^{-1}(b - v)$ in~\eqref{eqn:pflow_system_EPR_1} we deduce
\begin{align}
    \label{eqn:pflow_system_EPR_3}
    -\frac{d}{dt}\log\rho(\Rt(r)) &= -\left(b(\Rt(r))  - v(\Rt(r))\right)\cdot D^{-1}\cdot v(\Rt(r)) \nonumber\\
    &= \underbrace{|v(\Rt(r))|^2_{D^{-1}}}_{\mathclap{\sdottot(R_t(r))}}  - \underbrace{b(\Rt(r))\cdot D^{-1}v(\Rt(r))}_{\mathclap{\sdotm(R_t(r))}},
\end{align}
consistent with~\eqref{eqn:stoch_epr}.

\section{Reverse-time dynamics}
Here, we describe in greater detail the reverse-time SDE~\eqref{eqn:anti_sde} considered in the main text. Specifically, we establish:

\begin{proposition}
    \label{prop:reverse:t}
    Let $r_t$ denote the solution of the SDE~\eqref{eqn:microscopic_sde} with initial condition $r_0$ drawn from the stationary density~$\rho$. Then, given any $T>0$, the path $\{r_t\}_{t\in[0,T]}$ has the same law as the path $\{r^\rev_{T-t}\}_{t\in[0,T]}$, where $r^\rev_t$ satisfies
\begin{equation}
\label{eqn:trev:ito}
    dr_t^\rev = -b(r_t^\rev)dt +2D\nabla\log\rho(r_t^\rev)dt + \sqrt{2D}dW_t,
\end{equation}
solved with the initial condition $r^\rev_0$ drawn from the stationary density~$\rho$.
\end{proposition}

Since $-b(r) +2D\nabla\log\rho(r) = b(r)- 2 v(r)$,~\eqref{eqn:trev:ito} is \eqref{eqn:anti_sde} written as an It\^o SDE.  The statement of Proposition~\ref{prop:3} is equivalent to saying that the path $\{r_{T-t}\}_{t\in[0,T]}$ has the same law as the path $\{r^\rev_{t}\}_{t\in[0,T]}$, i.e. $r^\rev_t$ is $r_t$ played in reverse. We will establish Proposition~\ref{prop:reverse:t} as a corollary of Proposition~\ref{prop:3} below. 
This result holds whether or not $D$ is invertible.
\subsection{Justification from the FPE~\eqref{eqn:fpe}} It is useful to first give a simple justification of \eqref{eqn:trev:ito}.  Using the identity $\nabla\cdot\left(D\nabla\rho_t\right) = -\nabla\cdot\left(D\nabla\rho_t\right) + 2\nabla\cdot \left(D\nabla\log\rho_t\,\rho_t\right)$ we can re-write the FPE~\eqref{eqn:fpe} as an equivalent equation with anti-diffusion
\begin{equation}
  \label{eqn:anti_fpe}
  \begin{aligned}
  \partial_t\rho_t +\nabla\cdot \left(\left(b - 2D\nabla\log\rho_t\right)\rho_t\right) + \nabla\cdot\left(D\nabla\rho_t\right) = 0.
  \end{aligned}
\end{equation}
So that~\eqref{eqn:anti_fpe} remains well-posed, it must be solved backwards in time.
Letting $\rho^\rev_{t} = \rho_{T-t}$, equation~\eqref{eqn:anti_fpe} can be written as a proper FPE
\begin{equation}
  \label{eqn:anti_fpe:rev}
  \begin{aligned}
  \partial_t\rho^\rev_t =\nabla\cdot \left(\left(b - 2D\nabla\log\rho_{T-t}\right)\rho^\rev_t\right) + \nabla\cdot\left(D\nabla\rho^\rev_t\right).
  \end{aligned}
\end{equation}

The SDE associated with the FPE~\eqref{eqn:anti_fpe:rev} is
\begin{equation}
\label{eqn:trev:ito:t}
    dr_t^\rev = -b(r_t^\rev)dt +2D\nabla\log\rho_{T-t}(r_t^\rev)dt + \sqrt{2D}dW_t,
\end{equation}
where $W_t$ is a standard Wiener process. At stationarity, $\rho_t(r) = \rho(r)$, so that the SDE~\eqref{eqn:trev:ito:t} reduces to the SDE~\eqref{eqn:trev:ito}.

\subsection{Derivation of the reverse-time stochastic integrator} Given $T>0$ and $N\in \N$, let $h=T/N$ and let us use the following stochastic integrator for the SDE~\eqref{eqn:microscopic_sde}
\begin{equation}
    \label{eq:stoch:int}
    r_{(n+1)h} = r_{nh} + h b(r_{nh}) + \sqrt{2hD}  \xi_n, \qquad n = 0,\ldots, N-1
\end{equation}
where $\{\xi_n\}_{n=0}^{N-1}$ are independent standard normal random variables and $r_0$ is drawn from $\rho$. Similarly, let us use the following  reverse-time integrator for the SDE~\eqref{eqn:trev:ito} in which, for convenience, we work with $\tilde r_t = r^\rev_{T-t}$
\begin{equation}
    \label{eq:stoch:int:rev}
    \tilde r_{(n-1)h} = \tilde r_{nh} - h b(\tilde r_{nh}) + 2D \nabla \log\rho(\tilde r_{nh}) + \sqrt{2hD}  \xi'_n, \qquad n = N, \ldots, 1
\end{equation}
where $\{\xi_n'\}_{n=1}^{N}$ are independent standard normal random variables and $\tilde r_{Nh=T}$ is drawn from $\rho$.
The following proposition shows that the forward-time integrator~\eqref{eq:stoch:int} and the reverse-time integrator~\eqref{eq:stoch:int:rev} generate paths that are equal in law:

\begin{proposition}
\label{prop:3}
Given any $T>0$ and $N\in \N$, let $h = T/N$ and let $\{r_{nh}\}_{n=0}^N$ be the forward path generated using the forward integrator \eqref{eq:stoch:int} with an initial condition $r_0$ drawn from the stationary density $\rho$. Similarly, let $\{\tilde r_{nh}\}_{n=0}^N$ be the reverse path generated using the reverse integrator \eqref{eq:stoch:int:rev} with a final $r^\rev_{Nh}$ drawn from the stationary density $\rho$. Then, up to error of order $O(h^{1/2})$, the paths $\{r_{nh}\}_{n=0}^N$ and $\{\tilde r_{nh}\}_{n=0}^N$ have the same law.
\end{proposition}

Note that, since the errors are of order $O(h^{1/2})$, they have no impact in the limit as $h\to0$ ($N \to\infty$), i.e. we take the continuous-time limit of both paths and compare them on the time interval~$[0,T]$. That is, Proposition~\ref{prop:3} implies Proposition~\ref{prop:reverse:t}

\begin{proof}[Proof of Proposition~\ref{prop:3}]
  We will establish the result by considering the probability of a pair $(r_t,r_{t+h})$ generated  using~\eqref{eq:stoch:int}  with that of a pair $(\tilde r_t,\tilde r_{t+h})$  generated  using~\eqref{eq:stoch:int:rev}, with $t=nh$ and any $n\in\{0,\ldots, N\}$. The transition probability densities associated with~\eqref{eq:stoch:int} and \eqref{eq:stoch:int:rev} read respectively
\begin{equation}
    \label{eq:transition:pdf}
    \rho(r_{t+h}|r_t ) = C^{-1} \exp\Big(-\frac1{4h}|r_{t+h}-r_t-h b(r_t)|_{D^{-1}}^2 \Big)
\end{equation}
and
\begin{equation}
    \label{eq:transition:rev:pdf}
    \rho^\rev(r_t|r_{t+h} ) = C^{-1} \exp\Big(-\frac1{4h}|r_t-r_{t+h}-h b^\rev(r_{t+h})|_{D^{-1}}^2 \Big)
\end{equation}
where $C$ is a common normalization constant and we denote $b^\rev(r) = -b(r)-2D \nabla \log \rho(r)$.
Since both~\eqref{eq:stoch:int} and~\eqref{eq:stoch:int:rev} preserve the stationary PDF $\rho$ to $O(h)$, the joint PDF of $(r_t,r_{t+h})$ can  be expressed either in terms of $\rho(r_{t+h}|r_t)$ and the stationary density $\rho(r_t)$ as
\begin{equation}
    \label{eq:forwrad}
    \begin{aligned} 
    \rho(r_t,r_{t+h}) & =  \rho(r_{t+h}|r_t ) \rho(r_t) \equiv C^{-1}\exp(A)\\
    \end{aligned}
\end{equation}
with 
\begin{equation}
    \label{eq:A:def}
    A = -\frac1{4h}|r_{t+h}-r_t-h b(r_t)|_{D^{-1}}^2 + \log(\rho(r_t))
\end{equation}
or in terms of as $\rho^\rev(r_t|r_{t+h} ) $ and the stationary density $\rho(r_{t+h})$ as 
\begin{equation}
    \label{eq:reversed}
    \begin{aligned} 
    \rho(r_t,r_{t+h}) & = \rho^\rev(r_t|r_{t+h} ) \rho(r_{t+h}) \equiv C^{-1}\exp(A^\rev)\\
    \end{aligned}
\end{equation}
with 
\begin{equation}
    \label{eq:A:revdef}
    A^\rev =  -\frac1{4h}|r_t-r_{t+h}-h b^\rev(r_{t+h})|_{D^{-1}}^2 + \log(\rho(r_{t+h}))
\end{equation}
Since \eqref{eq:forwrad} and \eqref{eq:reversed} must coincide up to order $o(h)$ for the stochastic integrators to be consistent, we must have $A^\rev- A = O(h^{3/2})$. Using $\Delta r = r_{t+h} - r_t= O(\sqrt{h})$, we see that
\begin{equation}
    \label{eq:fact2}
    \begin{aligned}
        A^\rev-A &=   - \tfrac1{2} \Delta r \cdot D^{-1} b^\rev(r_t+\Delta r) 
        - \tfrac{1}{4}h  |b^\rev(r_t+\Delta r)|_{D^{-1}}^2 + \log(\rho(r_t+\Delta r))\\
        & \quad - \tfrac1{2} \Delta r \cdot D^{-1} b(r_t) + \tfrac{1}{4}h  |b(r_t)|_{D^{-1}}^2 - \log(\rho(r_t))\\
        &=   - \tfrac1{2} \Delta r \cdot D^{-1} b^\rev(r_t) - \tfrac1{2} \Delta r \cdot D^{-1} \nabla b^\rev(r_t) \Delta r 
        - \tfrac{1}{4}h  |b^\rev(r_t)|_{D^{-1}}^2 \\
        & \quad - \tfrac1{2} \Delta r \cdot D^{-1} b(r_t) + \tfrac{1}{4}h  |b(r_t|_{D^{-1}}^2 + \Delta r\cdot \nabla \log \rho(r_t) \\
& \quad + \tfrac12 \rho^{-1}(r_t) \Delta r\Delta r: \nabla \nabla \rho(r_t) - \tfrac12 |\Delta r\cdot \nabla \log \rho(r_t)|^2 + O(h^{3/2})
    \end{aligned} 
\end{equation}
In this last expression we can also use $\Delta r\Delta r^\T \stackrel{d}{=} 2h D +O(h^{3/2})$ to arrive at (after reordering the terms)
\begin{equation}
    \label{eq:fact2b}
    \begin{aligned}
        A^\rev-A & = - \tfrac1{2} \Delta r \cdot D^{-1} b^\rev(r_t)  - \tfrac1{2} \Delta r \cdot D^{-1} b(r_t) + \Delta r\cdot \nabla \log \rho(r_t) \\
& \quad    - h \nabla \cdot b^\rev(r_t) - \tfrac{1}{4}h  |b^\rev(r_t)|_{D^{-1}}^2+ \tfrac{1}{4}h  |b(r_t)|_{D^{-1}}^2  \\
& \quad + h \rho^{-1}(r_t) \nabla \cdot D\nabla  \rho(r_t) - h|\nabla \log \rho(r_t)|_{D^{-1}}^2 + O(h^{3/2})
\end{aligned} 
\end{equation}
Zeroing the terms of order $O(\Delta r)= O(\sqrt{h})$ confirms that we must have
\begin{equation}
    \label{eq:bR:def}
    b^\rev(r) = -b(r) + 2D \nabla \log \rho(r),
\end{equation}
and we are left with
\begin{equation}
    \label{eq:A1A2}
    \begin{aligned}
    A^\rev-A & = h \nabla b(r_t) + hb(r_t)\cdot \nabla \log(\rho(r_t)) - h\rho^{-1}(r_t) \nabla \cdot D \nabla \rho (r_t)+ O(h^{3/2}).
    \end{aligned}
\end{equation}
Since $\rho$ is the stationary density, it satisfies $0 = \nabla \cdot (b\rho - D \nabla \rho)$.
After distributing the $\nabla$ and dividing by $\rho$, this equation can also be written as $0 = \nabla \cdot b + b \cdot \nabla \log \rho -  \rho^{-1}\nabla \cdot D\nabla \rho$. 
Using this expression in~\eqref{eq:A1A2} indicates that $A^\rev-A = O(h^{3/2})$, as claimed. 
\end{proof}

This proof is formal, but it can be made rigorous as shown in~\cite{anderson1982reverse}.

\subsection{Some estimators of $v(r)$}
\label{sec:estim:v}
Here we discuss some pointwise estimators of $v(r_t)$ in terms of $r_t$ which can be evaluated using the forward and reverse-time integrators in Eqs.~\eqref{eq:stoch:int} and~\ref{eq:stoch:int:rev}. They are based in the following result:

\begin{proposition}
    \label{prop:2}
    Let $r_t$ be a solution to the SDE~\eqref{eqn:microscopic_sde}. Then, given any vector-valued test function $\phi(r)$, any $t\ge0$, and any $\alpha\in[0,1]$, we have
\begin{equation}
    \label{eq:v:cond:exp:disc:0}
    \begin{aligned}
      &\lim_{h\to0}\frac{1}{2h}\left\langle\left(\phi(r_{t+ \alpha h})+\phi(r_{t-(1-\alpha)h})\right)\cdot\left( r_{t+\alpha h}-r_{t-(1-\alpha)h}\right) | r_t=r\right\rangle\\
      &\qquad = \phi(r)\cdot v(r) + (1-2\alpha) \left( \text{\rm tr} [D \nabla \phi(r)] +  \phi(r) \cdot D\nabla \log \rho(r)\right),
    \end{aligned}
\end{equation}
where $\circ$ denotes the Stratonovich product and where $\langle \cdot | r_{t}=r\rangle$ denotes an expectation over both the noise $W_t$ in~\eqref{eqn:microscopic_sde} and the initial condition $r_0$ drawn from $\rho$, conditioned on the event $r_{t} = r$.
\end{proposition}

We stress that there is no contradiction between \eqref{eq:v:cond:exp:disc:0} and \eqref{eq:v:exp} since 
\begin{equation}
\label{eq:zero:abg}
    \begin{aligned}
      \int_\Omega \left( \text{tr} [D \nabla \phi(r)] +  \phi(r) \cdot D\nabla \log \rho(r)\right) \rho(r) dr  = \int_\Omega \left( -\phi(r) \cdot D \nabla \rho(r) + \phi(r)\cdot D \nabla \rho(r) \right) dr = 0,
     \end{aligned}
\end{equation}
where we used integration by parts for the first term under the integral and the identity $\nabla \log \rho \, \rho = \nabla \rho$ for the second. That is, the expectation of the right hand-side of \eqref{eq:v:cond:exp:disc:0} is $\int_\Omega v(r) \cdot \phi(r) \rho(r) dr$ for all $\alpha \in [0,1]$, consistent with~\eqref{eq:v:exp}.

The second term on the right-hand side of~\eqref{eq:v:cond:exp:disc:0} is  zero if we take $\alpha=1/2$, in which case this equation reduces to
\begin{equation}
    \label{eq:v:cond:exp:disc}
    \begin{aligned}
    \lim_{h\to0}\frac{1}{2h}\left\langle\left(\phi(r_{t+ h/2})+\phi(r_{t-h/2})\right)\cdot\left( r_{t+h/2}-r_{t-h/2}\right) | r_t=r\right\rangle= \phi(r)\cdot v(r).
    \end{aligned}
\end{equation}
We stress, however, that \eqref{eq:v:cond:exp:disc} cannot be used  to estimate $v$ directly.
Indeed,~\eqref{eq:v:cond:exp:disc} requires generating $r_{t-h}$ from $r_t$. As shown in the proof of Proposition~\ref{prop:2}, this requires use of the reverse-time integrator in~\eqref{eq:stoch:int:rev}, which itself involves the unknown $\nabla \log \rho$ entering $v$. 
The only way to use~\eqref{eq:v:cond:exp:disc:0} without prior knowledge of $\nabla \log \rho$ is when $\alpha=0$, but in that case it reduces to
\begin{equation}
    \begin{aligned}
    \lim_{h\to0}\frac{1}{2h}\left\langle\left(\phi(r_{t+ h})+\phi(r_{t})\right)\cdot\left( r_{t+h}-r_{t}\right) | r_t=r\right\rangle= \phi(r)\cdot b(r) + \text{\rm tr} [D \nabla \phi(r)],
    \end{aligned}
\end{equation}
which gives no information about $v$.

Finally, let us note that this second term on the right-hand side of~\eqref{eq:v:cond:exp:disc:0} is  zero for all $\alpha \in[0,1]$ if we take $\phi(r) = D^{-1} v(r)$, since it then reduces to $\text{tr} [D \nabla D^{-1}v] +  \nabla \log \rho \cdot D D^{-1} v = \nabla \cdot v + \nabla \log \rho \cdot v = 0$, where we used  $\nabla \cdot (v\rho)=0$ by the score-based formulation of the stationary FPE~\eqref{eqn:stationary_fpe_score}.  Interestingly the choice $\phi(r) = D^{-1} v(r)$ leads to the total EPR $\sdotsys(r)$, see \eqref{eqn:stochastic_entr_sde_4}.

\begin{proof}[Proof of Proposition~\ref{prop:2}]
Conditioned on $r_t=r$, we have
\begin{equation}
    \label{eq:for:a}
    r_{t+(1-\alpha) h} =  r + (1-\alpha) h b(r) + \sqrt{2(1-\alpha) h D}\, \eta + O(h^{3/2})
\end{equation}
and
\begin{equation}
    \label{eq:rev:a}
    r_{t-\alpha h} = r - \alpha h b(r) +2\alpha h \nabla \log \rho(r) + \sqrt{2\alpha h D}\, \eta' + O(h^{3/2})
\end{equation}
where  $\eta$ and $\eta'$ are independent standard Gaussian variables with mean zero and covariance identity: \eqref{eq:rev:a} is consistent with the reverse-time integrator given in \eqref{eq:stoch:int:rev}.
Eqs.~\eqref{eq:for:a} and~\ref{eq:rev:a} imply that
\begin{equation}
    \label{eq:diff:r}
    r_{t+(1-\alpha) h}-r_{t-\alpha h} = h b(r) - 2\alpha h \nabla \log \rho(r) + \sqrt{2(1-\alpha) h D}\, \eta - \sqrt{2\alpha h D}\, \eta' + O(h^{3/2})
\end{equation}
Eqs.~\eqref{eq:for:a} and~\ref{eq:rev:a} imply that, conditional on $r_t=r$, we have
\begin{equation}
    \label{eq:phi:f}
    \phi(r_{t-\alpha h})+\phi(r_{t+(1-\alpha) h}) = 2 \phi(r) + \nabla \phi(r) \, \big(\sqrt{2(1-\alpha) h D}\, \eta +\sqrt{2\alpha h D}\, \eta'\big)+ O(h)
\end{equation}
Inserting Eqs.~\eqref{eq:diff:r} and \ref{eq:phi:f} in \eqref{eq:v:cond:exp:disc:0} we arrive at
\begin{equation*}
    \begin{aligned}
    & \lim_{h\to0}\frac{1}{2h}\left\langle\left(\phi(r_{t+ \alpha h})+\phi(r_{t-(1-\alpha)h})\right)\cdot\left( r_{t+\alpha h}-r_{t-(1-\alpha)h}\right) | r_t=r\right\rangle\\
    & = \lim_{h\to0+} \frac1{2h} \Big\langle \big(2 \phi(r) + \nabla \phi(r) \, \big(\sqrt{2(1-\alpha) h D}\, \eta +\sqrt{2\alpha h D}\, \eta'\big)+ O(h)\big)\\
    & \qquad \qquad \qquad \cdot \big(h b(r) - 2\alpha h \nabla \log \rho(r) + \sqrt{2(1-\alpha) h D}\, \eta - \sqrt{2\alpha h D}\, \eta' + O(h^{3/2})\big))\Big\rangle\\
    & = \lim_{h\to0+} \frac1{2h} \left(2 h \phi(r)\cdot (b(r)-2\alpha D \nabla \log \rho(r)) + 2(1-\alpha) h  \text{tr}[D \nabla \phi(r)] -2\alpha h  \text{tr}[D \nabla \phi(r)] + O(h^{3/2})\right)\\
    & = \phi(r) \cdot b(r) - 2\alpha \phi(r)\cdot D \nabla \log \rho(r)  + (1-2\alpha)\text{tr}[D \nabla \phi(r)]\\
    & = \phi(r) \cdot v(r) + (1- 2\alpha)\left( \phi(r)\cdot D \nabla \log \rho(r)  + \text{tr}[D \nabla \phi(r)]\right)
    \end{aligned}
\end{equation*}
where in the last step we used $v=b-D\nabla \log \rho$.
\end{proof}

\section{Path KL and global EPR}
\subsection{Measure-theoretic (Girsanov) derivation}
We now consider the SDEs corresponding to~\eqref{eqn:microscopic_stochastic_general} (equivalent to~\eqref{eqn:microscopic_sde}) and~\eqref{eqn:anti_sde} (equivalent to~\eqref{eqn:trev:ito:t}), repeated here for convenience
\begin{equation}
    \label{eqn:fwd_rev}
    \begin{aligned}
        dr_t &= b(r_t)dt + \sqrt{2D}dW_t,\\
        dr_t^\rev &= b(r_t^\rev)dt -2v(r_t^\rev)dt + \sqrt{2D}dW_t.
    \end{aligned}
\end{equation}
We first observe that the difference in drifts is given by twice the current velocity $v(r)$.
Given this relation, we assume that $D$ is invertible and  Novikov's condition (see~\cite{oksendal2003sde}) holds, i.e. that
\begin{equation}
    \label{eqn:novikov}
    \left\langle\exp\left(\int_0^T |v(r_t)|_{D^{-1}}^2dt\right)\right\rangle < \infty,
\end{equation}
where the angular bracket denotes an expectation over the noise in the forward-time dynamics as well as over initial conditions drawn from the stationary density $\rho$. 
\eqref{eqn:novikov} guarantees that~\eqref{eqn:fwd_rev} fits the conditions of the Girsanov Theorem III, see~\cite{oksendal2003sde}.
Then, assuming a fixed initial condition $r_0 = r_0^\rev$ for both dynamics, we obtain the Radon-Nikodym derivative
between the measure $\calP^\rev$ of the time-reversed process $\phi_T^\rev = \{r^\rev_t\}_{t\in[0,T]}$ and the measure $\calP$ of the forward process $\phi_T = \{r_t\}_{t\in[0,T]}$
\begin{equation}
\label{eqn:Girsanov}
\frac{\calP^\rev(\phi_T | r_0)}{\calP(\phi_T | r_0)} = \exp\left(-\int_0^T |v(r_t)|_{D^{-1}}^2dt + \int_0^t v(r_t)\cdot D^{-1}(dr_t - b(r_t) dt)\right).
\end{equation}
Assuming that both the forward- and the reverse-time SDE~\eqref{eqn:fwd_rev} are soved with initial conditions drawn from the stationary density~$\rho$, we have that $\calP(\phi_T) = \calP(\phi_T | r_0)\rho(r_0)$ and $\calP^\rev(\phi_T) = \calP^\rev(\phi_T|r_0)\rho(r_0)$, so that
\begin{equation}
\begin{aligned}
    \dot{S}_{\text{tot}} &= \frac{1}{T}\left\langle\log\left(\frac{\calP^\rev(\phi_T)}{\calP(\phi_T)}\right)\right\rangle \\
    &= -\frac{1}{T}\left\langle\log\left(\frac{\calP^\rev(\phi_T | r_0)}{\calP(\phi_T | r_0)}\right)\right\rangle\\
    &=  \frac{1}{T}\left\langle\int_0^T|v(r_t)|_{D^{-1}}^2dt\right\rangle - \frac{\sqrt{2}}{T}\left\langle\int_0^T v(r_t)\cdot D^{-1/2}dW_t\right\rangle\\
    &= \frac{1}{T}\left\langle\int_0^T|v(r_t)|_{D^{-1}}^2dt\right\rangle\\
    &= \int_\domR |v(r)|_{D^{-1}}^2\rho(r)dr.
\end{aligned}
\end{equation}
Above, we used $dr_t - b(r_t) dt = \sqrt{2D} dW_t$ which follows from the SDE~\eqref{eqn:microscopic_stochastic_general}, the martingale property $\left\langle\int_0^T v(r(t))\cdot D^{-1/2}dW_t\right\rangle = 0$, and the fact that the density of $r_t$ is $\rho$ at all times $t>0$ if the initial conditions were drawn from $\rho$.

\subsection{Path integral (physics-style) derivation}
We now consider an analogous derivation to the previous section, instead making use of a path integral formulation~\cite{onsager_fluctuations_1953}.
We first write the path measures
\begin{equation}
    \label{eqn:path_measures}
    \begin{aligned}
    \calP(\phi_T) &= \frac{1}{Z}\exp\left(-\int_0^T \calL(\rt,\drt)dt\right) \rho(r_0),\\
    \calP^\rev(\phi_T) &= \frac{1}{Z}\exp\left(-\int_0^T \calL^\rev(r_t, \dot r_t)dt\right) \rho(r_0),
    \end{aligned}
\end{equation}
with $\{r_t:t\in[0,T]\}$ viewed as a dummy integration variable. We have defined the forward and reverse Lagrangians
\begin{equation}
    \label{eqn:actions}
    \begin{aligned}
        \calL(r, \dot r) &= \frac{1}{4}\left(\dot{r}_t-b(r_t)\right)\cdot D^{-1}\left(\dot{r_t} - b(r_t)\right)\\
        \calL^\rev(r, \dot r) &= \frac{1}{4}\left(\dot{r}_t - b(r_t) - 2v_t(r_t)\right)\cdot D^{-1}\left(\dot{r}_t - b(r_t) - 2v_t(r_t)\right)
    \end{aligned}
\end{equation}
Equation~\eqref{eqn:path_measures} is formal in the sense that the common normalization factor $Z$ is technically infinite, but it is common to both measures and cancels in the ratio defining the EPR.
It then follows that
\begin{equation}
\begin{aligned}
    \dot{S}_{\text{tot}} &= \frac{1}{T}\left\langle\log\left(\frac{\calP(\phi_T)}{\calP^\rev(\phi_T)}\right)\right\rangle\\
    &= \frac{1}{T}\left\langle\int_0^T\left(\calL(r_t,\dot r_t) - \calL^\rev(r_t,\dot r_t)\right)dt\right\rangle.
\end{aligned}
\end{equation}

Manipulating the expression for $\calL^\rev$,
\begin{equation}
\begin{aligned}
    \calL^\rev(r) &= \frac{1}{4}\left(\dot{r}_t - b(r_t)\right)\cdot D^{-1}\left(\dot{r}_t - b(r_t)\right) + |v(r_t)|_{D^{-1}}^2 - \left(\dot{r}_t - b(r_t)\right)\cdot D^{-1}v(r_t)\\
    &= \calL(r_t) + |v(r_t)|_{D^{-1}}^2 - \left(\dot{r}_t - b(r_t)\right)\cdot D^{-1}v(r_t)\\
    &= \calL(r_t) + |v(r_t)|_{D^{-1}}^2 - \sqrt{2D^{-1}}\eta(t) \cdot v(r_t).
\end{aligned}
\end{equation}
In the last line, we have used the identity $\dot{r}_t = b(r_t) + \sqrt{2D} \eta(t)$, which holds when taking an expectation over the forward paths.
Hence, we find that
\begin{equation}
\begin{aligned}
    \dot{S}_{\text{tot}} &= \frac{1}{T}\left\langle\int_0^T\left(|v(r_t)|_{D^{-1}}^2 - \sqrt{2D^{-1}}\eta(t) \cdot v(r_t)\right)dt\right\rangle\\
    &= \frac{1}{T}\left\langle\int_0^T |v(r_t)|_{D^{-1}}^2dt - \int_0^T\sqrt{2D^{-1}}\eta(t) \cdot v(r_t) \right\rangle\\
    &= \frac{1}{T}\left\langle\int_0^T |v(r_t)|_{D^{-1}}^2dt\right\rangle\\
    &= \int_\domR |v(r)|_{D^{-1}}^2\rho(r)dr.
\end{aligned}
\end{equation}
\section{Review of score matching}
In this section, we review the score matching loss~\eqref{eqn:score_matching} introduced by~\cite{hyvarinen_estimation_2005} and used heavily in the score-based diffusion literature~\cite{song_maximum_2021, song_score-based_2021}.

\subsection{Loss derivation}
We first note that to estimate the score a natural objective is the $\calL_2$ error
\begin{equation}
    \label{eqn:sm_l2}
    \tilde{\calL}_{\text{sm}}[\hat h] = \E_{\rho}\left[|\hat h - \nabla\log\rho|^2\right].
\end{equation}
Expanding the square,
\begin{equation}
    \label{eqn:sm_l2_expand}
    \tilde{\calL}_{\text{sm}}[\hat h] = \E_{\rho}\left[|\hat h|^2 - 2\nabla\log\rho\cdot \hat h + |\nabla\log\rho|^2\right].
\end{equation}
The cross term can be integrated by parts (Stein's identity)
\begin{align}
    \E_{\rho}\left[\nabla\log\rho\cdot \hat h\right] &= \int_\domR  \nabla\log\rho(r)\cdot \hat h(r)\rho(r)dr,\\
    &= \int_\domR \nabla\rho\cdot \hat h(r)dr,\\
    &= -\int_\domR \nabla\cdot \hat h(r)\rho(r)dr,\\
    &= \E_{\rho}\left[-\nabla\cdot \hat h\right].
\end{align}
Plugging this back in to~\eqref{eqn:sm_l2_expand}, we find
\begin{equation}
    \tilde{\calL}_{\text{sm}}[\hat h] = \E_{\rho}\left[|\hat h|^2 + 2\nabla\cdot \hat h + |\nabla\log\rho|^2\right].
\end{equation}
Neglecting the constant term $\E_{\rho}\left[|\nabla\log\rho|^2\right]$ gives the score matching loss $\calL_{\text{sm}}$ in~\eqref{eqn:score_matching}, repeated for completeness:
\begin{equation}
    \calL_{\text{sm}}[h] = \E_{\rho}\left[|h|^2 + 2\div h\right].
\end{equation}
Moreover, because~\eqref{eqn:sm_l2} is zero for $h = \nabla\log\rho$ and $\tilde\calL[h]\geq 0$ for all $h$, we find that 
\begin{equation}
    \label{eqn:sm_optimal}
    \calL_{\text{sm}}[\nabla\log\rho] = -\E_{\rho}\left[|\nabla\log\rho|^2\right]
\end{equation}
which is one of the quantitative tests for convergence introduced in the main text.

\subsection{Convexity and global minimizer}
It is clear that~\eqref{eqn:sm_l2} is convex in $\hat h$ and that its unique minimizer is $\hat h = \nabla\log\rho$.
We now show that the same holds for $\Lsm$ in~\eqref{eqn:score_matching}.
We first observe that the first variation with respect to $\hat h$ is given by
\begin{equation}
    \label{eqn:first_var}
    \frac{\delta}{\delta \hat h(r)}\calL_{\text{sm}}[\hat h] = 2\hat h(r)\rho(r) - 2\cdot \nabla\rho(r).
\end{equation}
Setting $\frac{\delta}{\delta \hat h(r)}\Lsm[h^*] = 0$ recovers the condition
\begin{equation}
    h^*(r) = \nabla\log\rho(r),
\end{equation}
after division by $\rho$.
Hence, the only extreme points $h^*(r)$ satisfy $h^*(r) = \nabla\log\rho(r)$.
Moreover, the second variation is given by
\begin{equation}
    \frac{\delta^2}{\delta \hat h(r)\delta \hat h(r')}\Lsm[h] = 2\rho(r) \delta(r-r') \geq 0
\end{equation}
so that $\Lsm$ is convex and $h^*(r) = \nabla\log\rho(r)$ is the unique global minimizer.

\subsection{Denoising loss}
\label{sec:denoising}
We now derive an equivalent of the loss $\Lsm$ that avoids computation of the divergence $\div \hat h$ by use of the transition probabilities for the SDE~\eqref{eqn:microscopic_stochastic_general} that generate the density $\rho$.
This loss was introduced by~\cite{doucet_score-based_2022}, but we provide a description here for completeness.
Consider a small timestep $\Delta t$, and note that
\begin{equation}
    \rho(r_{t+\Delta t}) = \int_\domR \rho(r_{t+\Delta t} | r_t)\rho(r_t)dr_t
\end{equation}
by stationarity of the dynamics.
It then follows that
\begin{equation}
    \nabla \rho(r_{t+\Delta t}) = \int_\domR \nabla \rho(r_{t+\Delta t} | r_t)\rho(r_t)dr_t,
\end{equation}
where $\nabla = \nabla_{r_{t+\Delta t}}$.
Hence, we may write that
\begin{equation}
\begin{aligned}
    \E_{\rho}\left[\nabla\log\rho\cdot \hat h\right] &= 
    \int_\domR  \nabla\log\rho(r_{t+\Delta t})\cdot \hat h(r_{t+\Delta t})\rho(r_{t+\Delta t})dr_{t+\Delta t}\\
    &= \int_\domR  \nabla\rho(r_{t+\Delta t})\cdot \hat h(r_{t+\Delta t})dr_{t+\Delta t}\\
    &= \int_\domR  \left(\int_\domR  \nabla \rho(r_{t+\Delta t}|r_{t})\rho(r_t)dr_t\right)\cdot \hat h(r_{t+\Delta t})dr_{t+\Delta t}\\
    &= \int_\domR  \left(\int _\domR \nabla \log \rho(r_{t+\Delta t}|r_{t})\rho(r_{t+\Delta t}|r_t)\rho(r_t)dr_t\right)\cdot \hat h(r_{t+\Delta t})dr_{t+\Delta t}\\
    &= \int_{\domR \times\domR }\nabla\log\rho(r_{t+\Delta t} | r_t)\cdot \hat h(r_{t+\Delta t}) \rho(r_{t+\Delta t}|r_t)\rho(r_t)dr_{t+\Delta t}dr_t\\
    &= \E_{r_{t+\Delta t}, r_t}\left[\nabla\log\rho(r_{t+\Delta t}|r_t)\cdot \hat h(r_{t+\Delta t})\right]
\end{aligned}
\end{equation}
Similarly,
\begin{equation}
\begin{aligned}
    \E_{\rho}\left[|\hat h|^2\right] &= \int_\domR  |\hat h(r_{t+\Delta t})|^2\rho(r_{t+\Delta t})dr_{t+\Delta t}\\
    &= \int_{\domR \times\domR } |\hat h(r_{t+\Delta t})|^2\rho(r_{t+\Delta t}|r_t)\rho(r_t)dr_{t+\Delta t}dr_t\\
    &= \E_{r_{t+\Delta t}, r_t}\left[|\hat h(r_{t+\Delta t})|^2\right].
\end{aligned}
\end{equation}
Hence, letting $\sim$ denote equivalent up to a constant with respect to $\hat h$, we find that
\begin{equation}
\begin{aligned}
    \E_{\rho}\left[|\hat h-\nabla\log\rho|^2\right] &= \E_{\rho}\left[|\hat h|^2\right] -2\E_{\rho}\left[\nabla\log\rho\cdot \hat h\right] \\
    &\qquad + \E_{r_{t+\Delta t}, r_t}\left[|\nabla\log\rho(r_{t+\Delta t}|r_t)|^2\right]\\
    &\qquad\qquad + \E_{\rho}\left[|\nabla\log\rho|^2\right] - \E_{r_{t+\Delta t}, r_t}\left[|\nabla\log\rho(r_{t+\Delta t}|r_t)|^2\right]\\
    &\sim \E_{\rho}\left[|\hat h|^2\right] -2\E_{\rho}\left[\nabla\log\rho\cdot \hat h\right] + \E_{r_{t+\Delta t}, r_t}\left[|\nabla\log\rho(r_{t+\Delta t}|r_t)|^2\right]\\
    &= \E_{r_{t+\Delta t}, r_t}\left[|\hat h(r_{t+\Delta t})|^2\right] - 2\E_{r_{t+\Delta t}, r_t}\left[\nabla\log\rho(r_{t+\Delta t}|r_t)\cdot \hat h(r_{t+\Delta t})\right]\\
    &\qquad +  \E_{r_{t+\Delta t}, r_t}\left[|\nabla\log\rho(r_{t+\Delta t}|r_t)|^2\right]\\%
    &= \E_{r_{t+\dt}, r_t}\left[|\hat h(r_{t+\dt}) - \nabla\log\rho(r_{t+\dt}|r_t)|^2\right].
\end{aligned}
\end{equation}
For an Euler-Maruyama discretization of~\eqref{eqn:microscopic_stochastic_general}, we have the exact relation for the transition probability
\begin{equation}
\begin{aligned}
    \rho(r_{t+\dt}|r_t) &= \N\left(r_t + \dt b(r_t), 2\dt D\right),\\
    &= \frac{1}{Z}\exp\left(-\tfrac{1}{4\dt}\left(r_{t+\dt} - r_t - \dt b(r_t)\right)\cdot D^{-1} \left(r_{t+\dt} - r_t - \dt b(r_t)\right)\right)
\end{aligned}
\end{equation}
Let us define the noise $\xi$ by the relation
\begin{equation}
    \sqrt{2\dt D}\xi_t = r_{t+\dt} - r_t - \dt b(r_t).
\end{equation}
Then we can compute the transition score exactly,
\begin{equation}
\begin{aligned}
    \nabla\log\rho(r_{t+\dt}|r_t) &= -\frac{1}{2\dt}D^{-1}\left(r_{t+\dt} - r_t - \dt b(r_t)\right)\\
    &= -\left(2\dt D\right)^{-1/2}\xi_t.
\end{aligned}
\end{equation}
Finally, we obtain the ``denoising'' loss function
\begin{equation}
    \Ld[\hat h] = \E_{r_{t+\dt}, r_t, \xi_t}\left[|\hat h(r_{t+\dt}) + \left(2\dt D\right)^{-1/2}\xi_t|^2 \right].
\end{equation}
By the preceding derivations, up to discretization errors already present in the dataset, $\Ld$ is equivalent to $\Lsm$.

\section{Transformer architecture}
In this section, we provide a more detailed description of the particle transformer architecture (Figure~\ref{fig:network}) introduced in this work.

\subsection{Multi-head attention}
A critical component of the transformer encoder block used in the particle transformer is the multi-head self-attention module.
We make use of the scaled dot-product attention introduced by~\cite{vaswani_attention},
\begin{align}
    \attn(Q, K, V) = \softmax\left(\frac{QK^\T}{\sqrt{d_k}}\right)V.
\end{align}
Above, $Q\in\R^{n_q \times d_q}$ denotes the matrix of queries, $K\in\R^{n_v \times d_q}$ denotes the matrix of keys, and $V\in\R^{n_v\times d_v}$ denotes the matrix of values.
Here, we use a multi-head self attention module with $h$ heads that sets the queries, keys, and values equal
\begin{equation}
\begin{aligned}
    \multihead(X) &= \concat(\head_1, \hdots, \head_h)W^O,\\
    \head_i &= \attn(XW_i^Q, XW_i^K, XW_i^V),
\end{aligned}
\end{equation}
so that $n_q = n_v$ and $d_q = d_v$.
Above, $W_i^Q, W_i^K, W_i^V \in \R^{d_q \times d_h}$, where $d_h$ denotes the dimension of each head.
Typically, $d_h = d_q / h$ with $d_q$ a multiple of $h$, and this is the convention we adopt in this work.
In our architecture, the tokens will correspond to particles, so we let $n_q = N$ and define $d_e = d_q$ to be the embedding dimension coming from the MLP embeddings (see Figure~\ref{fig:network}).

\subsection{Encoder block}
The encoder block is given by $L$ layers, each consisting of four data transformations applied in succession:
\begin{enumerate}
    \item A multi-head self-attention module with a residual connection.
    \item A LayerNorm~\cite{ba_layer_2016} block.
    \item A single-layer MLP applied to each particle individually with a residual connection.
    \item A LayerNorm block.
\end{enumerate}
Mathematically, we can thus describe each layer by the assignments
\begin{equation}
\begin{aligned}
    X &\leftarrow X + \multihead(X),\\
    X &\leftarrow \layernorm(X),\\
    X &\leftarrow X + \psi(X),\\
    X &\leftarrow \layernorm(X),
\end{aligned}
\end{equation}
where $\psi$ denotes a (shared) MLP applied to each row of $X$.
At the first layer, the input $X$ to the multi-head self-attention module is given by the matrix of particle state embeddings. 
That is, letting $\varphi(x_i, g_i) = \left(\varphi_x(x_i)^\T, \varphi_g(g_i)^\T\right) \in \R^{d_e}$,
\begin{equation}
    X = \begin{pmatrix} \varphi(x_1)^\T\\%
    \varphi(x_2)^\T \\ %
    \vdots \\ %
    \varphi(x_N)^\T
    \end{pmatrix}.
\end{equation}
The input to each following layer is given by the output of the preceding layer.
Importantly, the $\multihead(\cdot)$ operation is equivariant with respect to permutations of the rows of its input, i.e., permutations of the particles.
The output of the encoder block is then decoded by an MLP, again applied row-wise to the particle states, to produce the approximate score model $\hat{h}$.

\subsection{Rollout attention}
The output of $\multihead(\cdot)$ at each layer (the attention map) provides a way to visualize and interpret the prediction of the transformer.
Yet, it is well-known that the attention map becomes more difficult to interpret deeper in the encoder block because it describes higher-order interactions between the input tokens.
Rollout attention is a simple method that combines the attention map of each layer into a single map that can be visualized~\cite{abnar_quantifying_2020}.
Letting the attention map for layer $i$ be given by $A_i$, the formula used to obtain the rollout attention map $A$ visualized in Figure~\ref{fig:64_attention} is given by
\begin{equation}
   \label{eqn:rollout_attn} 
   A = \prod_{i=1}^L \tfrac{1}{2}\left(I + A_i\right),
\end{equation}
i.e., we take the successive product of the attention map in each layer, accounting for the residual connection by the addition of the identity matrix $I$.
The factor $\tfrac{1}{2}$ ensures that the resulting rollout attention map remains normalized.

\section{Numerical experiments}
\subsection{Smoothed potential}
\begin{figure}
    \centering
    \includegraphics[width=0.65\textwidth]{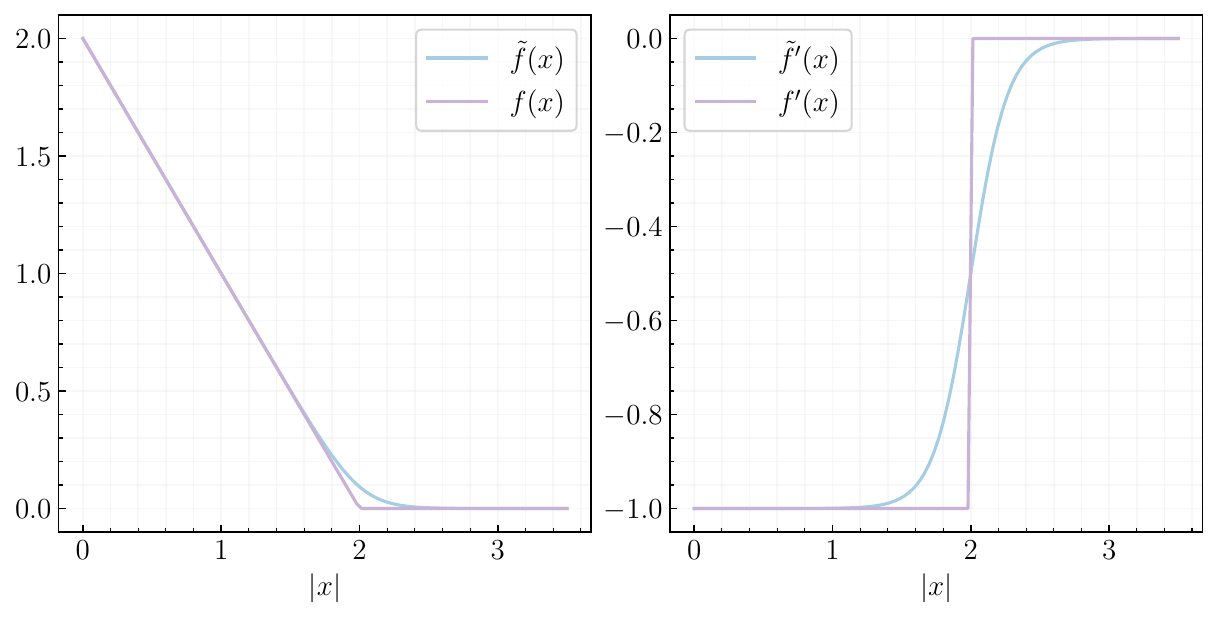}
    \caption{\textbf{Force smoothing.} (Left) Comparison of the smoothed force $\tilde{f}(x)$ given by~\eqref{eqn:smooth_force}, to the hard cutoff force $f(x)$ given by~\eqref{eqn:force_hard_cutoff} with $r=1$, as a function of the norm $|x|$.
    The function values are similar over most of the domain, except for the cutoff point at $|x|=2$.
    (Right) Comparison of the derivatives $\tilde f'(x)$ and $f(x)$ as a function of the norm $|x|$.
    While $f'(x)$ is discontinuous, which poses difficulties for learning, $\tilde{f}'(x)$ is smooth.}
    \label{fig:force}
\end{figure}
We found that the hard cutoff force listed in the main text
\begin{equation}
    \label{eqn:force_hard_cutoff}
    f(x) = (2r-|x|)\frac{x}{|x|}\Theta(2r-|x|)
\end{equation}
made it difficult to obtain quantitative accuracy with our learning algorithm.
This is likely because we use smooth activation functions in our networks, and $f$ has a discontinuous derivative.
To alleviate this difficulty, we used a smoothed version of~\eqref{eqn:force_hard_cutoff}
\begin{equation}
    \label{eqn:smooth_force}
    \begin{aligned}
    \tilde{f}(x) &= \softplus(2r-|x|)\frac{x}{|x|},\\
    \softplus(x) &= \tfrac{1}{\beta}\log\left(1 + \exp(\beta x)\right),
    \end{aligned}
\end{equation}
with $\beta = 7.5$. We found the resulting dynamics to be visually very similar, but the learning algorithm converged faster and to higher accuracy.
A comparison of $f(x)$ and $\tilde{f}(x)$ is shown in Figure~\ref{fig:force}.

\subsection{Two particles on a ring}
\begin{figure*}[t!]
\centering
\begin{tabular}{c}
\begin{overpic}[width=0.9\textwidth]{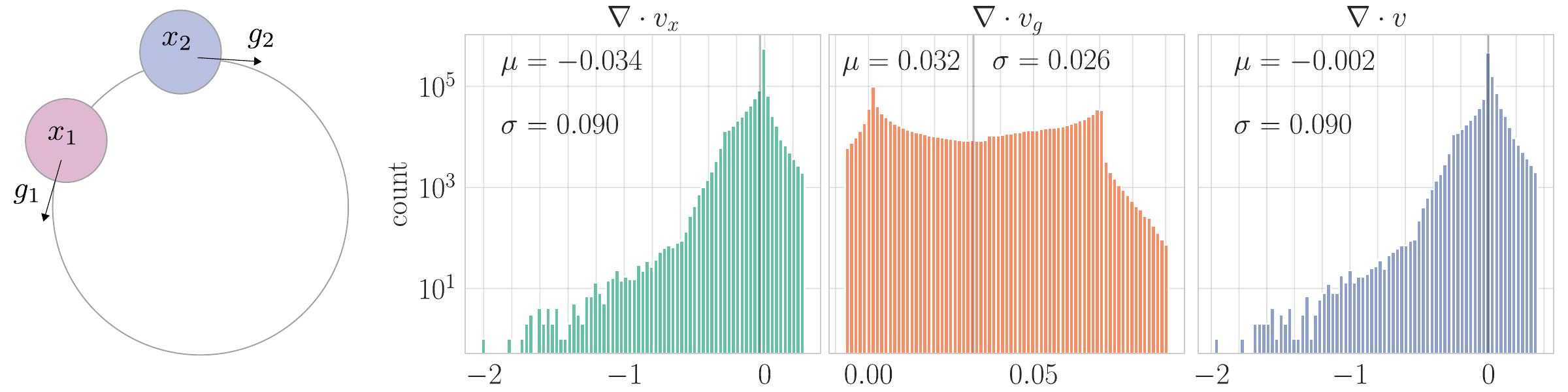}
\put(0, 25){\textbf{A}}
\put(25, 25){\textbf{B}}
\end{overpic}\\\\
\begin{overpic}[width=0.75\textwidth]{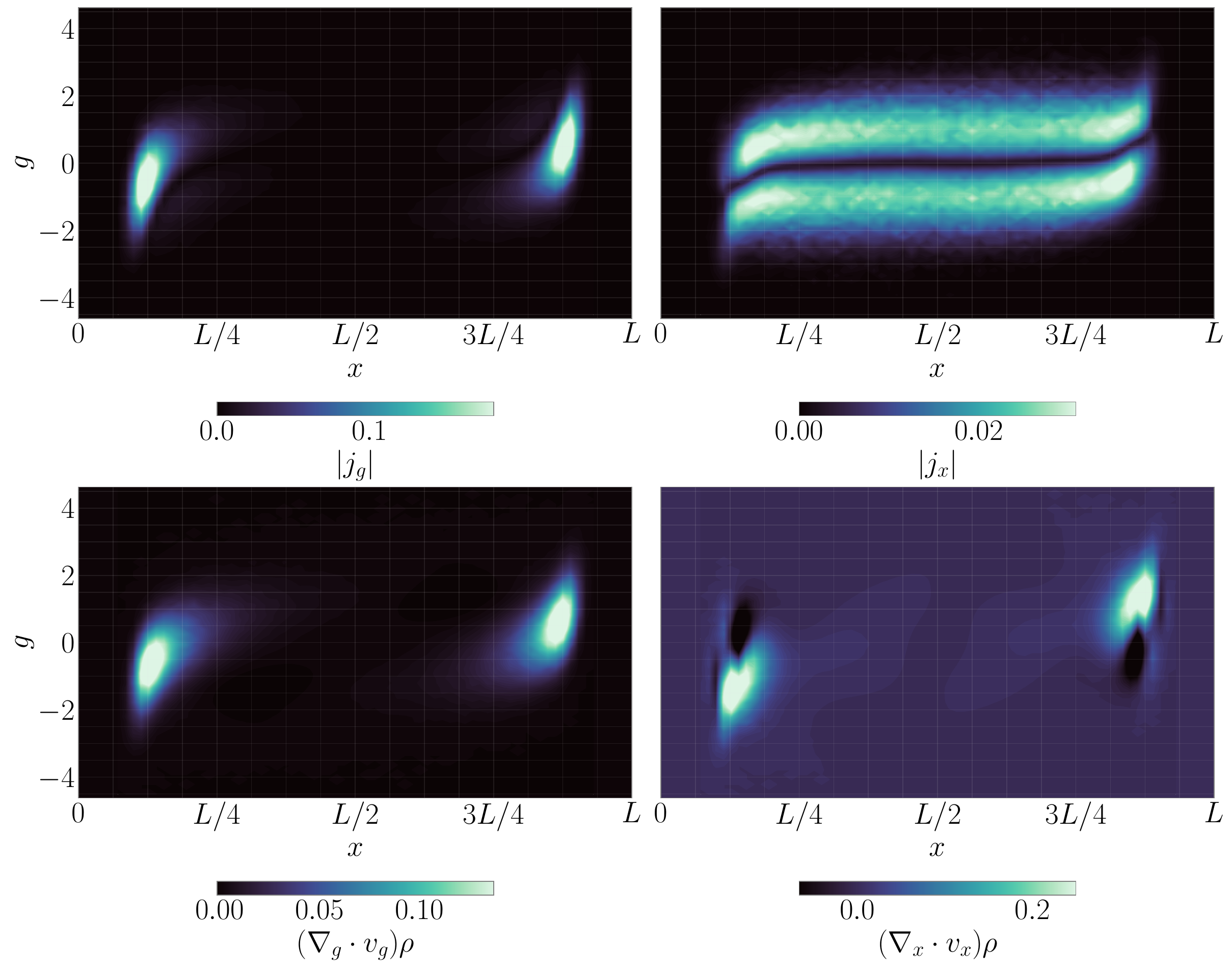}
\put(0, 75){\textbf{C}}
\put(0, 40){\textbf{D}}
\end{overpic}\\
\end{tabular}
\caption{\textbf{Two active particles on a ring.} (A) Diagram of the system. We study a system of two active swimmers in one dimension on the torus. The particles experience a pairwise repulsive force that prevents their overlap. (B) Statistics of the entropy production rate. (C) Decomposition of the probability current into orientational and translational degres of freedom. The orientational degrees of freedom have current dominated at the collision points, while the translational degrees of freedom have current at the collisions and during transitions between the two modes. (D) Decomposition of the system EPR into orientational and translational degrees of freedom. The orientational degrees of freedom have system EPR dominated by the collision point, while the translational degrees of freedom are of larger magnitude and hence similar to $\div v$ in Figure~\ref{fig:twop_div_v} from the main text.}
\label{fig:two_particle}
\end{figure*}

\subsubsection{Dynamics} We consider~\eqref{eqn:interacting_particles} for two particles ($N=2$) in one dimension ($d=1$) with periodic boundary conditions (Figure~\ref{fig:two_particle}A). In this case, the corresponding four-dimensional dynamics is simple enough to be written explicitly
\begin{equation}
    \label{eqn:two_particle}
    \begin{aligned}
    \dot x^1_t &= \mu f(x^1_t - x^2_t)dt  + v_0 g_t^1 + \sqrt{2\epsilon}\,\eta_{x}^1(t), \\
    \dot g^1_t &= -\gamma g^1_t  + \sqrt{2\gamma}\,\eta_{g}^1(t), \\
    \dot x^2_t &= \mu f(x^1_t - x^2_t)  + v_0 g_t^2  + \sqrt{2\epsilon}\,\eta_{x}^2(t), \\
    \dot g^2_t &= -\gamma g^2_t  + \sqrt{2\gamma}\,\eta_{g}^2(t),
    \end{aligned}
\end{equation}
where $f(x)$ is given by~\eqref{eqn:smooth_force}.

\subsubsection{Transformed dynamics} In displacement coordinates $x_t = x^1_t - x^2_t$ and $g_t = g^1_t - g^2_t$, we can write~\eqref{eqn:two_particle} as 
\begin{equation}
    \label{eqn:two_particle_displacement}
    \begin{aligned}
    \dot x_t &= 2\mu f(x_t) + v_0 g_t + 2\sqrt{\epsilon}\,\eta_x(t),\\
    \dot g_t &= -\gamma g_t + 2\sqrt{\gamma}\,\eta_g(t),
    \end{aligned}
\end{equation}
where $\eta_x(t) = \eta^1_x(t)-\eta^2_x(t)$ and $\eta_g(t) = \eta^1_g(t)-\eta^2_g(t)$ are two new independent white-noises. Equation~(\eqref{eqn:two_particle_displacement}) has the advantage that the phase space is two-dimensional, which enables us to visualize the probability current, probability flow, and the EPR globally as a function of $x$ and $g$, as well as to construct a phase portrait of the probability flow described by~\eqref{eqn:pflow}; these were depicted in Figures~\ref{fig:two_particle_intro}~\&~\ref{fig:twop_div_v}.

\subsubsection{Numerical implementation and training details} We simulated~\eqref{eqn:two_particle_displacement} using the Euler-Maruyama method with a time step of $\Delta t = 10^{-3}$, a thermal noise scale $\epsilon = 10^{-1}$, and a persistence parameter $\gamma = 10^{-1}$ over a horizon $T=100$ to generate a dataset of size $n = 10^6$.
Because there are only two particles, it suffices to use a simpler architecture than the particle transformer, and we adopted a fully connected network with four hidden layers, $256$ neurons per layer, and the GeLU activation function~\cite{hendrycks_gaussian_2023}.
We used the Adam optimizer with a learning rate of $10^{-6}$, a batch size of $4096$, and trained the network for $200000$ steps.
The learning rate was scheduled according to a cosine decay (without warmup), ending at a learning rate of $0$ after $200,000$ steps, according to the function \texttt{warmup\_cosine\_decay\_schedule} in the \texttt{optax} package.
Between each epoch, we re-sampled the dataset by taking $250$ steps of the dynamics~\eqref{eqn:two_particle_displacement}.
In~\eqref{eqn:loss}, we set $\lambda_1 = \lambda_2 = 1$.
We found best performance by learning the scores $s_x$ and $s_g$ separately, which decouples the learning in~\eqref{eqn:score_matching}.
The learning of the two scores becomes coupled through the loss term~\eqref{eqn:stationary_fpe_score_loss}.

\subsubsection{Quantitative statistics} 
\begin{figure*}[t]
\centering
\begin{overpic}[width=0.75\textwidth]{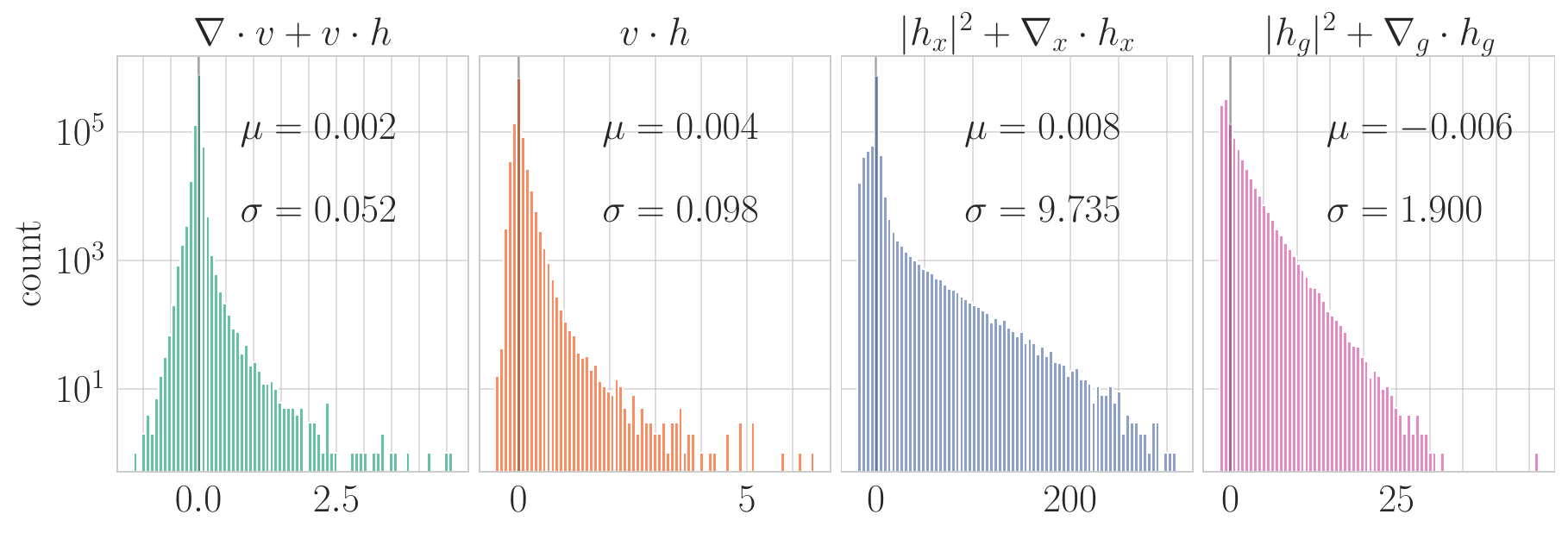}
\end{overpic}
\caption{\textbf{Two active particles on a ring: quantitative convergence statistics.}
 We consider four measures of convergence: the residual for the stationary score-based FPE~\eqref{eqn:stationary_fpe_score}, the condition $\E_{\rho}[v\cdot \nabla\log\rho] = 0$, obtained by integrating the condition $\E_{\rho}\left[\div v\right] = 0$ by parts, and the condition $\E_{\rho}\left[\nabla\log\rho + \Delta \log \rho\right] = 0$, applied separately to the individual components of the score.
 In all cases, the mean $\mu$ is small in comparison to the standard deviation $\sigma$.}
\label{fig:app:lowd_stats}
\end{figure*}

Figure~\ref{fig:two_particle}B shows histograms (over independent samples) of the system EPR, along with the contributions from the translational and orientational degrees of freedom.
The distribution of $\div v$ is centered around zero (mean $\sim 10^{-3}$), consistent with the requirement that $\E_\rho[\div v] = 0$ at stationarity.
The distribution of $\divx v_x$ is centered around a negative value and has a heavy tail, while the distribution of $\divg v_g$ is centered around a positive value of opposite sign and nearly equal magnitude.
The statistics of the residual for the score-based FPE~\eqref{eqn:stationary_fpe_score}, $v\cdot s$, $|s_x|^2 + \divx s_x$ and $|s_g|^2 + \divg s_g$ all demonstrate that both scores have been learned to high-accuracy (Figure~\ref{fig:app:lowd_stats}).

\subsubsection{Probability current}  Figure~\ref{fig:two_particle}C visualizes a decomposition of the probability current into contributions from the translational and orientational degrees of freedom $|j_g|$ and $|j_x|$, respectively, as a function of position in phase space $(x, g)$. 
This decomposition reveals an interesting structure that explains the appearance of limit cycles in the phase portrait (Figure~\ref{fig:two_particle_intro}D).
$|j_g|$ is largest near the collisions, which is where $|j_x|$ is smallest; these states correspond to a slow-moving particle with a value of $g$ that is changing sign. 
As the sign flips, the particle begins to move in the other direction.
While it does so, $g$ remains roughly constant (as indicated by a small or zero $|j_g|$), but $|j_x|$ is largest, indicating that the particle is in motion.
The system remains in this ``free particle'' state until another collision occurs on the other side of the ring, where the process repeats.
A similar line of reasoning explains the appearance of within-mode limit cycles.

\subsubsection{System EPR} 
\begin{figure*}[t]
\centering
\begin{overpic}[width=\textwidth]{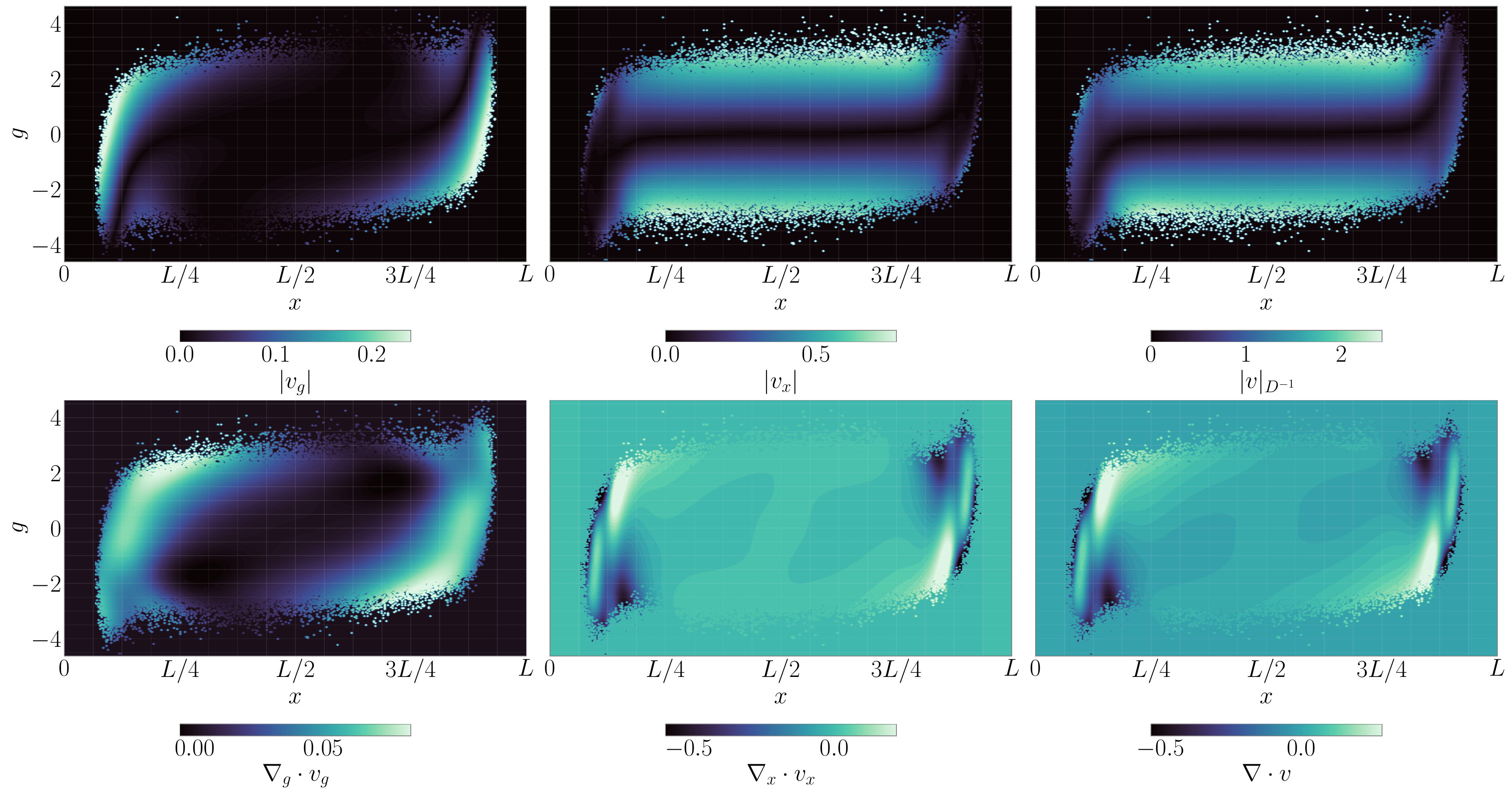}
\end{overpic}
\caption{\textbf{Two active particles on a ring: pointwise contributions.} 
Pointwise depiction of the total EPR (top) and the system EPR (bottom) over the phase space.
Unlike Figure~\ref{fig:two_particle}, these quantities are not weighted by the density $\rho$.
This gives a full picture of both definitions of the EPR, but is more dominated by outlier contributions.}
\label{fig:app:lowd_entropy_hist}
\end{figure*}

Figure~\ref{fig:two_particle}D visualizes a decomposition of the system EPR into translational and orientational contributions $\divx v_x$ and $\divg v_g$, as a function of phase space position.
We find that $\div v$ is dominated by the contribution from $\divx v_x$.
$\divg v_g$ displays features that are quite similar to $|v_g|$, and highlights the interface between the two particles. The regions of highest entropy production in the phase space correspond to when one particle exits the condensed phase.
$\divx v_x \approx \div v$ shows regions of entropy production and entropy consumption, corresponding to whether the system is entering or leaving a collision. Collectively these regions ensure that $\E\left[\div v\right] \approx 0$.
These quantities, as well as the orientational and translational contributions to the total EPR, are also visualized pointwise over phase space in Figure~\ref{fig:app:lowd_entropy_hist}, where outliers are more prominent.

\subsection{$N=64$ particles in a harmonic potential}

\subsubsection{Numerical implementation and training details} We simulated~\eqref{eqn:interacting_particles} with the potential~\eqref{eqn:trap_potential} (subject to the same smoothing as described in~\eqref{eqn:smooth_force}) using the Euler-Maruyama method with a time step of $\Delta t = 10^{-3}$, a thermal noise scale $\epsilon=10^{-1}$, and a persistence parameter $\gamma=10^{-1}$ over a horizon $T=200$ to generate a dataset of $n=10^7$ trajectories. 
We use the rectified Adam optimizer~\cite{liu_variance_2021} with a batch size of $64$ and train for $2.5\times 10^5$ iterations. 
We use the transformer architecture described in Figure~\ref{fig:network} with four layers in the transformer encoder block, four heads, and two hidden layers in the MLP embedding and decoder. 
Each MLP (embedding, decoder, and in the transformer encoder block) was selected to have $256$ neurons and the GeLU activation~\cite{hendrycks_gaussian_2023}.
The embedding dimension $d_e=256$ for the transformer encoder block.
We employed a cosine warmup and annealing schedule starting and ending at a learning rate of $0$ and peaking at a learning rate of $10^{-4}$ after $10^4$ iterations, again according to the function \texttt{warmup\_cosine\_decay\_schedule} in the \texttt{optax} package.
Similar to the previous system, we learn $s_x$ and $s_g$ separately, but couple the learning with $\lambda_1 = \lambda_2 = 1$.
\newchange{Each experiment was allowed to run for 48 hours on a single Nvidia a100 GPU.}

\subsubsection{Boltzmann-informed initialization} For $v_0 = 0$, the system reduces to a Langevin dynamics with stationary measure $\rho_{\mathsf{ss}}(x, g) = \frac{1}{Z} e^{-\Phi(x)/\epsilon}e^{-|g|^2/2}$ where $Z$ is the partition function.
In this case, the stationary score can be computed exactly, giving the relations
\begin{equation}
    \label{eqn:boltzmann_score}
    \begin{aligned}
       \nabla_x \log \rho_{\mathsf{ss}}(x, g) &= -\frac{1}{\epsilon}\nabla_x \Phi(x),\\ 
       \nabla_g \log \rho_{\mathsf{ss}}(x, g) &= -g.
    \end{aligned}
\end{equation}
~\eqref{eqn:boltzmann_score} shows that $v=0$ and $\div v = 0$ for $v_0 = 0$.
We expect that as $v_0$ is tuned smoothly away from zero, the score $\nabla\log\rho(x, g)$ can be written as a correction to $\nabla\log\rho_{\mathsf{ss}}(x, g)$.
In order to take this into account, we parameterize the score as $s_x = -\frac{1}{\epsilon}\nabla_x\Phi(x) + N_x$ and $s_g= -g + N_g$ with $N_x$ and $N_g$ learnable neural networks; this obviates the need to learn $\frac{1}{\epsilon}\nabla_x\Phi(x)$ explicitly.
Because the neural network weights are initialized randomly around zero, we can view this procedure as shifting the initialization of the network in function space to a ball around the Boltzmann score.
Empirically, we find significant improvements in quantitative accuracy metrics when making use of this Boltzmann-informed initialization.

\subsubsection{Additional results}
\begin{figure*}[t]
\begin{tabular}{c}
    \begin{overpic}[width=\textwidth]{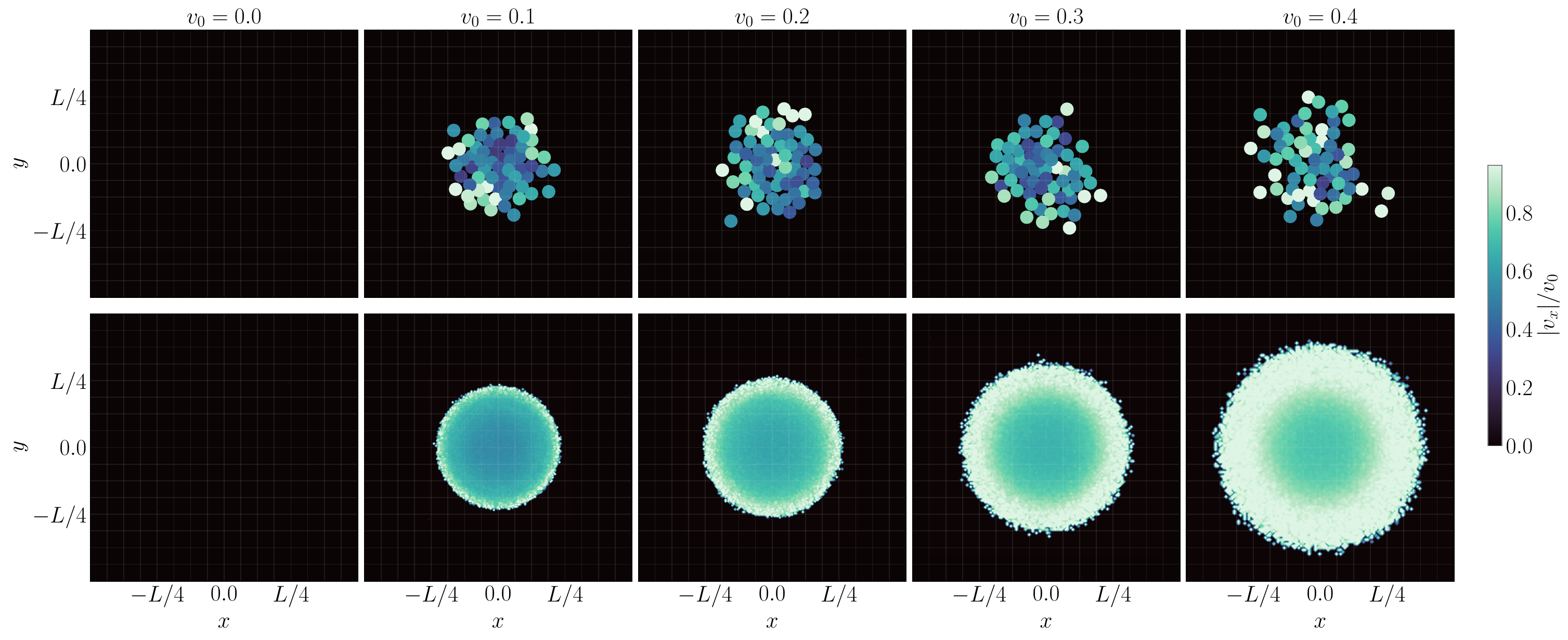}
    \put(3, 40){\textbf{A}}
    \end{overpic}\\
    \begin{overpic}[width=\textwidth]{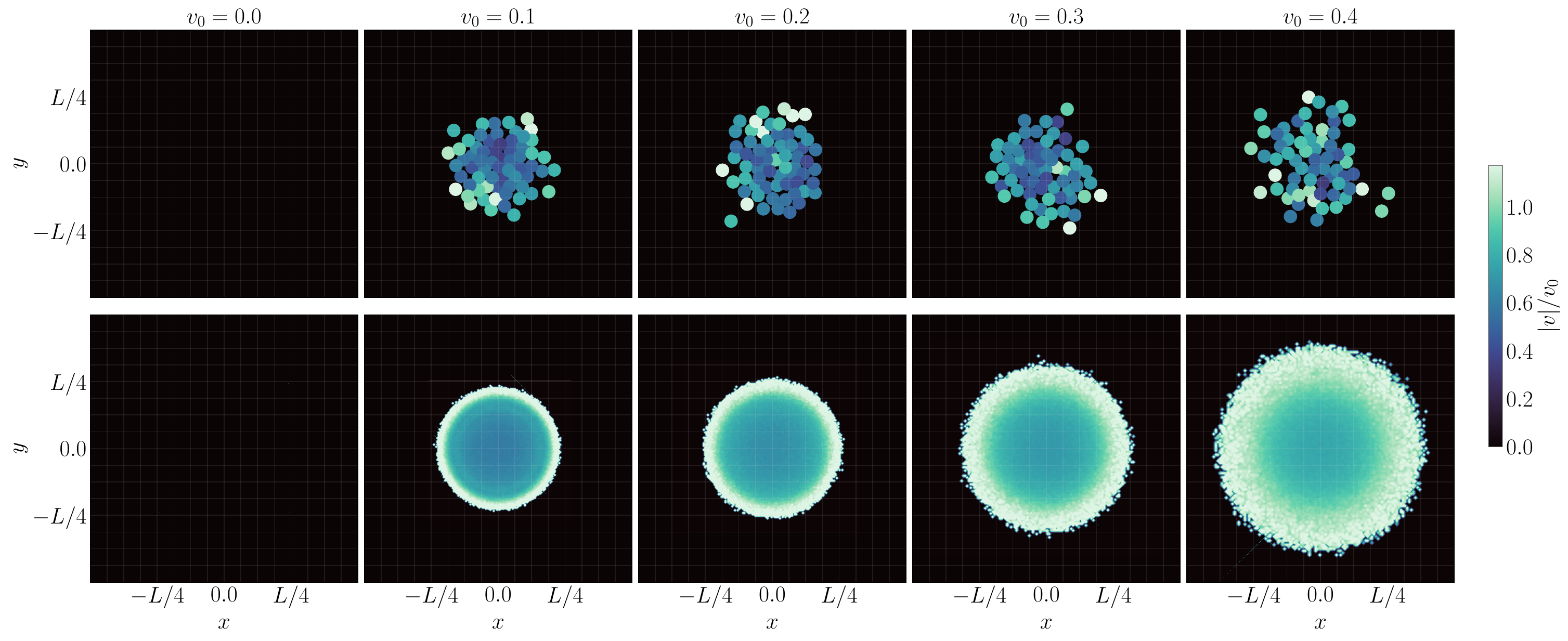}
    \put(3, 40){\textbf{B}}
    \end{overpic}
\end{tabular}
    \caption{\textbf{64 swimmers in a trap: total EPR} 
    (A) [Top] Individual particle contributions from the translational degrees of freedom $|\vxi|$ to the total EPR.
    [Bottom] Spatial map of the translational component of the total EPR $|v_x|$, obtained by averaging the data in the top row. Both the particle contributions and the spatial map display a signal concentrated on the interface, but both signals are less clear than the orientational degrees of freedom considered in the main text.
    (B) [Top] Individual particle contributions to the total EPR $|\vi|$. [Bottom] Spatial map, obtained by averaging the data in the top row. Similar to (A), there is a concentrated signal on the interface, but it less clear than the orientational contributions alone.}
    \label{fig:app:64_omitted_1}
\end{figure*}

\begin{figure*}[t]
\begin{tabular}{c}
    \begin{overpic}[width=\textwidth]{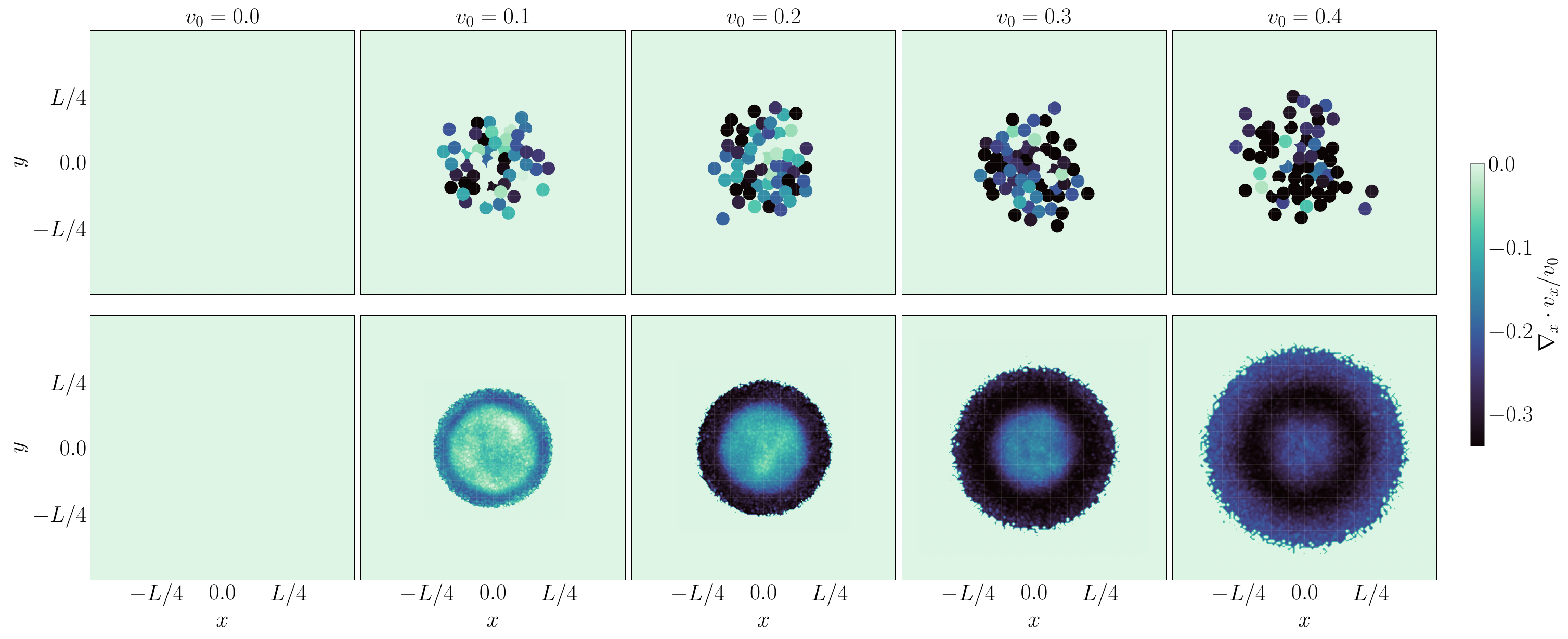}
    \put(3, 40){\textbf{A}}
    \end{overpic}\\
    \begin{overpic}[width=\textwidth]{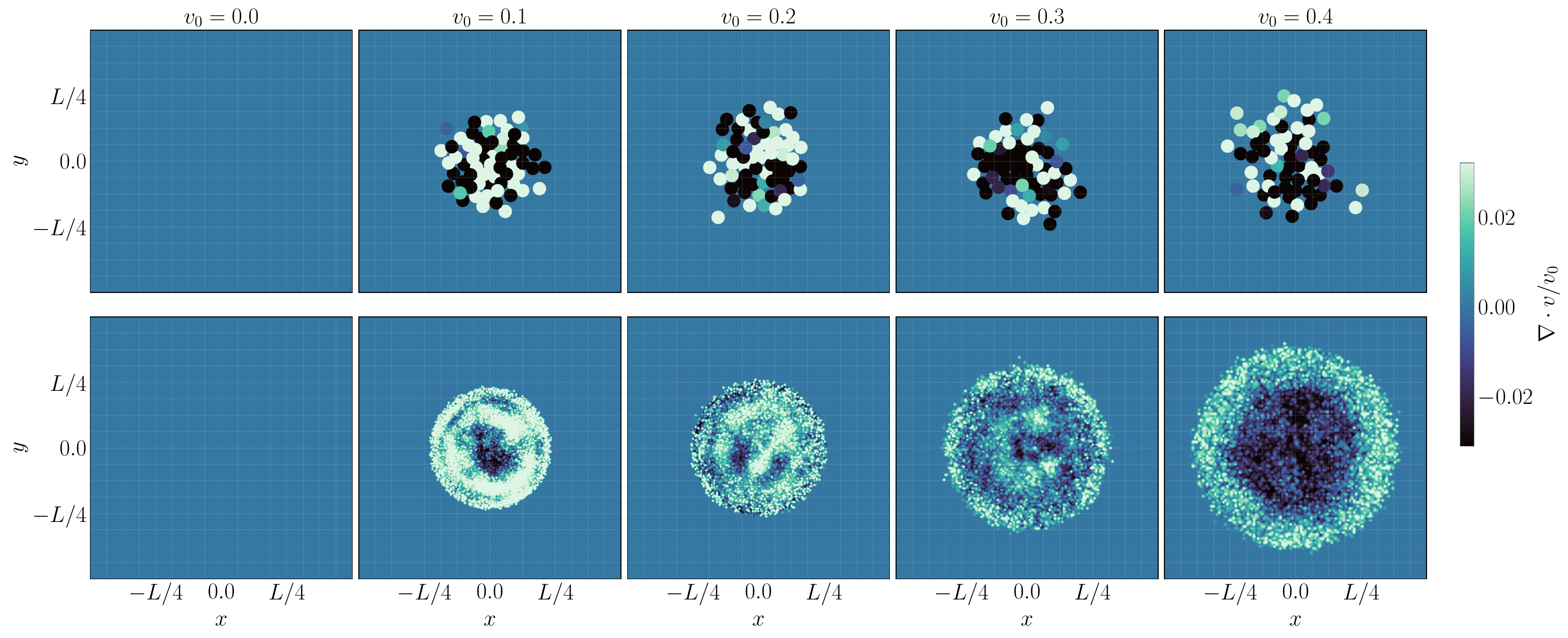}
    \put(3, 40){\textbf{B}}
    \end{overpic}
\end{tabular}
    \caption{\textbf{64 swimmers in a trap: system EPR} 
    (A) [Top] Individual particle contributions from the translational degrees of freedom $\divxi \vxi$ to the total EPR.
    [Bottom] Spatial map of the translational component of the system EPR $\divx \vx$, obtained by averaging the data in the top row. The signal is roughly inverted with respect to $\divg \vg$, so that $\divx \vx \approx -\divg \vg$; this is clearest at the level of the spatial map.
    (B) [Top] Individual particle contributions to the system EPR $\divi \vi$. [Bottom] Spatial map, obtained by averaging the data in the top row. Because $\divxi \vxi \approx - \divgi \vgi$, the particle contributions nearly vanish up to small-scale fluctuations around zero. These fluctuations are an order of magnitude lower than the values of $\divgi \vgi$ and $\divxi \vxi$ (see colorbar).}
    \label{fig:app:64_omitted_2}
\end{figure*}

\begin{figure*}[!b]
\centering
\begin{tabular}{c}
\begin{overpic}[width=0.9\textwidth]{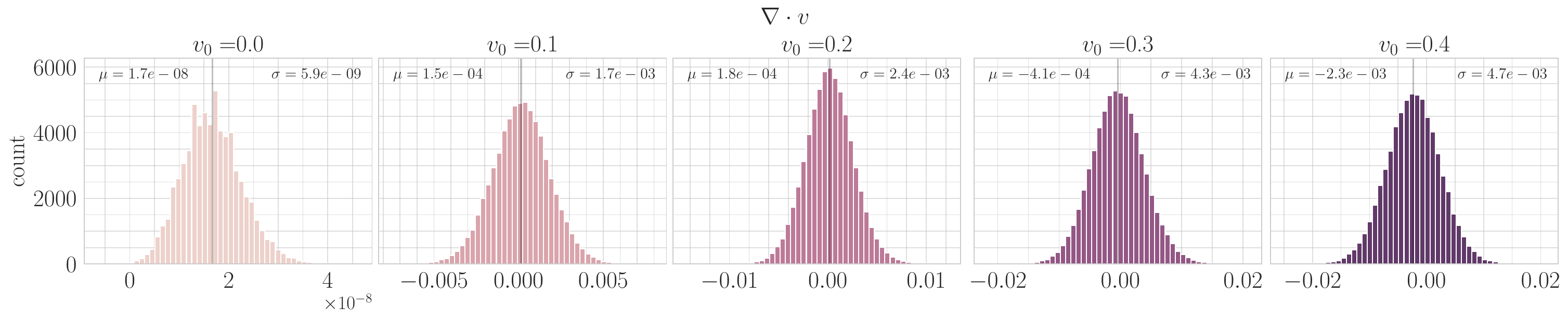}
\put(3, 19){\textbf{A}}
\end{overpic}\\
\begin{overpic}[width=0.9\textwidth]{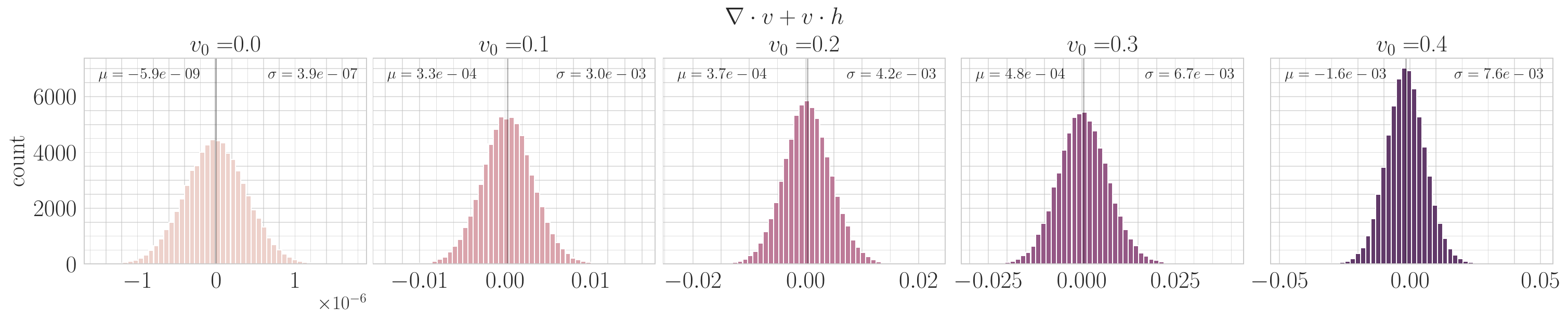}
\put(3, 19){\textbf{B}}
\end{overpic}\\
\begin{overpic}[width=0.9\textwidth]{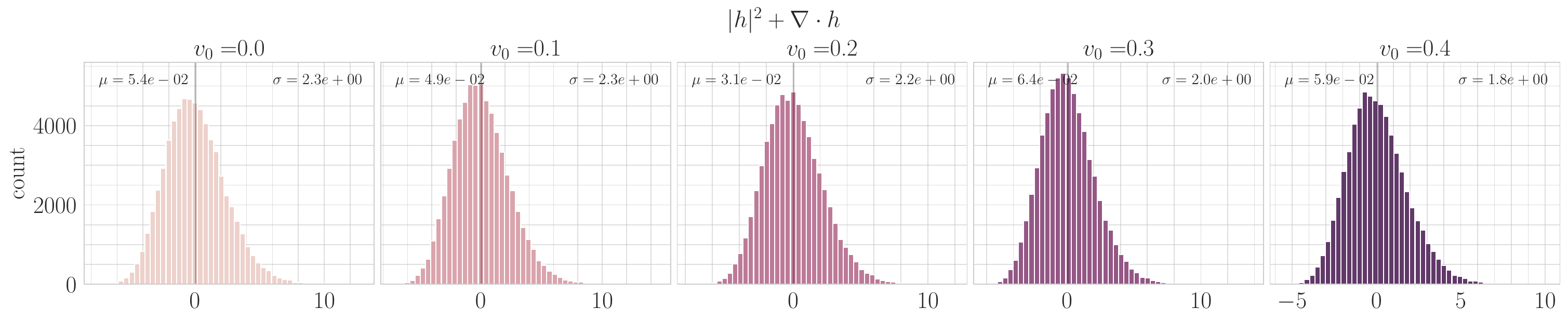}
\put(3, 19){\textbf{C}}
\end{overpic}\\
\begin{overpic}[width=0.9\textwidth]{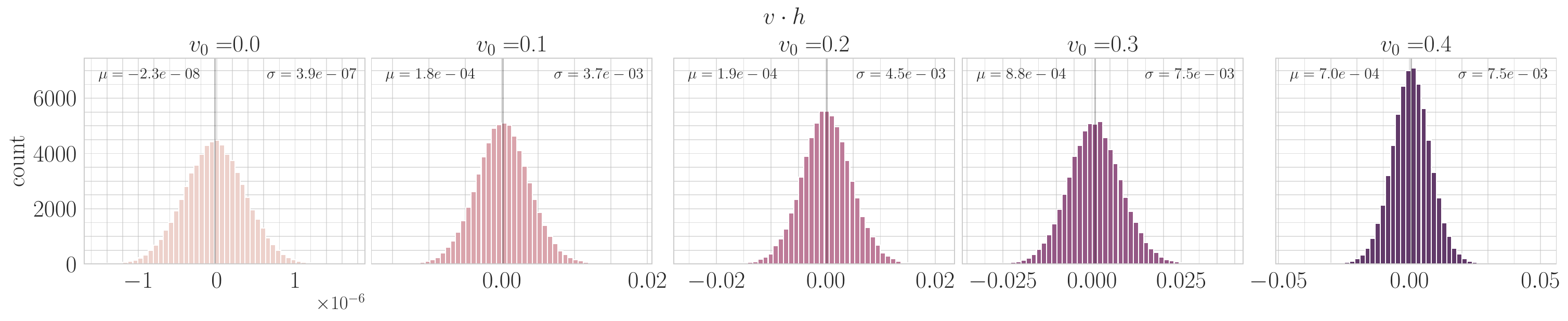}
\put(3, 19){\textbf{D}}
\end{overpic}
\end{tabular}
\caption{\textbf{64 Particles: quantitative convergence statistics.}
Because each quantity considered here consists of a sum of $N$ terms, we normalize by the number of particles.
(A) Distribution of $\nabla\cdot v = \sum_{i=1}^N \divi \vi$ over snapshots of the system. By stationarity of the Gibbs entropy, $\partial_t \E_{\rho}\left[\log \rho\right] = 0$, we require that $\E_{\rho}\left[\nabla\cdot v\right] = 0$. For each value of $v_0$, we find a small mean value $\mu$ and a small standard deviation $\sigma$. This indicates that $\E_{\rho}[\div v] \approx 0$. Interestingly, the small value of $\sigma$ indicates $\div v \approx 0$ on each snapshot, which is not strictly required for convergence.
(B) Distribution of the score-based stationary FPE (\eqref{eqn:stationary_fpe_score}) residual, $\nabla\cdot v + v\cdot s$, over system snapshots. To satisfy the stationary FPE, this quantity should be zero pointwise. We find low mean values $\mu$ and low standard deviations $\sigma$ in all cases, indicating that this is approximately satisfied.
(C) Distribution of $|\nabla\log\rho|^2 + \Delta \log\rho$. By integration by parts, we require that $\E_{\rho}\left[|\nabla\log\rho|^2 + \Delta \log\rho\right] = 0$. We find a low mean value in all cases. The standard deviation is orders of magnitude higher, indicating that this quantity does not vanish pointwise (which is not required for convergence).
(D) Distribution of $v \cdot \nabla\log\rho$ over snapshots. By integration by parts, $\E_{\rho}\left[\div v\right] = \E_{\rho}\left[-v \cdot \nabla\log\rho\right] = 0$, so that this quantity should vanish in expectation by stationarity of the Gibbs entropy. Consistent with (A), we find a low $\mu$ and a low $\sigma$, indicating that it approximately vanishes both in expectation and pointwise.}
\label{fig:64_stats}
\end{figure*}

\begin{figure*}[!t]
    \includegraphics[width=\textwidth]{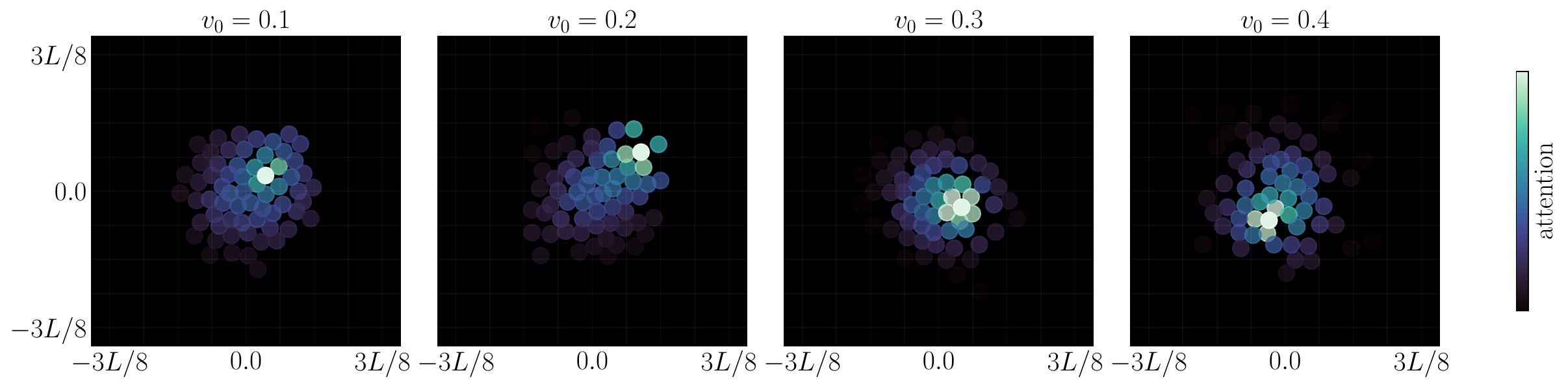}
    \caption{\textbf{64 swimmers in a harmonic trap: attention map.} The attention map, which quantifies the relevance of other particles for the score of a given particle, for several example swimmers (circled in white) as a function of $v_0$.
    The network learns a spatially-local attention pattern in each case, indicating that the score for each particle is primarily determined by its local neighborhood.
    Despite this spatial-locality, the learned attention pattern is longer-range than the physical interaction potential.}
    \label{fig:64_attention}
\end{figure*}

Figures~\ref{fig:app:64_omitted_1}~\&~\ref{fig:app:64_omitted_2} display the quantities $|v_x|, \divx v_x, |v|$ and $\div v$ that were omitted from the main text, while Figure~\ref{fig:64_stats} displays the quantitative statistics we used to assess convergence (see captions for discussion).
Figure~\ref{fig:64_attention} shows the attention map of the learned network (see discussion in main text).

\subsection{Motility-induced phase separation}
\begin{figure*}[t]
\centering
\begin{tabular}{c}
\begin{overpic}[width=\textwidth]{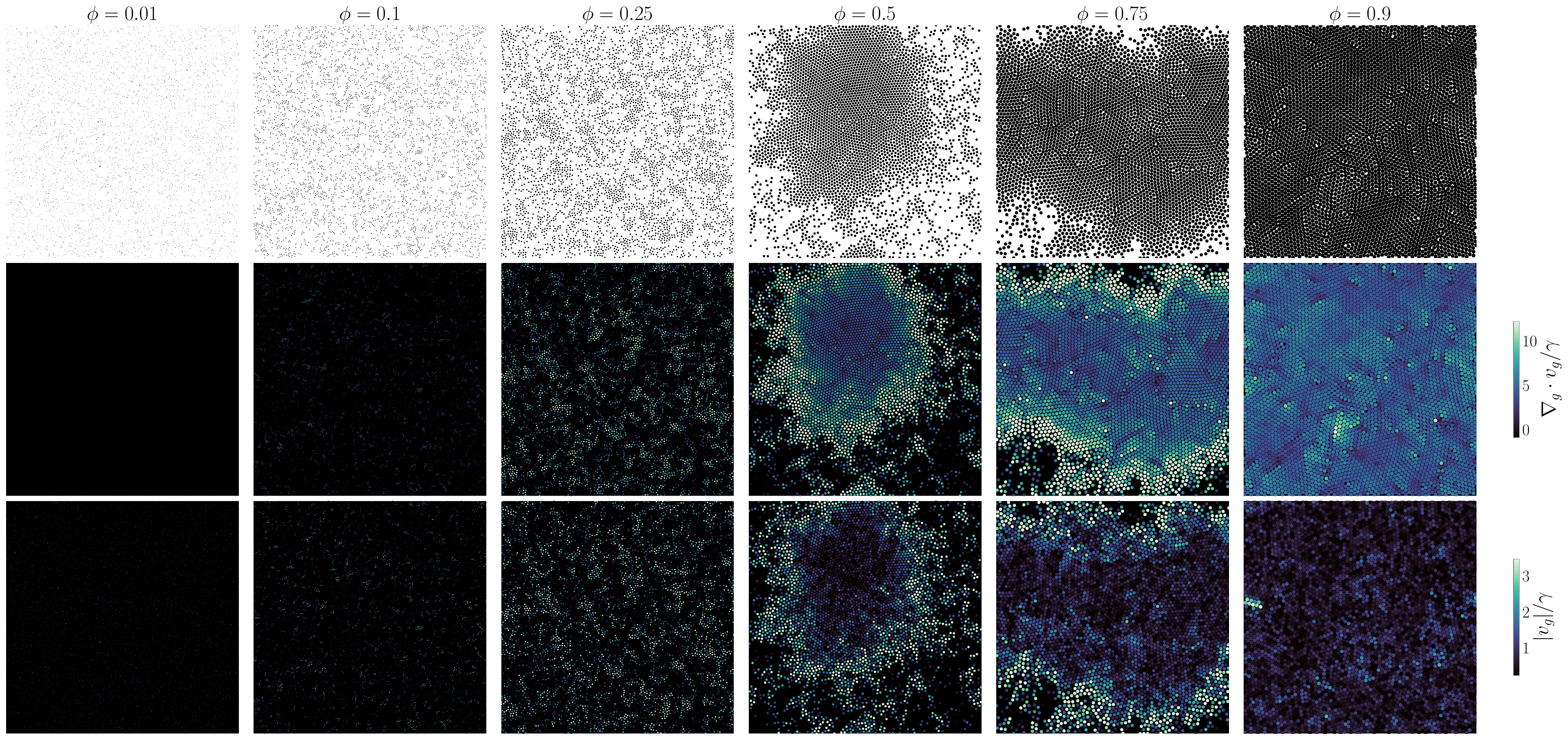}
\end{overpic}
\end{tabular}
\caption{\textbf{Motility-induced phase separation. Packing fraction generalization.} Packing fraction generalization with $N=4096$.}
\label{fig:mips_phi_transfer_N4096}
\end{figure*}

\begin{figure*}[t]
\centering
\begin{tabular}{c}
\begin{overpic}[width=\textwidth]{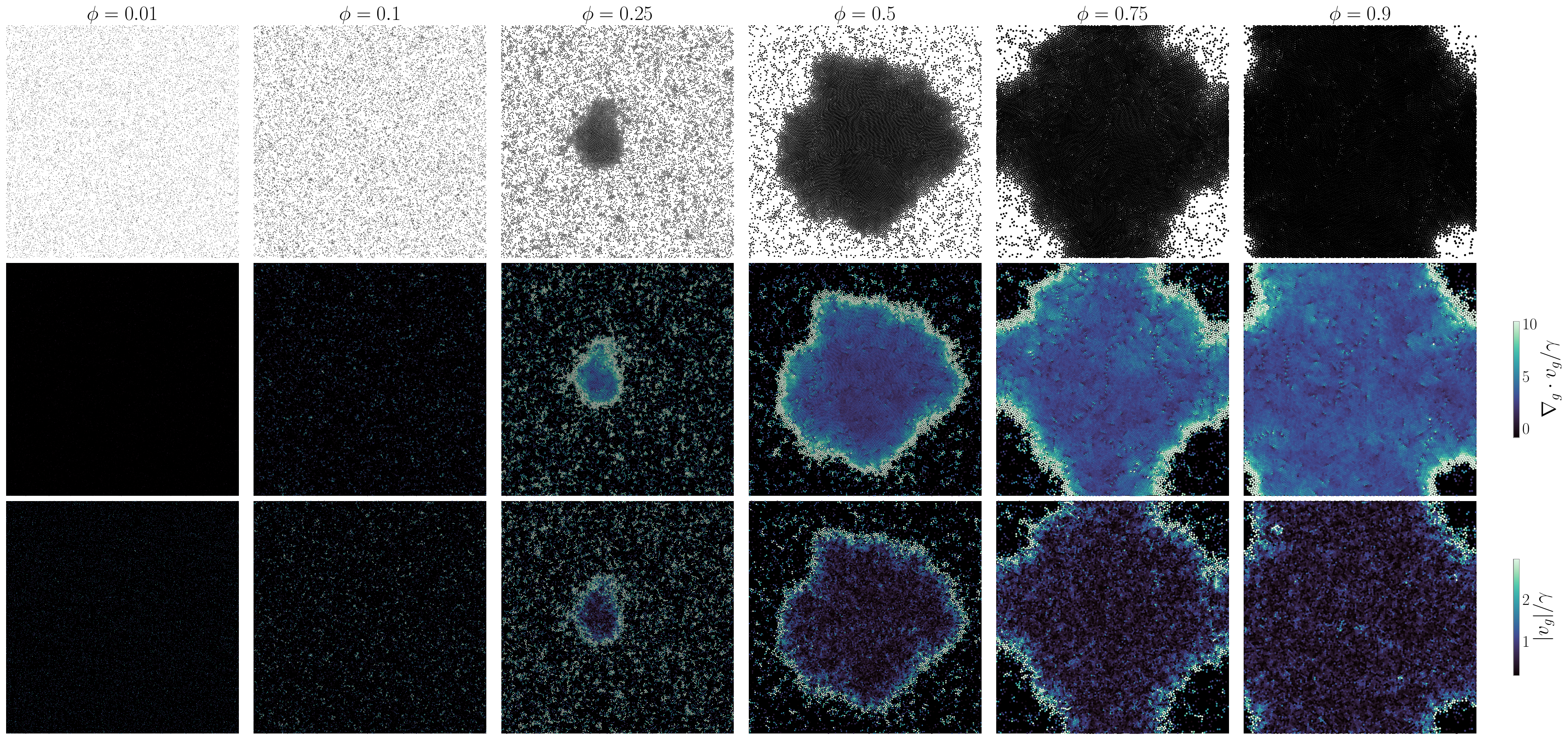}
\end{overpic}
\end{tabular}
\caption{\textbf{Motility-induced phase separation. Packing fraction generalization.} Packing fraction generalization with $N=16384$.}
\label{fig:mips_phi_transfer_N16384}
\end{figure*}

\begin{figure*}[t]
\centering
\begin{tabular}{c}
\begin{overpic}[width=\textwidth]{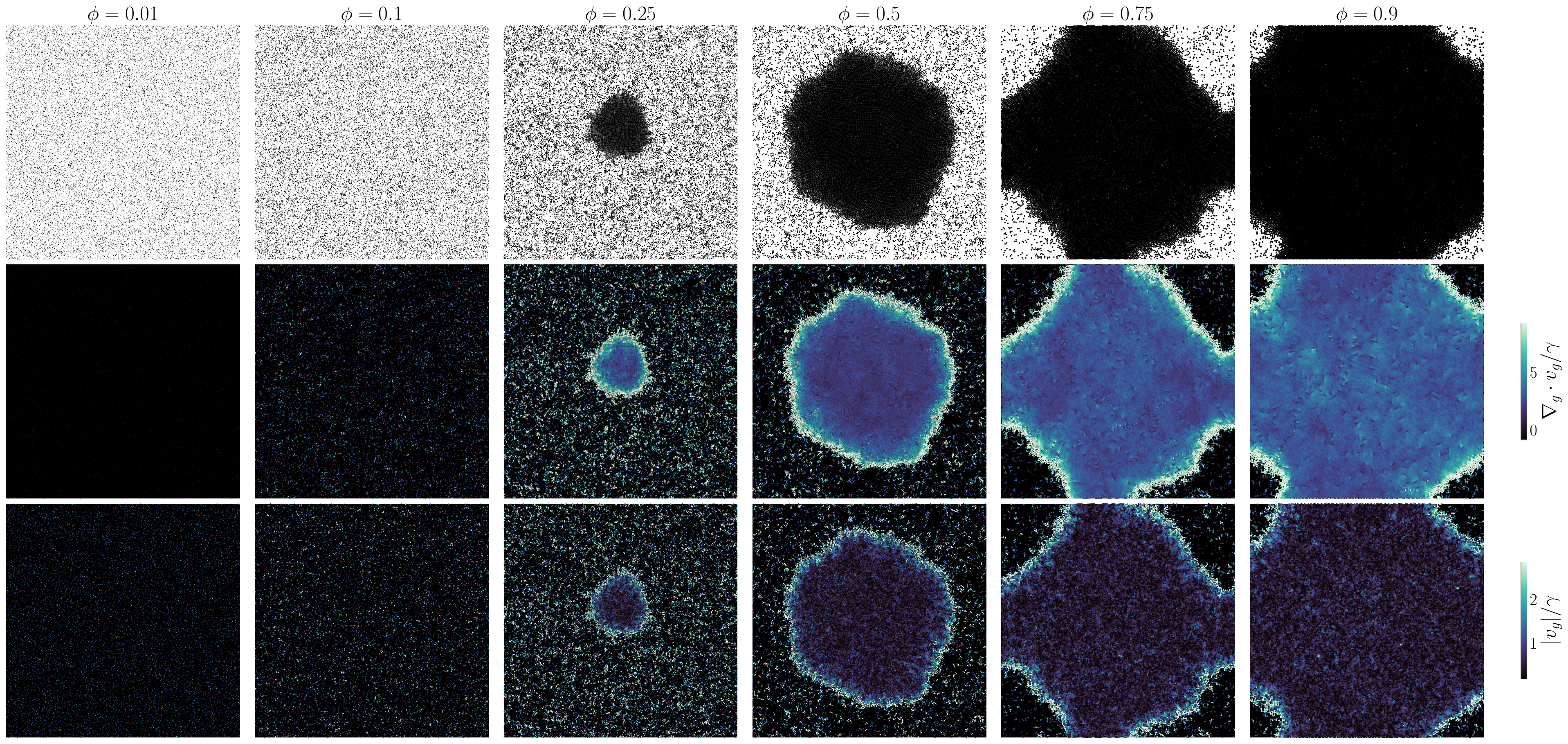}
\end{overpic}
\end{tabular}
\caption{\textbf{Motility-induced phase separation. Packing fraction generalization.} Packing fraction generalization with $N=32768$.}
\label{fig:mips_phi_transfer_N32768}
\end{figure*}

\subsubsection{Dataset generation}
We simulated~\eqref{eqn:interacting_particles} using the Euler-Maruyama method in the athermal regime with $\epsilon = 0$, a persistence parameter $\gamma = 10^{-4}$, a timestep of $\Delta t = 10^{-1}$, a self-propulsion velocity $v_0 = 0.025$ and a packing fraction of $\phi=0.5$. 
These parameters were chosen to ensure the system was in the motility-induced phase separation regime.
Due to the large number of interacting particles, it is computationally expensive to generate a large dataset of thermalized independent samples, as was done in the previous two examples.
Instead, we simulated $12$ independent initial conditions until time $1/\gamma$ and verified that a cluster had formed.
Then we generated a trajectory of $10,000$ samples per initial condition separated by a time lag $\tau = 0.02/\gamma = 20$.
\newchange{Each trajectory was generated in around 24 hours on a single Nvidia RTX8000 GPU.}

\subsubsection{Network architecture} We define the score $s^i$ for each particle $i$ using the architecture in Figure~\ref{fig:network}; to reduce time and memory complexity, the input to the encoder block is \textit{only} the embedding for particle $i$ along with the embeddings for the $N_{\text{neighbors}}$ nearest neighbors.
The output of the network is then a matrix of size $(N_{\text{neighbors}}+1) \times d$.
These represent the score $s^i$ along with a set of rich features for each neighboring particle used to compute $s^i$; the latter are discarded, so that as $N_{\text{neighbors}} \rightarrow N$, we recover the architecture used for $N=64$ defined at the level of the full system.
To ensure equivariance amongst particles, we share weights in the embedding, encoder, and decoder between all particles.

\subsubsection{Scale separation and singularity} 
In the athermal limit $\epsilon=0$, we found it more challenging to maintain the zero system EPR condition $\E_\rho\left[\div v\right] = 0$ than in the thermal regime.
To see analytically why this may be the case, note that
\begin{equation}
\begin{aligned}
    v_x(x, g) &= -\nabla_x\Phi(x),\\
    v_g(x, g) &= -\gamma g - \gamma \nabla_g \log\rho(x, g).
\end{aligned}
\end{equation}
for $\epsilon = 0$.
Hence, 
\begin{equation}
    \div v = -\Delta_x \Phi(x) - \gamma d - \gamma \Delta_g \log \rho(x, g),
\end{equation}
so that the condition $\E_{\rho}\left[\div v\right] = 0$ reads
\begin{equation}
    \E_\rho\left[-\Delta_x\Phi(x)\right] = \gamma d + \gamma \E_\rho\left[\Delta_g \log \rho(x, g)\right].
\end{equation}
The Laplacian term $-\Delta_{x}\Phi(x)$ (for the hard cutoff potential) is given by
\begin{equation}
    -\Delta_x \Phi(x) = \sum_{i, j=1}^N \left(\frac{2r(d-1)}{|x^i-x^j|} - d\right)\Theta\left(2r-|x^i-x^j|\right).
\end{equation}
In the solid, $|x^i-x^j| \sim 2r$, so that each term in the sum is $\calO(1)$, and the overall Laplacian can be as large as $\calO(N)$.
For the small $\gamma \sim 10^{-4}$ values needed to induce MIPS, this creates a difficult scale-separation problem for a learning algorithm, as $\Delta_g\log \rho(x, g)$ must be of order $\mathcal{O}(1/\gamma)$ to cancel the contribution from $\Delta_x\Phi(x)$.
Moreover, because the stationary density for $v_0 = 0$ is singular, 
\begin{equation}
    \rho_{\text{ss}}(x, g) = \frac{1}{Z}\delta\left(\nabla\Phi(x)\right)\exp\left(-\frac{1}{2}|g|^2\right),
\end{equation}
it is not clear how to initialize to avoid this problem. By contrast, in the thermal setting
\begin{equation}
\begin{aligned}
    v_x(x, g) &= -\nabla_x\Phi(x) - \epsilon \nabla_x \log \rho(x, g),\\
    v_g(x, g) &= -\gamma g - \gamma \nabla_g\log\rho(x, g).
\end{aligned}
\end{equation}
Stationarity of the Gibbs entropy $\E_{\rho}\left[\div v \right] = 0$ becomes
\begin{equation}
\E_{\rho}\left[-\Delta_x \Phi(x)\right] -\epsilon\E_{\rho}\left[\Delta_x\log\rho(x, g)\right] = - \gamma d - \gamma\E_{\rho}\left[\Delta_g\log\rho(x, g)\right].
\end{equation}
Here, the scale-separation problem is offloaded to the score in $x$.
This can be seen most clearly by observing that the stationary score is now well-defined for $v_0 = 0$,
\begin{equation}
\begin{aligned}
    \nabla_x\log\rho_{\text{ss}} &= -\frac{1}{\epsilon}\nabla\Phi(x),\\
    \nabla_g\log\rho_{\text{ss}} &= -g.
\end{aligned}
\end{equation}
We exploited this observation in the $N=64$ system via the introduction of a Boltzmann-informed initialization.
However, to ensure a well-defined MIPS cluster for finite~$\epsilon$, we must take~$\epsilon$ very small.
While the Boltzmann-informed initialization avoids the need to \textit{learn} a term of order $1/\epsilon$, such a term is still included in the score parameterization, which makes the loss function poorly conditioned.

\subsubsection{Loss function}
To sidestep these issues, we make use of the denoising loss function described in Section~\ref{sec:denoising} to learn only the score in~$g$.
Importantly, we add noise in the $x$ variables, which corresponds to taking a step along a thermal dynamics (where the $x$ score is nearly singular, but the $g$ score is well-behaved) starting from athermal data.
This approach ensures that there is a well-defined boundary of the cluster, avoids the need to learn a term of large magnitude, and is structured so that cancellation of the equilibrium force term $-\nabla_x\Phi(x)$ to ensure $\E\left[\div v\right] = 0$ will happen in the score in $x$, which we do not learn.
We have already seen in the $N=64$ example that $\divx v_x \approx -\divg v_g$ and that there is a rich signal contained in $v_g$ alone, which further justifies restricting to $v_g$.

\subsubsection{Training details}
We set $N_{\text{neighbors}} = 64$ and otherwise use identical network hyperparameters as in the example of $64$ active swimmers.
We use a cosine warmup and annealing schedule, peaking at a learning rate of $10^{-6}$ after $10,000$ steps and ending at a learning rate of $0$ after $500,000$ steps.
Because all terms in the denoising loss split into contributions from individual particles, and because the network is defined at the particle level, we consider the particles as datapoints and batch over the particles themselves (ensuring that batches never include particles from two separate snapshots of the system).
We use a batch size of $1024$ particles.
In the denoising loss, we use a noise with standard deviation $\sigma = 0.01$ and neglect the drift term $b$ in the denoising loss because it is higher-order in $\Delta t$.
\newchange{The training was allowed to run for 24 hours on a single Nvidia a100 GPU.}

\subsubsection{Additional results}
Figures~\ref{fig:mips_phi_transfer_N4096}-\ref{fig:mips_phi_transfer_N32768} show the ability of the learned network (trained with $N=4096$ and packing fraction $\phi=0.5$) to generalize to other values of $\phi$ for $N=4096, N=16384$, and $N=32768$; only $N=8192$ was considered in the main text. In all cases, the predictions are physically reasonable.

\clearpage
\bibliographystyle{unsrtnat}
\bibliography{refs}

\end{document}